\documentclass[fleqn,usenatbib]{mnras}

\usepackage{newtxtext,newtxmath}
\usepackage[T1]{fontenc}

\usepackage{graphicx}
\usepackage{amsmath}	       
\usepackage{amssymb}	        
\usepackage{xspace}

\newcommand{\mcore}{\ensuremath{M_{\rm core}}\xspace}
\newcommand{\menv}{\ensuremath{M_{\rm env}}\xspace}
\newcommand{\xc}{\ensuremath{X_{\rm c}}\xspace}
\newcommand{\yc}{\ensuremath{Y_{\rm c}}\xspace}
\newcommand{\msun}{\ensuremath{\mathrm{M}_{\odot}}\xspace}
\newcommand{\aov}{\ensuremath{\alpha_{\rm ov}}\xspace}
\newcommand{\teff}{\ensuremath{T_{\rm eff}}\xspace}
\newcommand{\amlt}{\ensuremath{\alpha_{\rm mlt}}\xspace}
\newcommand{\fhshell}{\ensuremath{F_{\rm H-shell}}\xspace}
\newcommand{\mtotal}{\ensuremath{M_{\rm total}}\xspace}
\newcommand{\mucore}{\ensuremath{\mu_{\rm core}}\xspace}
\newcommand{\asc}{\ensuremath{\alpha_{\rm sc}}\xspace}
\newcommand{\lnuc}{\ensuremath{L_{\rm nuc}}\xspace}
\newcommand{\ltotal}{\ensuremath{L_{\rm total}}\xspace}
\newcommand{\lsun}{\ensuremath{\mathrm{L}_{\odot}}\xspace}
\newcommand{\rsun}{\ensuremath{\mathrm{R}_{\odot}}\xspace}
\newcommand{\epsnuc}{\ensuremath{\epsilon_{\rm nuc}}\xspace}
\newcommand{\tc}{\ensuremath{T_{\rm central}}\xspace}
\newcommand{\lhshell}{\ensuremath{L_{\rm H-shell}}\xspace}
\newcommand{\lcore}{\ensuremath{L_{\rm core}}\xspace}
\newcommand{\mtrans}{\ensuremath{M_{\rm transition}}\xspace}
\newcommand{\logg}{\ensuremath{\log g}\xspace}

\newcommand{\fovcore}{\ensuremath{f_{\rm ov,\,core}}\xspace}
\newcommand{\fovenv}{\ensuremath{f_{\rm ov,\,env}}\xspace}
\newcommand{\fovshell}{\ensuremath{f_{\rm ov,\,shell}}\xspace}

\newcommand{\logllsun}{\ensuremath{\log \mathrm{L}/\lsun}}
\newcommand{\logrrsun}{\ensuremath{\log \mathrm{R}/\rsun}}
\newcommand{\logteff}{\ensuremath{\log T_{\rm eff}}}
\newcommand{\mesa}{\textsc{mesa}\xspace}
\newcommand{\snap}{\textsc{snapshot}\xspace}

\title[{\sc snapshot}: Stellar Structure Models]{{\sc snapshot}: Connections between Internal and Surface Properties of Massive Stars}

\author[Farrell et al.]{
Eoin J. Farrell,$^{1}$\thanks{E-mail: efarrel4@tcd.ie}
Jose H. Groh,$^{1}$
Georges Meynet,$^{2}$
J.J. Eldridge,$^{3}$
\newauthor
Sylvia Ekstr{\"o}m,$^{2}$
Cyril Georgy$^{2}$
\\
$^{1}$School of Physics, Trinity College Dublin, The University of Dublin, Dublin, Ireland\\
$^{2}$Geneva Observatory, University of Geneva, Chemin des Maillettes 51, 1290 Sauverny, Switzerland\\
$^{3}$Department of Physics, Private Bag 92019, University of Auckland, New Zealand
}

\date{Accepted XXX. Received YYY; in original form ZZZ}

\pubyear{2020}

\begin{document}
\label{firstpage}
\pagerange{\pageref{firstpage}-\pageref{lastpage}}
\maketitle

\begin{abstract}
We introduce \snap, a technique to systematically compute stellar structure models in hydrostatic and thermal equilibrium based on 3 structural properties -- core mass \mcore, envelope mass \menv and core composition. This approach allows us to connect these properties of stellar interiors to the luminosity and effective temperature \teff in a more systematic way than with stellar evolution models. We compute core-H burning models with total masses of $\mtotal = 8$ to $60 \msun$ and central H mass fractions from 0.70 to 0.05. Using these, we derive an analytical relationship between \mcore, \mtotal and central H abundance that can be readily used in rapid stellar evolution algorithms. In contrast, core-He burning stars can have a wide range of combinations of \mcore, \menv and core compositions. We compute core-He burning models with $\mcore = 2 - 9 \msun$, $\menv = 0 - 50 \msun$ and central He mass fractions of 0.50 and 0.01. Models with $\mcore/\mtotal$ from 0.2 to 0.8 have convective envelopes, low \teff and will appear as red supergiants. For a given \mcore, they exhibit a small variation in luminosity (0.02 dex) and \teff ($\sim 400\,\mathrm{K}$) over a wide range of \menv ($\sim 2 - 20\,\msun$). This means that it is not possible to derive red supergiant masses from luminosities and \teff alone. We derive the following relationship between \mcore and the total luminosity of a red supergiant during core He burning: $\log \mcore \simeq 0.44\logllsun - 1.38$. At $\mcore/\mtotal \approx 0.2$, our models exhibit a bi-stability and jump from a RSG to a BSG structure. Our models with $\mcore/\mtotal > 0.8$, which correspond to stripped stars produced by mass loss or binary interaction, show that \teff has a strong dependence on \menv, \mcore and the core composition. We constrain the mass of one of these stripped stars in a binary system, HD 45166, and find it to be less than its estimated dynamical mass. When a large observational sample of stripped stars becomes available, our results can be used to constrain their \mcore, \menv, mass-loss rates and the physics of binary interaction.
\end{abstract}

\begin{keywords}
stars: evolution -- stars: massive -- stars: interiors -- stars: atmospheres
\end{keywords}

%__________________________________________________________________

\section{Introduction}

The evolution of massive stars is dominated by the effects of mass loss, rotation, convection and binary interaction. Because of  uncertainties in these physical mechanisms, we have a limited understanding on how massive stars evolve, die, and affect their host galaxies. Mass loss, convection, rotation and binary interaction affect the evolution of massive stars by modifying their internal and surface properties. Depending on their impact, a star can evolve to have different combinations of core mass, envelope mass and core composition. These are important properties of stellar structure and they affect the observable surface properties of massive stars. This connection is the main topic of this paper.

Mass loss by stellar winds has a dramatic effect on the evolution of massive stars \citep[e.g.,][]{Chiosi:1986, Maeder:2012, Smith:2014}. Depending on the assumptions in a stellar evolution model for mass-loss rates, a star can evolve to very different regions of the Hertzsprung-Russell (HR) diagram \citep{Meynet:1994, Vanbeveren:2007} leading to different outcomes at the end of their lives \citep{Eldridge:2004, Georgy:2012, Meynet:2015, Renzo:2017}. While a theoretical framework for computing mass-loss rates exists for the main sequence (MS; \citealt{Vink:2001}), mass-loss rates are uncertain for other regions of the HR diagram \citep[e.g.,][]{DeBeck:2010, Mauron:2011, Beasor:2018}. Additionally, there is increasing theoretical and observational evidence that massive stars may undergo eruptive and explosive mass-loss events during the late nuclear burning stages before core-collapse \citep[e.g.,][]{Kotak:2006, Smith:2007, Pastorello:2007, Gal-Yam:2009, Fraser:2013, Gal-Yam:2014, Smith:2014, Groh:2014a, Fuller:2017, Yaron:2017, Boian:2018}. Despite their potential significance, the nature of these eruptions is very uncertain and they are not usually accounted for in stellar evolution models.

Another important, yet uncertain component of stellar evolution models is the behaviour of their convective cores. Convection is usually parameterised by an overshooting parameter, \aov, which is calibrated from observations. Different values for \aov are adopted in different models \citep[e.g.,][]{Brott:2011, Ekstrom:2012, Choi:2016, Higgins:2019}, depending on the calibration method. Furthermore, the dependence of \aov on mass and metallicity is also uncertain \citep{Castro:2014a}. For a given initial mass, the choice of \aov can significantly affect the mass of the convective core during the MS phase and the mass of the He-core during subsequent burning stages. As well as the value adopted for \aov, the exact implementation of core-overshooting in the stellar evolution code can also modify a star's evolution \citep{Martins:2013}.

Stellar rotation can also significantly affect the evolution of massive stars \citep[e.g.,][]{Maeder:2000, Meynet:2000, Heger:2000, Heger:2005, Maeder:2009, Brott:2011, Brott:2011a, Chieffi:2013}. Massive stars exhibit a range of rotational velocities \citep{Huang:2010, Hunter:2008, Ramirez-Agudelo:2013, Ramirez-Agudelo:2015, Dufton:2019}. For sufficiently low velocities, rotation is expected to have a qualitatively similar effect on the evolution of the convective core mass to core overshooting. The behaviour of stellar structure under extreme rotation has recently been suggested to be significantly different than previously thought, with consequences for the late stages of stellar evolution and the explosion \citep{Aguilera-Dena:2018}.

The possible evolutionary outcomes for massive stars is greatly complicated by the possibility of binary interaction \citep[e.g.,][]{Paczynski:1971, Podsiadlowski:1992, Eldridge:2009, deMink:2013, Yoon:2015, Eldridge:2017, Zapartas:2019}. Observations indicate that a large fraction of stars exist in binary systems \citep[e.g.,][]{Garmany:1980, Kobulnicky:2007, Chini:2012, Kobulnicky:2014, Sota:2014, Dunstall:2015, Almeida:2017}. Binary systems can have a wide range of orbital periods, mass-ratios and orbital eccentricities \citep{Moe:2017}. In particular, a significant fraction of massive stars are observed to reside in close binary systems and will interact with a companion at some point during their life \citep{Sana:2012, Sana:2013, Moe:2017}. Despite the significant progress in our understanding of binary interaction, the outcomes of this process are not always fully understood. This contributes to the overall uncertainty in our understanding of the evolution of massive stars.

Mass loss, convection, rotation and binary interaction have important effects on the evolution of stars. One of their main effects is to modify the core and envelope masses (\mcore and \menv, respectively) of a star at a given evolutionary stage, i.e. for a given core composition. For example, higher mass-loss rates produce a lower \menv, and in some cases also produce a lower \mcore. Higher convective core-overshooting (i.e. a larger \aov) produces a higher convective core mass and a corresponding lower \menv during the MS, and a higher \mcore during the post-MS evolution. Binary interaction can dramatically change both \mcore and \menv. Envelope masses can be decreased as a result of stripping of the envelope by Roche-Lobe overflow (RLOF), increased by accretion from a companion star, and can be modified through common-envelope evolution or a merger event \citep{vanBever:1998, Dominik:2012, deMink:2014, Ivanova:2018, Fragos:2019}.

The main purpose of this work is to systematically produce a grid of stellar structure models in hydrostatic and thermal equilibrium, based on three important structural properties: \mcore, \menv and the core composition (described in this paper by \xc and \yc). We refer to these stellar structure models as \snap models. The models are constructed in a way that allows us to vary the value of one structural property, while keeping the other properties constant, e.g. varying \menv while keeping \mcore, \xc and \yc constant. This allows us to isolate the effect of each structural property on the surface properties of the star. 

This type of approach to stellar models was first used by \citet{Cox:1961}. It was further developed by \citet{Giannone:1967a} and \citet{Giannone:1968} to study core-Helium burning stars at low and intermediate masses, and by \citet{Lauterborn:1971a} and \citet{Lauterborn:1971} to study the occurrence of blue loops of intermediate mass stars in the HR diagram. However, to our knowledge, this method has never been used in a systematic and comprehensive way as in our work. Moreover some of these approaches used very simple internal structures. In this work, we take advantage of the advancements in stellar evolution and computational capabilities over the last decades to produce state-of-the-art models of stellar structure. 

We outline some of the advantages of the \snap models we present in this work below.

\begin{enumerate}
    \item The surface properties of a star depend on its internal structure. This approach allows us to make direct connections between the structural properties (\mcore, \menv, \xc and \yc) and the surface properties (luminosity, effective temperature, \teff, and surface gravity, \logg). For example, with this approach we can directly compare between two core-He burning stars with exactly the same Helium core, but with different \menv. These types of comparisons are difficult to obtain from classical stellar evolution models without fine-tuning of processes such as mass loss, overshooting etc.
    \item It is often difficult to disentangle cause and effect in stellar evolution. The equations of stellar structure and evolution are highly non-linear and processes such as rotation, convection etc. can combine and interact to produce complex effects on the evolution of a star. It is often challenging to connect these evolutionary effects to a particular physical process or combination of processes. Our approach using \snap stellar models is not subject to such complex evolutionary effects, which allows us to more easily disentangle connections between internal and surface properties. These connections can then be used to help establish cause and effect in stellar evolution models.
    \item We can produce a wider range of stellar structures than are currently obtained in stellar evolution calculations. Stellar evolution calculations always include prescriptions for computing effects such as mass loss and effects of close binary interactions such as Roche-Lobe Overflow. These prescriptions may limit the range of stellar structures that are produced in evolutionary calculations. Our method allows us to compute stellar structures that may not be produced in stellar evolution calculations and to see if they correspond to observations.
    \item \snap models can help us to study how many different internal structures could correspond to a given set of observed properties (such as luminosity, \teff and surface gravity). This is very important, as it will allow us to determine the degree of degeneracy between observable and non-observable stellar properties. For instance, we often deduce the actual/initial mass of a stars from its observed position in the HR diagram using stellar evolution models. The result from this procedure depends on the set of stellar evolution models that are used and their assumptions about convection, binary interaction etc. The \snap model approach will allow us to better estimate the degree of uncertainty of these deductions and to determine the regions where it gives reliable and unreliable results. It may also provide some hints into what kinds of additional observations could help to reduce the degeneracy. 
    \item The results from our \snap models can be useful for improving the approximations used in rapid stellar evolution algorithms to compute single and binary population synthesis models \citep[e.g.,][]{Eggleton:1989, Pols:1995, Tout:1996, Tout:1997, Hurley:2000, Hurley:2002}.
\end{enumerate}
    
\snap stellar structure models offer a complementary approach to stellar evolution models to study the lives of stars. Stellar evolution models allow us to obtain quantities such as age, to study connections between different evolutionary stages, to analyse short phases of stellar evolution such as those when the star is out of thermal equilibrium and to derive the frequency of evolutionary outcomes assuming a set of initial conditions. Our hope is that \snap models can be used alongside stellar evolution models to help to progress our understanding of how stars evolve.

We describe our method for producing \snap stellar structure models in detail in Sec. \ref{sec:method}. We analyse our results for core-H burning stars and core-He burning stars in Sections \ref{sec:results:ms} and \ref{sec:results:hecore} respectively. In Sec. \ref{sec:disc}, we discuss connections between our \snap stellar structure models and the evolution of massive single and binary stars. We conclude our analysis in Sec. \ref{sec:conclusions}.

%__________________________________________________________________

\section{Method}\label{sec:method}

\begin{figure}
	\centering
	\includegraphics[width=\hsize]{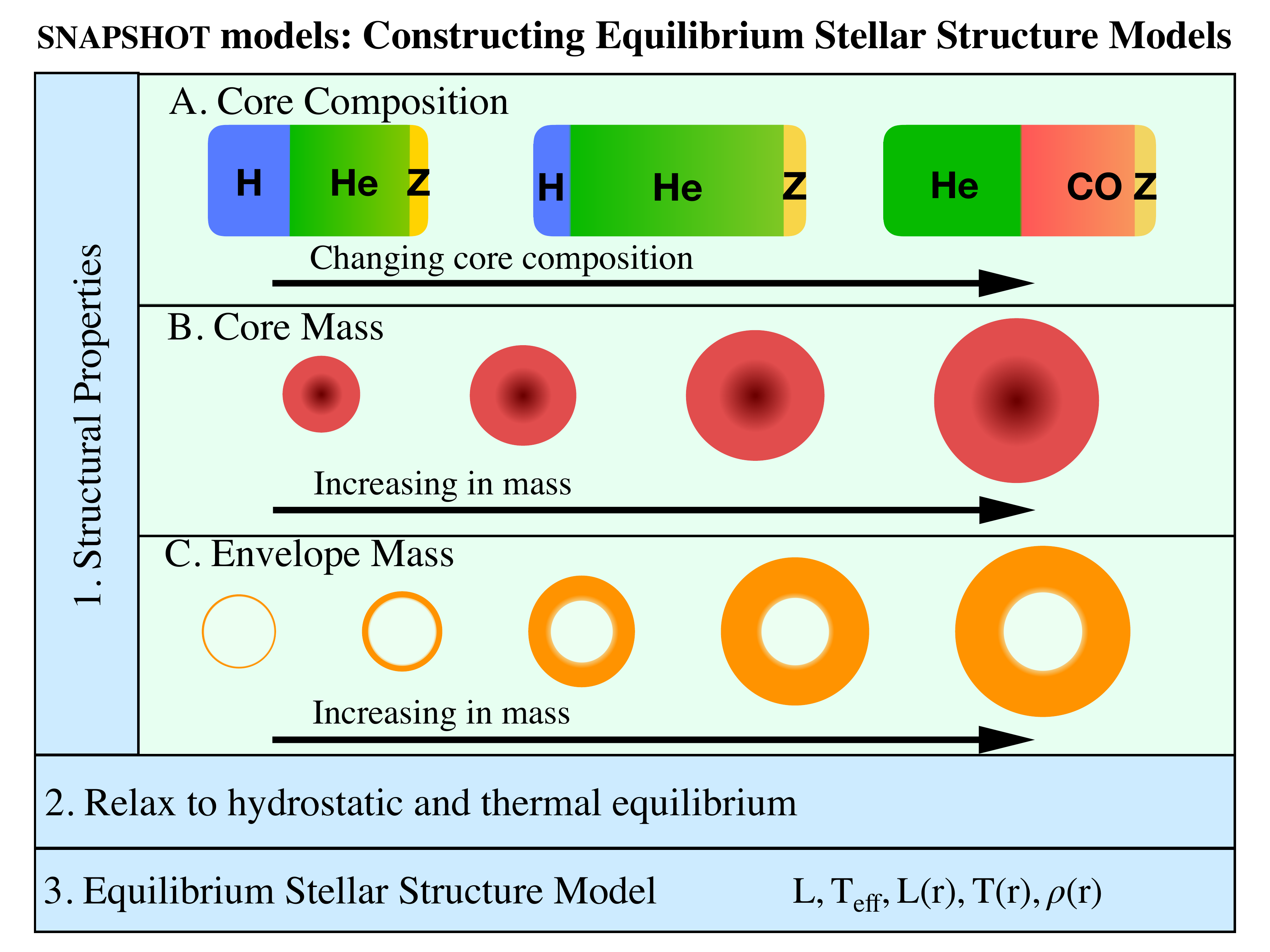}
	\caption{Schematic outline of the construction of \snap stellar structure models with different core compositions, core masses and envelope masses. We first select a core composition (1A). Secondly, we select a core mass to go with this core composition (1B). Finally, we select an envelope mass (1C). Using a stellar evolution code, the models are allowed to relax to a state of hydrostatic and thermal equilibrium (2). The final stellar model produces both the surface properties, e.g. luminosity and effective temperature, as well as the interior profiles of the standard quantities, e.g. temperature, luminosity, and density (3). See Sec. \ref{sec:method} for more details.}
	\label{fig:method_summary}
\end{figure}

A systematic method to compute stellar structure models with a range of core compositions, core masses and envelope masses requires an alternative approach to the usual methods for computing stellar evolution models\footnote{In principle, it is possible to compute a stellar evolution model and fine-tune the input parameters governing various physical processes e.g. mass loss, convective core-overshooting or binary interaction to achieve a desired combination of core composition, \mcore and \menv. However, this would be difficult to do in a systematic way as these processes often have complex interactions and feedback effects that affect the evolution of the star. For example, larger convective core overshooting produces a more massive core, a higher luminosity and higher mass-loss rates which can then affect the mass of the core.}. In Fig. \ref{fig:method_summary}, we provide a schematic outline of our approach to compute \snap stellar structure models. We use the \mesa software instrument \citep[r10398;][]{Paxton:2011, Paxton:2013, Paxton:2015, Paxton:2018, Paxton:2019} to compute our models. Our method can be summarised in the following three steps:

\begin{enumerate}
   \item To produce the initial stellar structures with a given \mcore, \menv and core composition, we compute a stellar evolution model with \mesa from the zero-age main sequence until the end of core-He burning.
    \item Using the stellar structures produced in (i), we modify some of the input controls to allow us to change the total mass of the star, without the star evolving.
    \item We allow the modified structures from (ii) to relax to hydrostatic and thermal equilibrium.
\end{enumerate}

We would like to emphasise that our approach is very flexible and can easily be updated to include new physics. Secondly, we will provide all our input files online and, as \mesa is open-source, the \snap method can easily be implemented by others. Thirdly, our method also benefits from the active development in the \mesa code in improving and updating the physical ingredients in the models.

The first step involves computing a stellar evolution model with \mesa from the zero-age main-sequence until the end of core-He burning. For these models, we adopt similar physical inputs as in the MIST grid of stellar evolution models \citep{Choi:2016}, which were also computed using \mesa. We discuss the potential effects of the input physics in Sec. \ref{sec:caveats}. We choose to apply this technique to study massive stars ($> 8 \msun$) at solar metallicity for this paper, but the same technique can also be applied to intermediate and low mass stars and also to stars at different metallicities. This simply requires computing a suitable initial stellar evolution model. Our core-He burning stellar structure models have Helium core masses of 2 to $9 \msun$. These cores correspond to stellar evolution models with initial masses of 8 to $25 \msun$ assuming our physical inputs detailed below. We explore a wide range of \menv from 0 to $50 \msun$, so that the total masses of our core-H and core-He burning models range from 2 to $59 \msun$. We explore core-H burning stellar structure models with values of \xc from 0.70 to 0.05 and core-He burning models with $\yc = 0.50$ and 0.01. We summarise the physical ingredients that we use below.

\begin{enumerate}
    \item We adopt a solar metallicity of Z = 0.020 (with the solar abundance scale from \citet{Grevesse:1998}) for all models, with an initial He abundance of 0.26. The exact value of the metallicity, C, N, O and Fe abundances affect the stellar properties.
    \item For mass loss, we use the `Dutch' wind scheme in \mesa with the default scaling factor of 1.0. This wind scheme combines mass-loss rates from \citet{Vink:2001} and \citet{Nugis:2000} for hot stars ($>10^4\mathrm{K}$) and from \citet{deJager:1988} for cool stars ($<10^4\mathrm{K}$).
    \item We use the Ledoux criterion for convective stability, with a semi-convective efficiency of \asc = 0.1.
    \item We use a time-dependent, diffusive convective core-overshooting parameter \citep{Herwig:2000, Paxton:2011}. We adopt the same overshooting parameters as in the MIST models \citep{Choi:2016} with core overshooting of \fovcore = 0.016 (roughly equivalent to \aov = 0.2 in the step overshoot scheme), and \fovshell = \fovenv = 0.0174.
    \item For most of the models, we use the standard mixing-length theory to model convective mixing, with a mixing-length parameter of \amlt = 1.82. For some of the higher mass models, in order to allow the models to converge it was necessary to use a modified treatment of convection known as MLT++ \citep{Paxton:2013}. MLT++ reduces the temperature gradient in some radiation-dominated convective regions to make it closer to the adiabatic gradient. This boosts the efficiency of energy transport which allows the model to run with reasonable timesteps.
    \item As a surface boundary condition, we use the \texttt{simple\_photosphere} option in \mesa \citep{Paxton:2011}.
    \item We adopt the \texttt{mesa\_49.net} nuclear network in \mesa which tracks and solves for the abundances of 49 species.
    \item The models are all non-rotating.
\end{enumerate}

For each stellar evolution model, we save snapshots at two points during core-H burning (for $\xc = 0.35$ and 0.05) and two points during core-He burning (for $\yc = 0.50$ and 0.01). In principle, any core composition can be studied, as long as the star is in thermal equilibrium. We take the snapshot models we saved from the stellar evolution models and change the value of \menv without the star evolving. To do this, we modify the following inputs in \mesa to effectively pause the evolution of the star:

\begin{enumerate}
    \item We turn off changes in abundances of the chemical elements due to nuclear burning by setting  \texttt{dxdt\_nuc\_factor = 0} in \mesa. This allows the nuclear energy generation rates to remain the same, but prevents the chemical abundances from changing.
    \item We turn off element diffusion by setting \texttt{do\_element\_diffusion = 0} and turn off all other mixing by setting \texttt{mix\_factor = 0}.
    \item We turn mass loss off.
\end{enumerate}

While the evolution of the star is paused, we use a routine in \mesa called \texttt{mass\_change} which allows an arbitrary mass-loss rate or mass-accretion rate from/to the surface of the star. We set the rate of mass loss/accretion to be $10^{-12} \msun$/yr. This low value ensures that the star will remain in thermal equilibrium while it is accreting or losing mass. We set the chemical abundance of the accreted material to be the same as the surface abundances. For each of the core masses and core compositions, we allow the star to lose mass until the H-rich envelope is stripped entirely, and to accrete mass until the star reaches an envelope mass of $50 \msun$ (as defined when we start to modify the mass). For each core, we save models for a range of \menv.

We allow every model to relax to hydrostatic and thermal equilibrium by restarting the evolution, with mass loss turned off but with all other physical inputs unchanged from the initial stellar evolution. This allows the convective core to readjust to the modified structure. We set a stopping criterion for these models based on \lnuc/\ltotal, the ratio of the total nuclear energy generation to the total luminosity of the star. The time for the models to mix by convection and relax to thermal equilibrium is typically on the order of $\sim 1$ kyr or less.

% -----------------------------------------------------------------

\section{Application to Main Sequence Stars}\label{sec:results:ms}

\begin{figure*}
	\centering
	\includegraphics[width=\hsize]{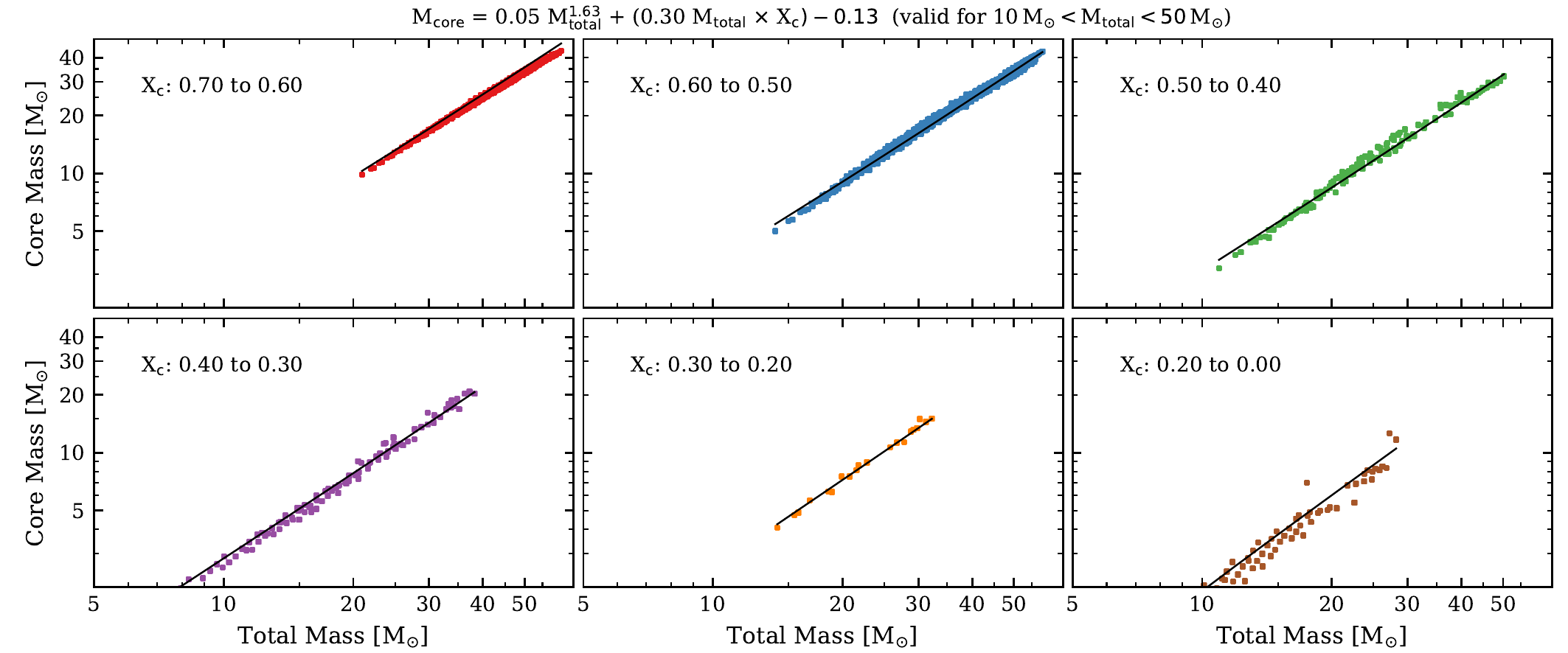}
	\caption{Convective core mass as a function of the total mass for core-H burning models. The models are divided up into bins according to \xc as indicated in each plot. The best fit line for each \xc is plotted in black and corresponding equations relating \mcore and the total mass $M_{\rm total}$ are indicated in the upper left of each subplot.}
	\label{fig:parameter_space_ms2}
\end{figure*}

In this section, we apply our \snap method to MS stars i.e. core-H burning stellar structures. We analyse our core-H burning models in terms of three structural properties, the convective core mass (\mcore), the envelope mass (\menv) and the central hydrogen abundance (\xc). We define the envelope as the rest of the star above the convective core. For models to which we have added mass, we find that the size of the convective core adjusts to the new stellar structure during the relaxation procedure. Material from the H-rich envelope mixes with H-depleted material in the core, the convective core increases in mass and the value of \xc increases. This so-called ``rejuvenation'' phenomenon in MS stars, in which a star can accrete from its companion and end up with a higher value of \xc than before the mass transfer episode, has been well studied in various stellar evolution contexts \citep[e.g.,][]{Hellings:1983, Hellings:1984, Schneider:2014}. Whether or not rejuvenation will take place depends on the treatment of convective stability and the choice of the semi-convective efficiency \asc \citep{Braun:1995}.

Our core-H burning structure models have \mcore ranging from 1 to $45 \msun$, \menv from 6 to $18 \msun$ and \xc from 0.70 to 0.05. We find that the combinations of \mcore, \menv and core composition are quite limited (Figs. \ref{fig:parameter_space_ms1} and \ref{fig:parameter_space_ms2}). We indicate the values of \mcore, \menv and \xc in Fig. \ref{fig:parameter_space_ms1} in Appendix \ref{app:parameter_space_h}, where we plot the values of \menv and \mcore for different values of \xc. In Fig. \ref{fig:parameter_space_ms2}, we plot the relationship between \mcore and \mtotal. For a core-H burning star, the value of \mcore is determined by the total mass of the star and the value of \xc. We fit the following function relating the total stellar mass $M_{\rm total}$, the convective core mass \mcore and \xc:

\begin{equation}
    \mcore = a M_{\rm total}^d + b M_{\rm total} \xc + c
\end{equation}

We obtain best fit values of a = 0.045, b = 0.304, c = -0.133 and d = 1.627. For a star of a given total mass, the convective core mass is lower for models with lower \xc, as expected from stellar evolution models. Furthermore, the dependence of \mcore on \xc is steeper for larger stellar masses. While the value of the convective core mass for a given total mass depends on our assumptions for convective overshooting, the overall trends observed in Fig. \ref{fig:parameter_space_ms2} do not depend on overshooting assumptions.

% Relationship between core and total mass for H burning stars
The mass of the convective core in core-H burning stars as a function of mass has been studied before for models at the beginning of the main sequence \citep{Schwarzschild:1958, Schwarzschild:1961, Stothers:1970, Stothers:1974, Maeder:1980, Maeder:1981, Maeder:1987, Maeder:1988, Maeder:1989,Langer:1989, Pols:1998}, by \citet{Stothers:1985} at the beginning and end of the MS, and by many others as in the context of stellar evolution models. Here, we make a connection between the \mcore, the total stellar mass and \xc (Fig. \ref{fig:parameter_space_ms2}), providing fits for the convective core mass for a range of stellar masses and core compositions that can be easily used by the community. These calculations are useful for a number of reasons, for instance in rapid binary stellar evolution algorithms for computing population synthesis models \citep[e.g.,][]{Belczynski:2002, vanBever:1998, Hurley:2000, Hurley:2002, Izzard:2006, Belczynski:2008, Eldridge:2008, Eldridge:2009, deMink:2015, Eldridge:2017} where stellar properties need to be updated after a mass transfer episode.

We have compared the location of our core-H burning \snap models in the HR diagram to standard \mesa stellar evolution models \citep{Choi:2016} and they are consistent, as expected. For the sake of brevity we do not include these comparisons here, as it is a well known result \citep[e.g][]{Henyey:1959}. Further tests show that the luminosity and \teff of core-H burning \snap stellar structure models not depend very much on the amount of overshooting. Models with no overshooting and with the the overshooting assumptions we adopt in this paper differ by about 0.01 dex in \logllsun and \logteff despite the differences in convective core masses.

% ______________________________________________________________________________________________

\section{Application to Post-Main Sequence Stars} \label{sec:results:hecore}

\begin{figure}
	\centering
	\includegraphics[width=\hsize]{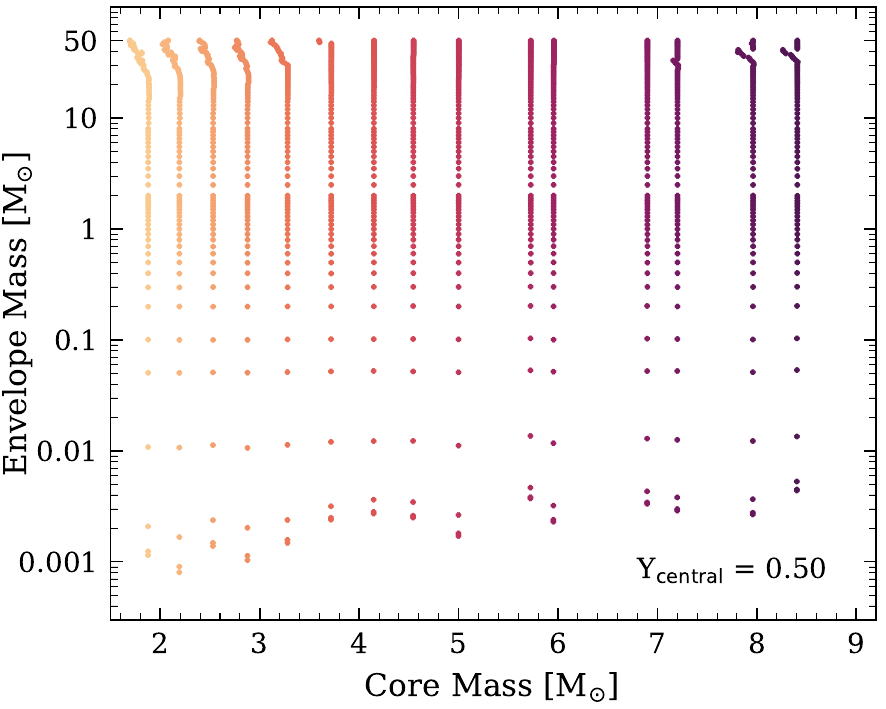}
	\caption{Envelope mass as a function of the core mass for our core-He burning models with $\yc = 0.50$. The colours indicate structure models with the same core mass.}
	\label{fig:parameter_he}
\end{figure}

In this section, we apply our \snap technique to post-MS core-Helium burning stars. We construct core-He burning \snap structure models based on three structural properties: the Helium core mass \mcore, the H-rich envelope mass \menv and the central He abundance \yc. The He-core is defined as the central Hydrogen depleted region where $X<10^{-4}$. The H-envelope is defined as the rest of the star above the He-core. Our models have He-core masses ranging from $\mcore = 2$ to $9 \msun$\footnote{These He-core masses correspond to initial masses of $8 - 25 \msun$, assuming single star evolution with our physical ingredients.}, envelope masses of $\menv = 0.001$ to $50 \msun$ and a central Helium abundance of $\yc = 0.50$ and 0.01.

Figure \ref{fig:parameter_he} shows the values of \mcore and \menv for our core-He burning models with $\yc = 0.50$. Each point corresponds to an individual \snap stellar structure model. A similar figure for models with $\yc = 0.01$ is included in Appendix \ref{app:parameter_space_he}. In contrast to core-H burning stars (Figs. \ref{fig:parameter_space_ms2} and \ref{fig:parameter_space_ms1}), core-He burning stars in thermal equilibrium can have a wide range of combinations of \mcore, \menv and \yc (Fig. \ref{fig:parameter_he}). The difference in the variety of stellar structures for core-He and core-H burning stars is due to the fact that there are (usually) two nuclear burning regions in core-He burning stars (the He-core and H-shell), while there is only one burning region in core-H burning stars. 
To understand the difference, we can consider what happens if we begin with a star in hydrostatic and thermal equilibrium with a given \mcore and \menv, and then very slowly increase \menv, such that the star remains in thermal equilibrium. The star must respond by readjusting its structure to support the increased mass. With only one nuclear burning region, a core-H burning star can respond only by producing more energy in its core. An increased mass changes the mechanical equilibrium structure of the star. The pressure and temperature gradients inside the star must increase to support the increased mass. As a result, the central temperature increases and hence the nuclear reaction rates in the core increase.This will generally cause an increase in the mass of the convective core, depending on the assumptions for mixing, particularly semi-convection. 
In contrast, a core-He burning star, with two nuclear burning regions (the He-core and the H-shell), could respond to the increased mass by modifying its mechanical equilibrium in a way that results in an increase in the the energy production in either the He-core, the H-shell or some combination. Our models show that, in almost all cases, core-He burning stars respond to an increase in the total mass in a way that increases the energy production in the H-shell. The value of \menv can change over a wide range without significantly modifying the conditions in the He-core, such as the central temperature (Appendix \ref{app:logtc_massratio}). This leads to the wide variety of combinations of \mcore and \menv for core-He burning stars.

\begin{figure}
	\centering
	\includegraphics[width=\hsize]{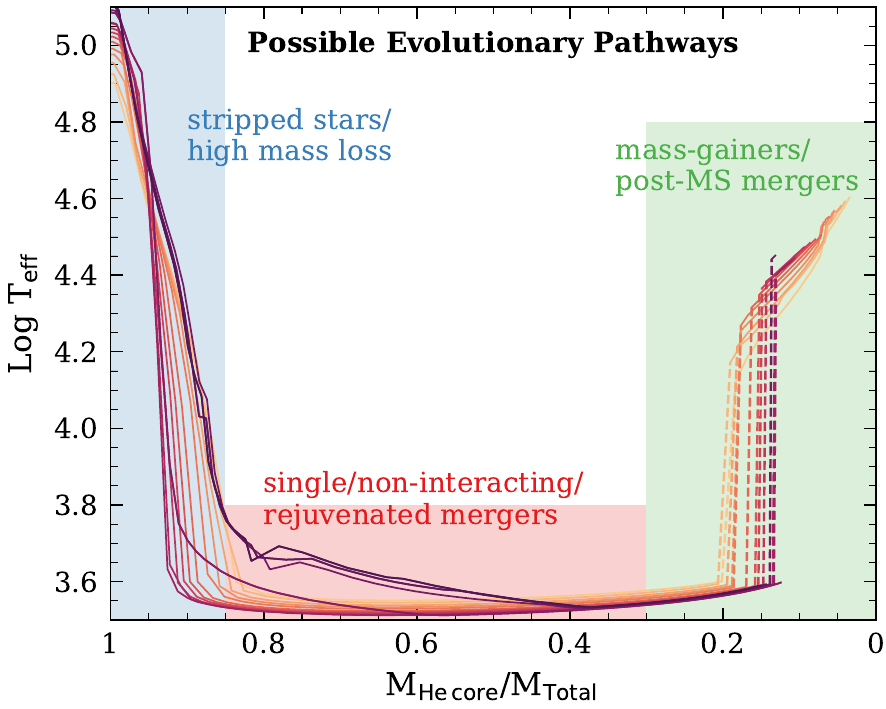}
	\caption{Effective temperature as a function of the core mass ratio for core-He burning models with constant \mcore, $\yc = 0.01$ and \menv varying from 0 to $50 \msun$ (same models as in panel d of Fig. \ref{fig:CC_massratio}). We roughly sketch possible binary evolutionary pathways which may produce the same stellar structures as these models.}
	\label{fig:discussion:binary}
\end{figure}

In Fig. \ref{fig:discussion:binary}, we summarise the results we obtain for our core-He burning models by plotting the value of \teff as a function of the core mass ratio. Each line consists of models with constant \mcore, a constant $\yc = 0.50$ and \menv varying from 0 to $50 \msun$ (same models as Fig. \ref{fig:CC_massratio}d). Dashed lines indicate a sharp transition from red supergiants to hotter, more luminous stellar structures due to a bi-stability in the stellar structure equations. To put our results in context, we discuss connections between our \snap models and stellar evolution pathways.

Stellar structures with high core mass ratios (shaded in blue in Fig. \ref{fig:discussion:binary}) correspond to stripped stars with low values of \menv. Most stripped stars are expected to form in binary systems when the primary star expands after the MS, fills its Roche-Lobe and is stripped of its envelope by the secondary \citep{Podsiadlowski:1992}. Some may also form due to high mass loss from a single star \citep{Groh:2013a}. These stars correspond to Regime I of the core-He burning models discussed in Sec. \ref{sec:results:env}.
Stellar structures with intermediate core mass ratios (shaded in red in Fig. \ref{fig:discussion:binary}) are mostly red supergiants. They are expected to be formed by single stars, non-interacting stars in binary systems or stars that rejuvenate after accreting mass. These stars correspond to Regime II in Sec. \ref{sec:results:env}. 
Stars with lower core mass ratios $\lesssim 0.2$ probably only form in binary systems, either as the product of mass-accretion or a post-MS merger \citep[e.g.,][]{Eldridge:2017,Zapartas:2019}. For example, a merger between a core-He burning star and a relatively massive main sequence companion could produce a star with a small He-core and a very high mass H-rich envelope \citep[e.g.,][]{Justham2014}. These stars may resemble OB-type stars or blue supergiants (BSGs) and they correspond to Regime III from Sec. \ref{sec:results:env}. They lie to the right of the MS in the HR diagram which may help to explain observations of a large number of stars in this location in the HR diagram \citep{Castro:2014a}.

In the following sections, we discuss our core-He burning models in detail and describe the connections between internal and surface properties when varying \menv (Sec. \ref{sec:results:env}), \mcore (Sec. \ref{sec:results:core}), and \yc (Sec. \ref{sec:results:corecomp}).

\subsection{Effect of Envelope Mass} \label{sec:results:env}

\begin{table*}
	\centering
	\caption{Summary of surface properties for Models A -- G with a He-core mass of $4.1 \msun$ and a central Helium abundance of 0.50. \menv indicates the mass of the H-rich envelope above the He-core. \fhshell refers to the fraction of the total nuclear energy that is generated by H-Shell burning.}
	\label{tab:surface_properties}
	\begin{tabular}{rrrrrrrrrr}
		\hline
        Model & \mcore [$M_{\odot}$] & \menv [$M_{\odot}$] & \mtotal [$M_{\odot}$] & \logteff [K] & \teff [K] &  \logllsun & \logg &  \logrrsun & \fhshell \\
        A & 4.1 & 0.0 & 4.1 & 4.96 & 91700 & 4.37 & 5.48 & -0.21 & 0.00 \\
        B & 4.1 & 0.5 & 4.6 & 4.45 & 28000 & 4.53 & 3.32 & 0.89 & 0.17 \\
        C & 4.1 & 2.0 & 6.1 & 3.53 & 3390 & 4.58 & -0.28 & 2.76 & 0.28 \\
        D & 4.1 & 6.0 & 10.1 & 3.54 & 3460 & 4.59 & -0.04 & 2.74 & 0.30 \\
        E & 4.1 & 17.0 & 21.1 & 3.58 & 3800 & 4.61 & 0.42 & 2.67 & 0.34 \\
        F & 4.1 & 18.0 & 22.1 & 4.29 & 19620 & 5.04 & 2.87 & 1.46 & 0.75 \\
        G & 4.1 & 50.0 & 54.1 & 4.51 & 32590 & 5.73 & 3.44 & 1.36 & 0.95 \\
    	\hline
	\end{tabular}
\end{table*}

\begin{figure*}
	\centering
	\includegraphics[width=\hsize]{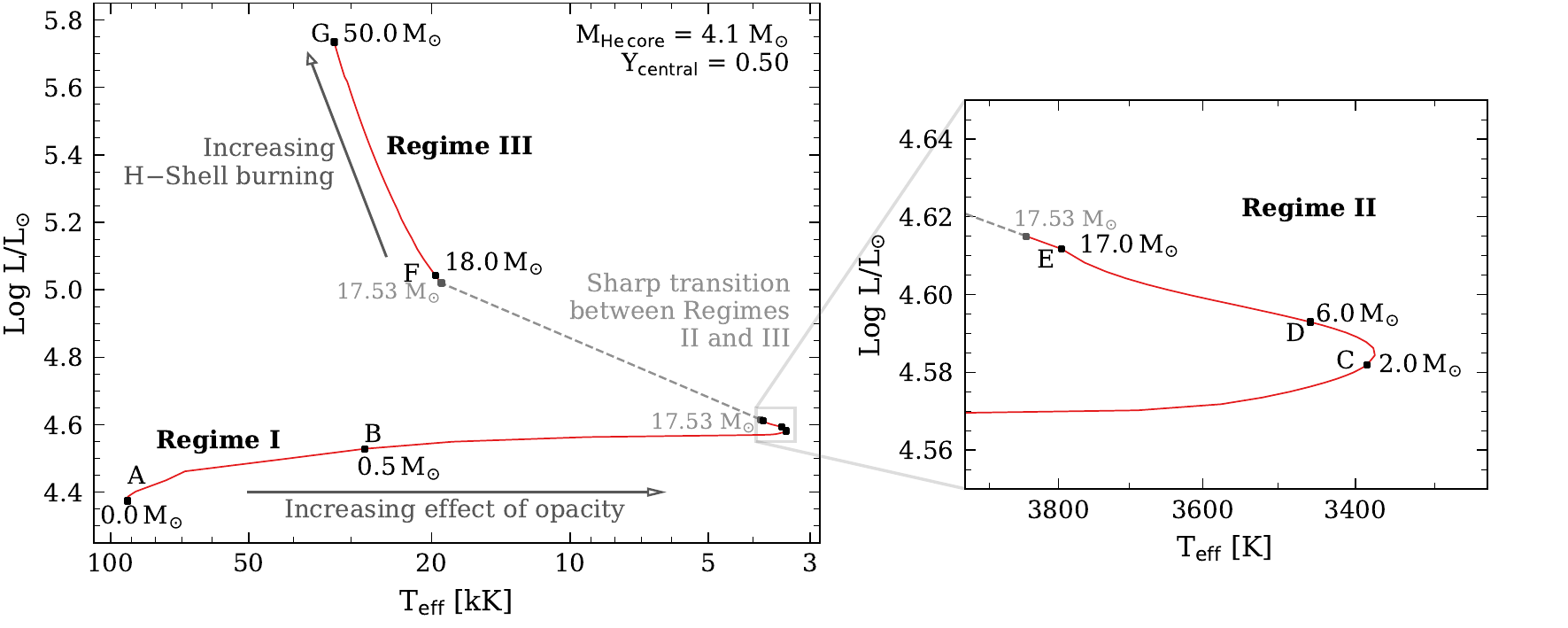}
	\caption{HR diagram showing core-He burning models with $\mcore= 4.1 \msun$, $\yc = 0.50$ and varying H-rich envelope mass in the range $0 - 50 \msun$ (\menv indicated in brackets). We label 7 representative models (A -- G) to discuss the trends in the surface properties as a function of envelope mass (Sects. \ref{sec:results:regime1}, \ref{sec:results:regime2} and \ref{sec:results:regime3}). The right panel shows a zoom-in of the RSG region (models C, D and E).}
	\label{fig:hrd_example_C4d1}
\end{figure*}

\begin{figure}
	\includegraphics[width=\hsize]{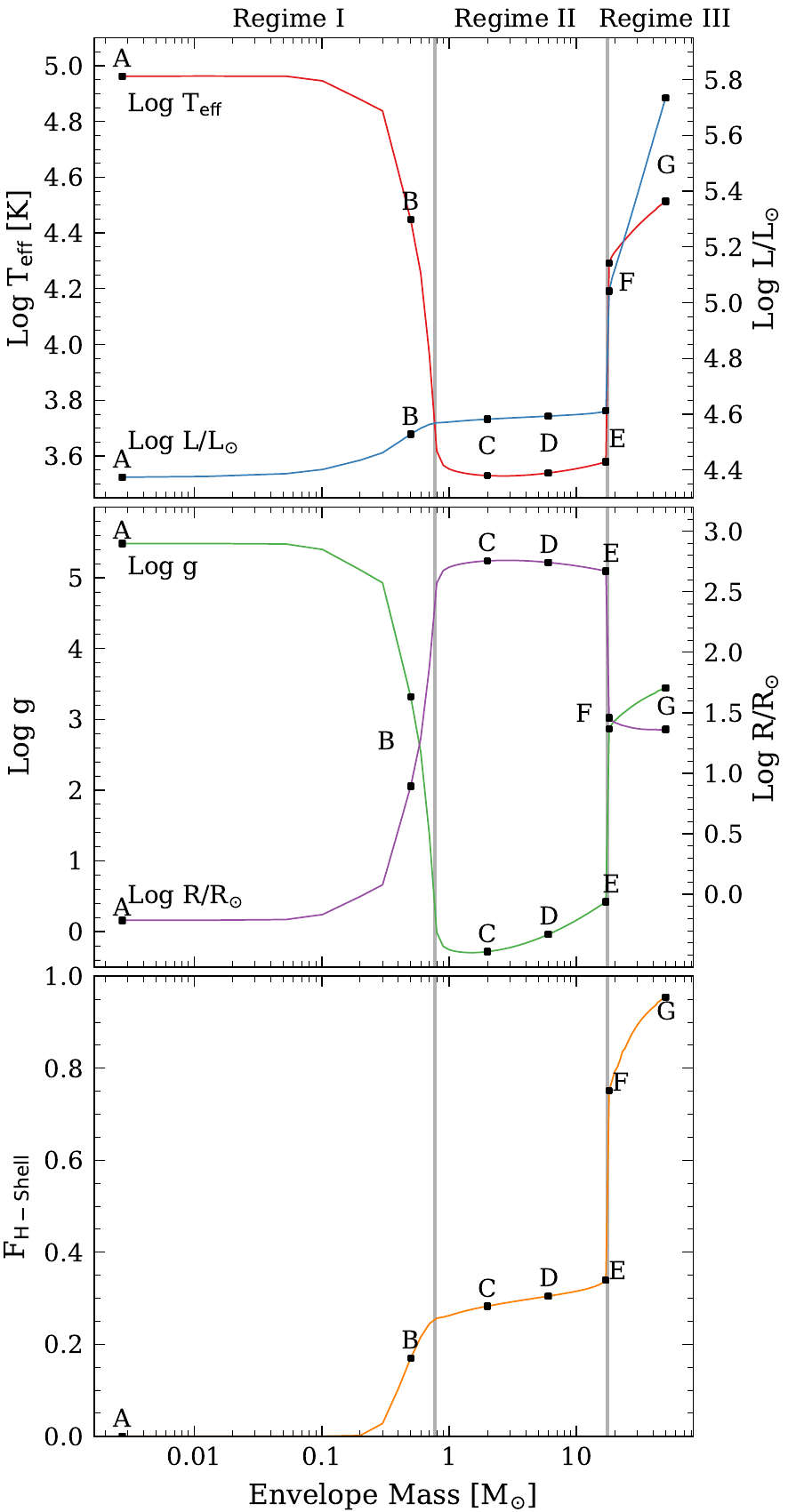}
    \caption{\textit{Upper Panel}:  Effective temperature (red curve) and luminosity (blue) as a function of \menv for the same models as in Fig. \ref{fig:hrd_example_C4d1} (a constant He-Core mass of $4.1 \msun$, $\yc = 0.50$ and envelope mass varying from $0 - 50 \msun$). Models A -- G from Fig. \ref{fig:hrd_example_C4d1} are indicated with black dots and labelled.
    \textit{Middle Panel}: Surface gravity, \logg and the stellar radius as a function of envelope mass for the same models as in the upper panel.
    \textit{Lower Panel}: Proportion of the total nuclear energy that is produced by the H-burning shell as a function of envelope mass for the same models as in the upper panel.}
     \label{fig:surface_example_C4d1}
\end{figure}

\begin{figure}
	\includegraphics[width=\hsize]{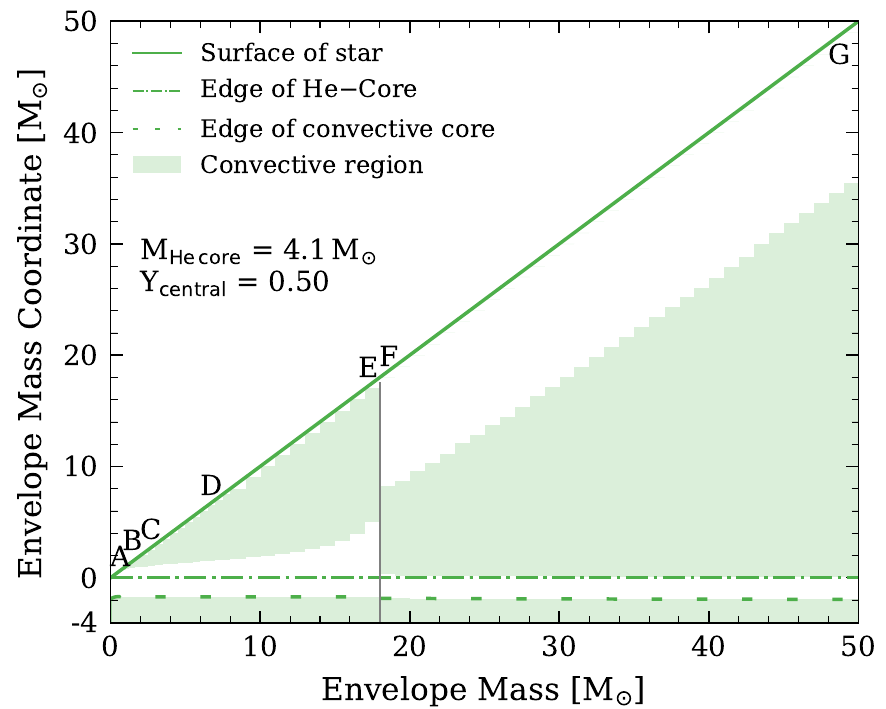}
    \caption{Kippenhahn-like diagram for the models in Figs. \ref{fig:hrd_example_C4d1} and \ref{fig:surface_example_C4d1} with the envelope mass on the x-axis and the envelope mass coordinate on the y-axis, where 0 corresponds to the edge of the He-core and $-4.1$ corresponds to the centre of the $4.1 \msun$ core. The convective regions are shaded in green and the boundary of the convective core, the He-core and the surface of the star are indicated with dashed, dash-dot, and solid lines respectively. A grey vertical line indicates the transition between Regimes II and III. Moving from left to right in this figure corresponds to increasing envelope mass.}
	\label{fig:kippenhahn_example_C4D1}
\end{figure}

\begin{figure*}
	\centering	
	\includegraphics[width=\hsize]{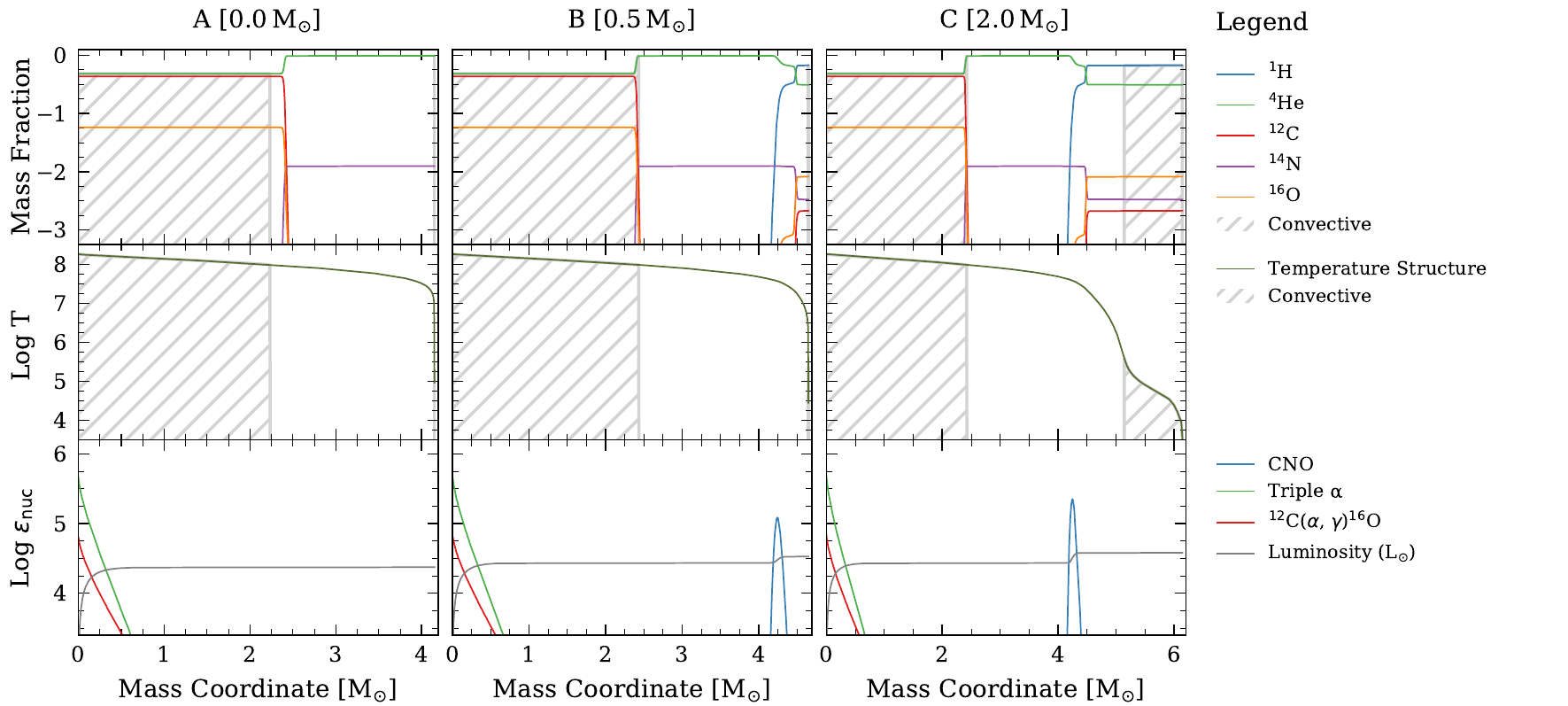}
    \includegraphics[width=\hsize]{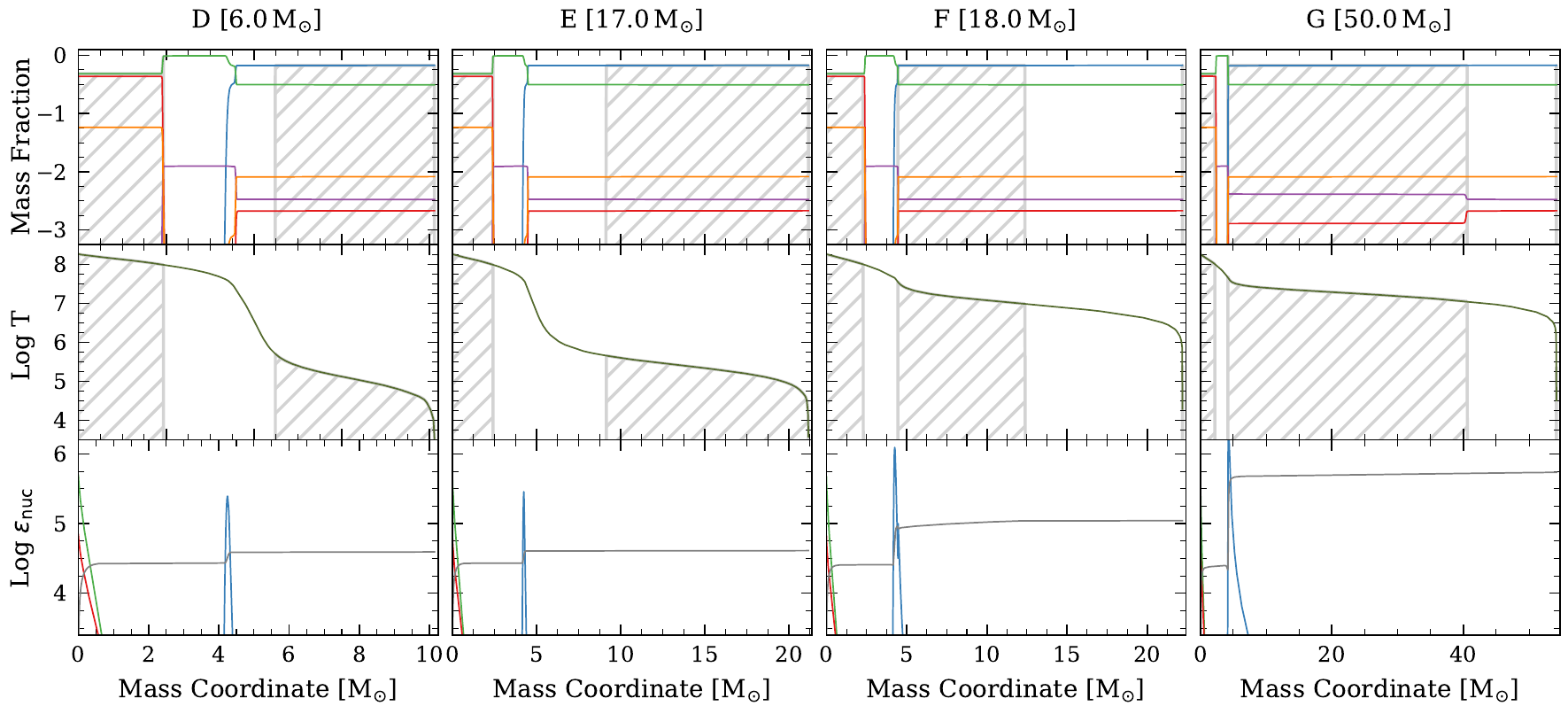}
    \caption{Internal structure of models from points labelled A to G in Figs. \ref{fig:hrd_example_C4d1} and \ref{fig:surface_example_C4d1}. The models have \mcore=$4.1 \msun$ and \yc=0.50. The values of \menv are indicated above each sub-figure. The convective regions are hashed in the upper and middle panels. The upper panels show the internal abundance profile of H (blue), He (green), C (red), N (purple) and O (orange). The middle panels show the internal temperature structure (solid line). The lower panels show  $\log \epsnuc$ in units of erg g$^{-1}$s$^{-1}$ for different reactions, and $\log L$, in units of $\lsun$.}
	\label{fig:interior_example_C4D1}
\end{figure*}

To analyse how the surface properties of a core-He burning star depend on the mass of the H-envelope, we discuss a representative set of models with the same He-core mass and composition ($\mcore = 4.1 \msun$ and $\yc = 0.50$) and different envelope masses (\menv ranges from 0.0 to $50.0 \msun$). We choose this particular set of models as its core properties are representative of massive core-He burning stars. A He-core mass of $4.1 \msun$ corresponds to an initial mass of $\sim 14 \msun$, assuming single star evolution and our physical ingredients. Additionally, $\yc = 0.50$ corresponds to the middle of core-He burning phase.

In Fig. \ref{fig:hrd_example_C4d1}, we plot this set of models in the HR diagram. From these, we select 7 models, labelled A -- G, which represent the qualitative trends in surface properties as a function of \menv. We indicate the location in the HR diagram and the value of \menv for each of the models A-G in Fig. \ref{fig:hrd_example_C4d1}. We summarise the internal and surface properties for models A-G in Table \ref{tab:surface_properties}.

Based on the location in the HR diagram, we divide the models plotted in Fig. \ref{fig:hrd_example_C4d1} into Regimes I, II and III. Regime I consists of stars with no H-envelope or low envelope masses and with mostly high effective temperatures \teff. We define the transition between Regimes I and II at $\log(\teff) = 3.7$ or $\teff = 5\,011$ K. Regime II consists of stars with intermediate envelope masses, a radiative H-burning shell and a convective outer envelope. They are located near the Hayashi line in the HR diagram. The transition between Regime II and III is defined by an abrupt change in the solution of the stellar structure equations from a cool star with a convective outer envelope to a hotter, more luminous star with a convective H-burning shell and a radiative outer envelope. Stars with envelope masses above this transition are defined to be in Regime III.

In Fig. \ref{fig:surface_example_C4d1}, we plot the surface luminosity, \teff, surface gravity (\logg), radius ($R$) and \fhshell (the fraction of the total nuclear energy that is generated in the H-burning shell) as a function of \menv for the same set of models in Fig. \ref{fig:hrd_example_C4d1}. The transitions between Regimes I, II and III are indicated by grey vertical lines.  We also label models A-G in each panel. Models A and B are in Regime I with envelope masses of 0.0 and $0.5 \msun$ respectively. Models C, D and E are in Regime II, close to the Hayashi line, with envelope masses of 2.0, 6.0 and $17.0 \msun$ respectively. Models F and G are in Regime III with envelope masses of 18.0 and $50.0 \msun$ respectively. 

To investigate the presence of convective regions as a function of \menv, we plot in Fig. \ref{fig:kippenhahn_example_C4D1} the envelope mass on the x-axis and the Lagrangian envelope mass coordinate on the y-axis for the same representative set of models as in Fig. \ref{fig:hrd_example_C4d1}. This is similar to a normal Kippenhahn plot, with convective regions indicated in solid color. In Fig. \ref{fig:interior_example_C4D1}, we plot internal abundance profiles, internal temperature profiles and internal nuclear burning profiles for models A -- G. Convective regions are hashed in the abundance and temperature profile plots.

In Sections \ref{sec:results:regime1} -- \ref{sec:results:regime3} below, we discuss Figs. \ref{fig:hrd_example_C4d1}, \ref{fig:surface_example_C4d1}, \ref{fig:kippenhahn_example_C4D1} and \ref{fig:interior_example_C4D1} in detail. We analyse the trends in surface properties as a function of envelope mass and establish connections between the internal and surface properties.

\subsubsection{Regime I -- Stripped Stars (Models A and B)}\label{sec:results:regime1}
We begin by discussing a model with a $4.1 \msun$ He-core and no envelope (Model A). It has a high \teff of $92\,000\,\mathrm{K}$, is highly compact with a radius of $R = 0.62 \rsun$ and has a high surface gravity of $\logg = 5.48$ (middle panel of Fig. \ref{fig:surface_example_C4d1}). The $4.1 \msun$ He-core is composed of a $2.2 \msun$ convective core, with $\yc = 0.50$, and a $1.9 \msun$ Helium rich shell (Fig. \ref{fig:interior_example_C4D1}).

For only a modest increase in the envelope mass from $\menv = 0$ to $0.5 \msun$ (Model B), the value of \teff drops sharply from $92\,000\,\mathrm{K}$ to $28\,000\,\mathrm{K}$. This is because the effect of opacity in the envelope increases with increasing \menv. The increased effect of opacity produces a larger stellar radius and a lower \teff. While this effect has been identified before in stellar evolution models \citep[e.g.,][]{Groh:2014,Meynet:2015}, with our models we can investigate this behavior in a more systematic way as a function of \menv, \mcore, and \yc. We find that the luminosity increases slightly when increasing \menv from 0 (model A) to 0.5 \msun (model B). This is due to the presence of a second nuclear energy generation region, i.e. the H-burning shell. In model B, 17 per cent of the total energy is generated in the H-shell, compared to 0 per cent in model A (see Fig. \ref{fig:surface_example_C4d1} and Table \ref{tab:surface_properties}). The presence of the H-burning shell and its contribution to the total luminosity can also be seen by comparing the energy generation profiles for models A and B in Fig. \ref{fig:interior_example_C4D1}.

\subsubsection{Regime II -- Red Supergiants (Models C, D, E)} \label{sec:results:regime2}
As the value of \menv increases, the star responds to the increased mass by increasing the energy generation in the H-burning shell. As the value of \menv increases from 2 to $17 \msun$ (i.e. from model C to E), \fhshell increases from 0.28 to 0.34 (Fig. \ref{fig:surface_example_C4d1}). As well as the increased energy generation in the H-shell, the mass of the convective region in the envelope increases greatly from model C to E (Figs. \ref{fig:kippenhahn_example_C4D1} and \ref{fig:interior_example_C4D1}). Over a wide range of \menv, the mass of the convective region in the outer envelope increases with \menv almost as fast as \menv (Fig. \ref{fig:kippenhahn_example_C4D1}). These changes in internal properties as a function of \menv can help to explain the trends in the HR diagram as a function of \menv.

The right panel of Fig. \ref{fig:hrd_example_C4d1} shows a zoom-in of Regime II in the HR diagram. For $\menv < 2.5 \msun$, \teff decreases with increasing \menv. It reaches a minimum of $\teff = 3\,380\mathrm{K}$ at $\menv = 2.5 \msun$ and increases with increasing \menv for $\menv > 2.5 \msun$. Conversely, the radius increases with \menv to a maximum at $\menv = 2.5$ and then decreases with further increasing \menv (Fig. \ref{fig:surface_example_C4d1}). The value of \teff is affected by two factors in this regime. Firstly, the effect of opacity increases with increasing \menv because there is more material in the envelope. This produces a larger radius and lower \teff. Secondly, as \menv and \fhshell increases, the interior temperature profile changes and a larger mass of material is convective (Fig. \ref{fig:kippenhahn_example_C4D1}). For stars on the Hayashi track, we expect the stellar radius to decrease with increasing mass \citep{Eggleton:2006}, resulting in a higher \teff. The combination of these factors causes \teff to decrease to a minimum and subsequently increase.

For a given \mcore and \yc, our models show that the surface properties of a RSG change very little over a wide range of envelope masses (Figs. \ref{fig:hrd_example_C4d1}, \ref{fig:surface_example_C4d1}, \ref{fig:CC_HRD}, \ref{fig:CC_envmass}). For example, for $\mcore = 4.1 \msun$, as \menv increases from 2 to $17 \msun$, the value of \teff increases from 3390 K to 3800 K and the luminosity increases from Log(L/\lsun) = 4.58 to 4.61 (Fig. \ref{fig:hrd_example_C4d1}). This means that there is a lot of degeneracy in the value of a stellar mass derived from a particular luminosity and \teff for RSGs.

Our models indicate that the minimum envelope mass required to produce a RSG with $\teff < 5000$ K is $\menv = 0.6 \msun$ for $\mcore = 1.9 \msun$ and $\menv = 1.7 \msun$ for $\mcore = 8.9 \msun$. In addition, they show that over a wide range of \mcore, the minimum \teff and maximum radius of a RSG occurs for a core mass ratio of $\mcore/\mtotal \approx 0.60$.

Although we only plot the convective regions for models with one core mass ($\mcore = 4.1 \msun$) in Fig. \ref{fig:kippenhahn_example_C4D1}, our models show that the mass of the convective region in the envelope depends only on \menv and is independent of \mcore for a wide range of core masses, from $\sim 2 - 7\msun$.

\subsubsection{Transition between Regime II and III}\label{sec:results:regime2_to_3}
As we keep increasing \menv, we find a bi-stability in the solution of the stellar structure equations at an envelope mass of $\mtrans = 17.532 \msun$ (the exact value depends on the input physics). For \menv < \mtrans, the star has a RSG structure with a radiative H-burning shell and a convective envelope (Figs. \ref{fig:kippenhahn_example_C4D1} and \ref{fig:interior_example_C4D1}). As \menv increases towards \mtrans, the pressure and temperature in the H-burning shell increase. This is accompanied by slightly increased nuclear energy generation in the H-burning shell (comparing models D and E in Fig. \ref{fig:interior_example_C4D1}). At \mtrans, the base of the envelope becomes unstable to convection due to the increased H-shell nuclear burning and the solution of the stellar structure changes. For $\menv > \mtrans$, the star is more condensed and hotter, with a convective H-burning shell and a radiative envelope (see Fig. \ref{fig:kippenhahn_example_C4D1} and compare models E and F in Fig. \ref{fig:interior_example_C4D1}). Although we have only included structure models at intervals of $1 \msun$ in Figs. \ref{fig:hrd_example_C4d1}, \ref{fig:surface_example_C4d1} and \ref{fig:kippenhahn_example_C4D1}, we observe in our paused models that the star `jumps' from a RSG solution to a BSG solution with no change in mass.

\subsubsection{Regime III -- High Envelope Mass (Models F and G)}\label{sec:results:regime3}
For $\menv \geq 18 \msun$, our models present a hot, convective H-shell above the He-core and a radiative envelope (models F and G). The presence of a convective region at the base of the envelope results in a modified internal structure with a greatly increased temperature and nuclear energy generation rate in the H-shell. This increased energy production in the H-shell causes a much higher surface luminosity as compared to models with lower \menv in Regime II. The modified structure of the envelope also results in a much smaller stellar radius and a higher \teff compared to Regime II.

The energy generation of models in Regime III is dominated by H-shell burning (see Fig. \ref{fig:surface_example_C4d1} and energy generation profiles in Fig. \ref{fig:interior_example_C4D1}). More than 75 per cent of the total nuclear energy generation occurs in the H-shell, compared to Regime II stars which have $\fhshell \sim 0.30$. The mass of the region that is convective increases with increasing \menv (Fig. \ref{fig:kippenhahn_example_C4D1}). For $\menv = 50 \msun$ (model G), the envelope dominates the structure and surface properties of the star. The value of $\fhshell \sim 0.90$ and the star has a similar structure to a core-H burning star but with a $4.1 \msun$ He-core in the centre. Regime III stars will appear as blue stars in the HR diagram with $\logg$ of $1.8 - 3.9$ dex. Models with lower core masses can produce blue stars with lower values of \logg. Some may resemble OB-type stars and others may resemble blue supergiants.

%_________________________________________________________________

\subsection{Effect of Core Mass} \label{sec:results:core}

\begin{figure*}
	\centering
	\includegraphics[width=\hsize]{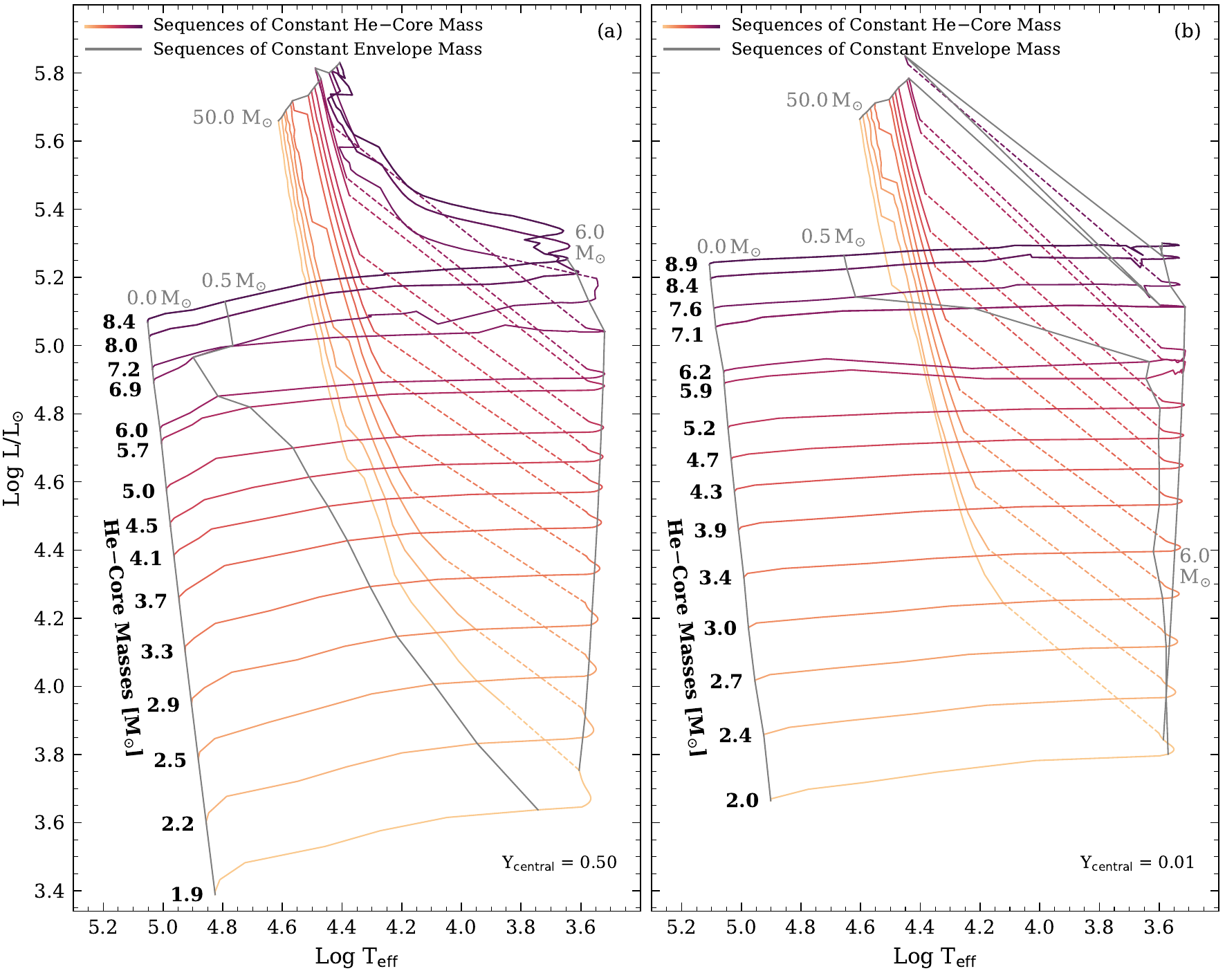}
	\caption{\textit{Left Panel:} Coloured lines indicate constant He-core mass, constant $\yc = 0.50$ and envelope masses varying from 0.0 to $50.0 \msun$ (colours indicate different He-core masses). The He-core masses are indicated in bold along the left-hand side. We also plot four lines of constant envelope mass and varying He-core mass in grey, with the envelope mass in \msun indicated in grey text above each line. As in Fig. \ref{fig:hrd_example_C4d1}, the dashed coloured lines indicate a transition between two stable solutions of the stellar structure equations. \textit{Right Panel:} Same as left panel but for models with $\yc = 0.01$.}
	\label{fig:CC_HRD}
\end{figure*}

\begin{figure*}
	\includegraphics[width=\hsize]{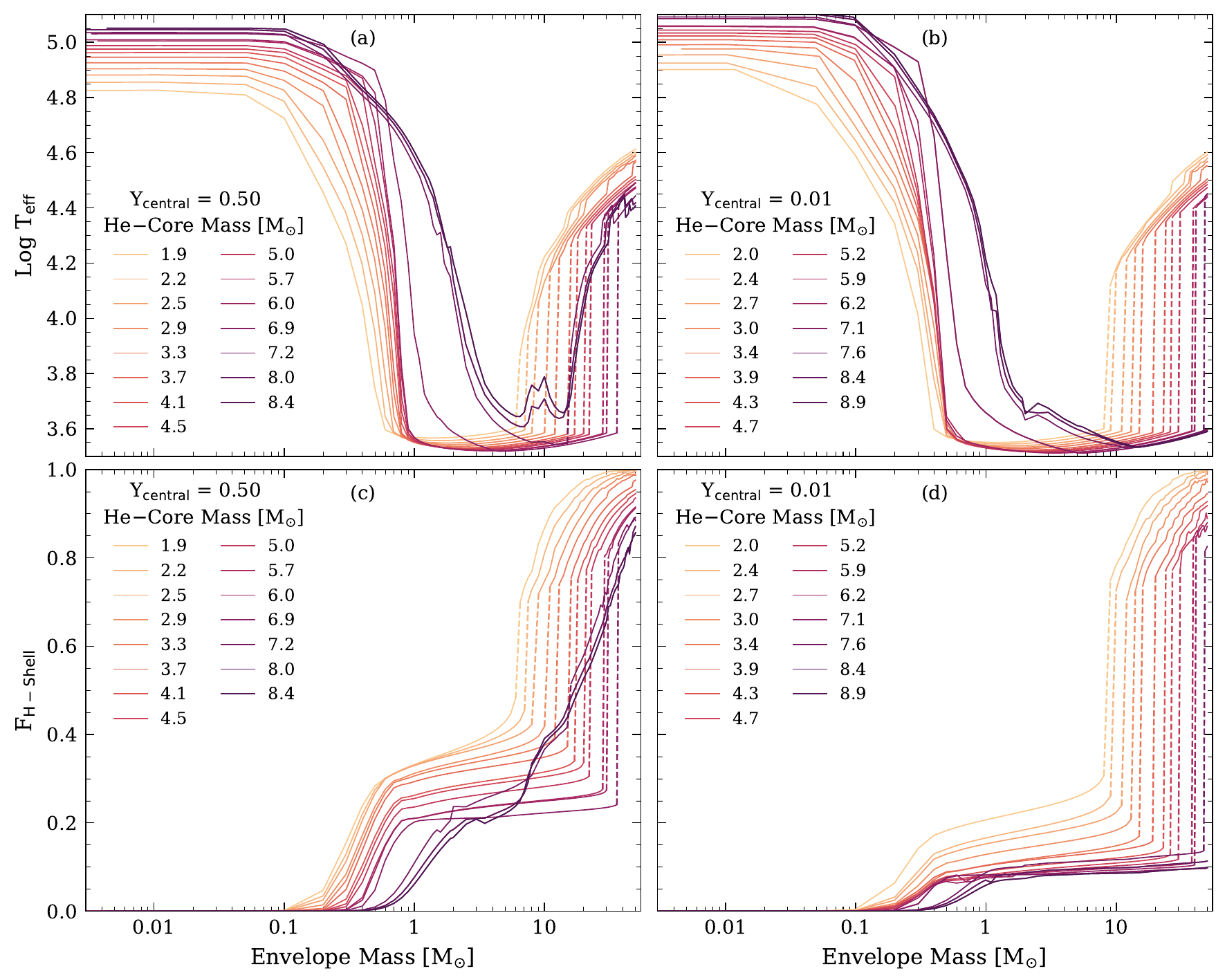}
    \caption{\textit{(a)}: Effective temperature vs. envelope mass for models with constant He-core mass and $\yc = 0.50$. The dashed lines indicate the bi-stability between Regimes II and III.
    \textit{(b)}: Same as (a) but for models with $\yc = 0.01$. 
    \textit{(c)}: The fraction of the total nuclear energy generated in the H-Shell (\fhshell) vs. envelope mass for the same models as in (a).
    \textit{(d)}: Same as (c) but for models with $\yc = 0.01$.}
	\label{fig:CC_envmass}
\end{figure*}

\begin{figure}
	\includegraphics[width=\hsize]{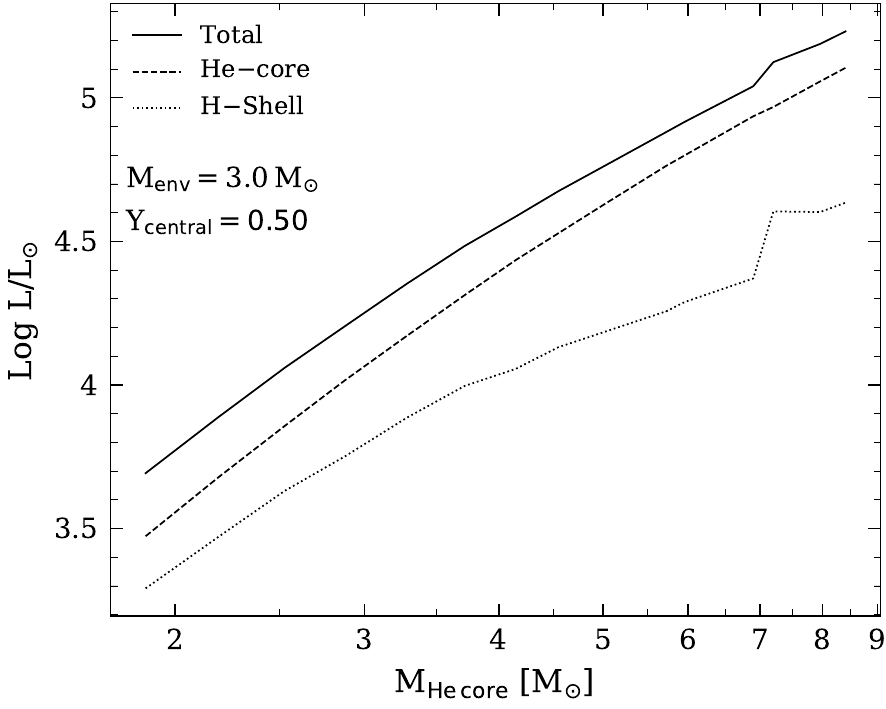}
    \caption{Luminosity vs. core mass for models with the same $\yc = 0.50$ and $\menv = 3.0 \msun$ and with core masses ranging from $\mcore = 2$ to $8 \msun$. The luminosities of the He-core, the H-shell and the total luminosity are indicated by a dash-dot, dotted and solid lines respectively.}
	\label{fig:core_comp_lum}
\end{figure}

\begin{figure*}
	\includegraphics[width=\hsize]{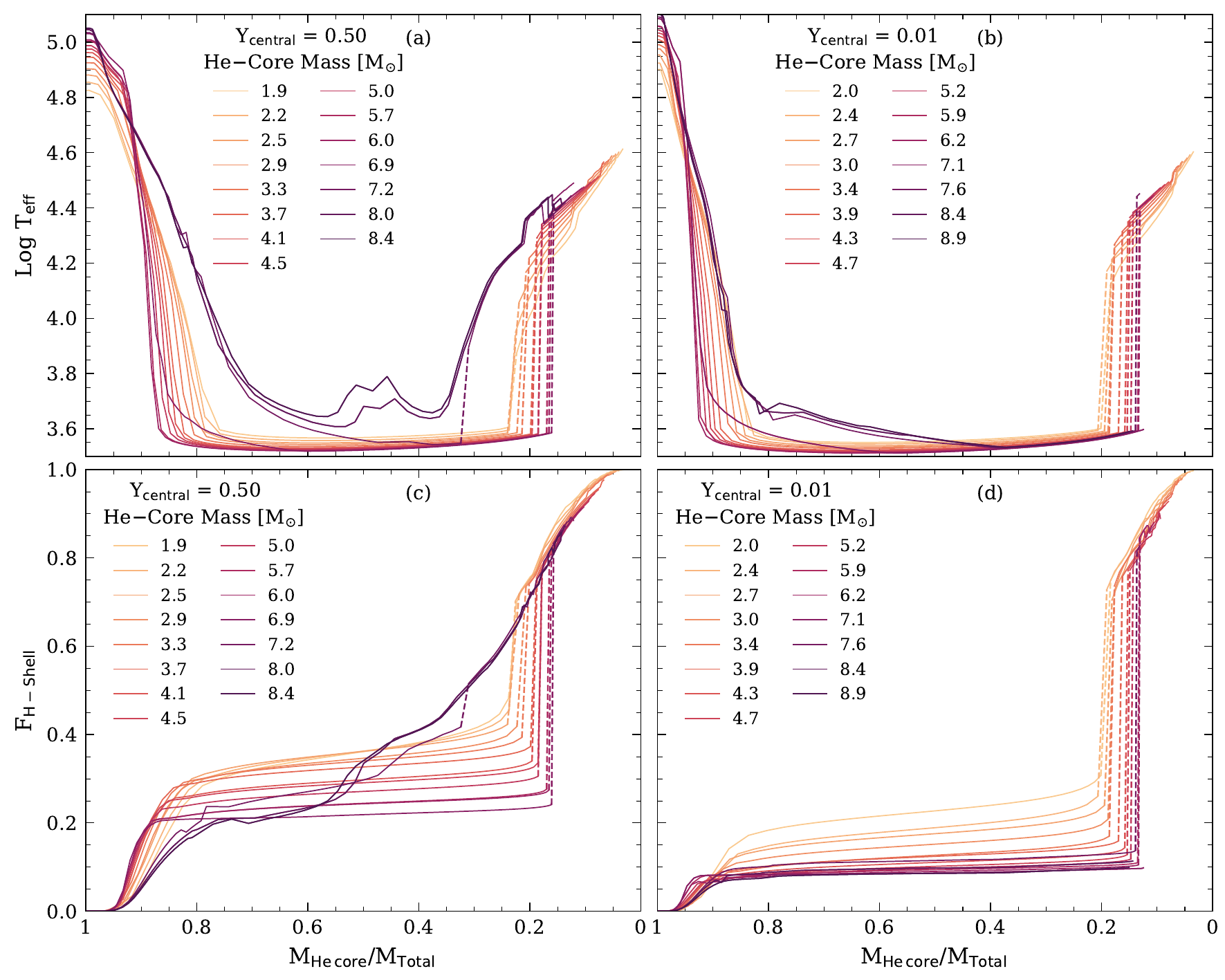}
    \caption{\textit{(a)}: Effective temperature vs. core mass ratio ($\mcore / \mtotal$) for models with constant He-core mass and $\yc = 0.50$. The dashed lines indicate the bi-stability between Regimes II and III.
    \textit{(b)}: Same as (a) but for models with $\yc = 0.01$. 
    \textit{(c)}: The fraction of the total nuclear energy generated in the H-Shell (\fhshell) vs. envelope mass for the same models as in (a).
    \textit{(d)}: Same as (c) but for models with $\yc = 0.01$.}
	\label{fig:CC_massratio}
\end{figure*}

% Overview
We now discuss stellar structure models with \mcore ranging from 1.9 to $8.9 \msun$, \menv ranging from 0.0 to $50.0 \msun$, and \yc of 0.50 (corresponding to the middle of core-He burning).

We plot these models in the HR diagram in Fig.~\ref{fig:CC_HRD}a, and Fig.~\ref{fig:CC_envmass}a shows the effect of \menv on \teff for models with the same \mcore and $\yc = 0.50$. Qualitatively, models with $\mcore < 7 \msun$ show similar trends as a function of \menv as the model with $\mcore = 4.1 \msun$ (Sec. \ref{sec:results:env}). We discuss these models below and examine models with $\mcore > 7 \msun$ in Sec. \ref{sec:results:highmcore}.

% Low Menv
For models with $\menv \lesssim 0.5 \msun$, the value of \teff increases with \mcore at constant \menv (Fig. \ref{fig:CC_envmass}a). This is due to an increase in luminosity with \mcore at constant \menv, mostly due to higher energy generation in the He-core. For some models, there is also an increase in energy generation in the H-Shell with increasing \mcore at constant \menv, which also contributes to the higher luminosity.

% Transition Menv
As the envelope mass increases from 0.1 to $1 \msun$, the value of \teff decreases sharply. This is due to the increasing effect of opacity with increasing \menv. The value of \menv at which \teff begins to decrease increases with \mcore (Fig. \ref{fig:CC_envmass}a). Additionally, the rate of decrease of \teff as a function of \menv increases with \mcore. The value of \teff depends on the structure of the envelope and, in particular, on the presence and mass of any convective regions in the outer envelope. Models with higher \mcore can support a higher \menv before the outer envelope becomes convective. This allows the envelope to remain compact (i.e. higher \teff) up to a higher \menv.

% Intermediate Menv
For intermediate envelope masses from $\sim 1$ to $10 \msun$, models with higher \mcore have lower \teff for the same \menv (Fig. \ref{fig:CC_envmass}a). Models with higher \mcore have more extended, lower density envelopes, larger radii which results in a lower \teff.

% Bi-stability Menv
For each value of \mcore, the location of the bi-stability of the stellar structure equations at the transition between Regimes II and III is indicated in Fig. \ref{fig:CC_envmass}a by dashed lines. At this point, the solution of the stellar structure transitions from a RSG structure with a convective envelope and radiative H-shell to a structure with a radiative envelope and a convective H-shell. The value of \menv at which the transition occurs increases with \mcore. Models with higher \mcore are able to support a higher \menv before the base of the envelope becomes unstable to convection.

% High Menv
Models with higher envelope mass generally converge towards a similar \teff as a function of \menv, independent of \mcore. In this regime, $\gtrsim 80$ per cent of the mass of the star is contained in the envelope and the surface properties are dominated by the nature of the H-shell burning, which depends mostly on \menv.

% F H-shell
To assess the relative contributions of the core and the envelope to the overall structure of the star, we compute the fraction of the total nuclear energy that is generated in the H-shell (\fhshell) as a function of \menv for models with constant \mcore (Fig. \ref{fig:CC_envmass}c). In all models, the value of \fhshell is $\approx 0$ for $\menv \lesssim 0.1 \msun$. For stars with these envelope masses, very little burning takes place in the H -shell. As the value of \menv increases from $\sim 0.1 - 1 \msun$, \fhshell increases sharply from 0 up to $\fhshell \approx 0.1 - 0.3$ for $\menv = 1 \msun$. As \menv further increases, the star must respond to support the extra mass. It does this by producing more energy in the H-shell. For models with \menv of $1 - 10$, the value of \fhshell increases only slightly (by a factor of $\sim 0.25$) over a large range of \menv (a factor of $\sim 10$). 
The value of \fhshell decreases with \mcore at constant \menv (Fig. \ref{fig:CC_envmass}c). The luminosity generated in the core and in the H-shell both increase with increasing \mcore at constant \menv. However, the luminosity generated in the core increases at a higher rate as a function of core mass than the H-shell (see Sec. \ref{sec:results:hshell} and Fig. \ref{fig:core_comp_lum}). As a consequence, for models with higher \mcore, a lower fraction of the overall energy comes from the H-shell. 

For models in Regime III (high envelope mass), $> 70$ per cent of the total energy production occurs in the H-shell (Fig. \ref{fig:CC_envmass}c). The envelope has a structure similar to a massive core-H burning star with a nuclear burning region at the base, and a radiative outer region. The structure of the star is dominated by the H-shell burning at the base of the envelope, and the energy generated in the H-shell increases with increasing \menv.

% Analysis in terms of core mass ratio
In Fig. \ref{fig:CC_envmass}a, c, we see that the bi-stability between RSG structures and more luminous, blue stars occurs over a very wide range of \menv for different \mcore. To further explore this transition, we plot \teff and \fhshell as a function of the core mass ratio, i.e. \mcore divided by total mass (Fig. \ref{fig:CC_massratio}a, c). Moving from left to right in this figure corresponds to increasing envelope mass.

In Fig. \ref{fig:CC_massratio}a, \teff shows similar trends as a function of \mcore/\mtotal for each \mcore. The value of \teff decreases sharply from a core mass ratio of 1.0 to about 0.8. For core mass ratios $\approx 0.8 - 0.2$, most models have a RSG structure (Regime II). For models with $\mcore < 7 \msun$, the transition from Regime II to III occurs at a core mass ratio of $\approx 0.2$. We can also see the different behaviour of models with $\mcore > 7.0 \msun$, the value of \teff decreases much more slowly than models with lower \mcore.

% Bumps for extreme core mass ratios
Looking in more detail at Fig. \ref{fig:CC_envmass}a, we notice that for models with $\mcore \lesssim 3 \msun$ and very high envelope masses of $\menv \approx 20 - 50 \msun$, there are some small sharp increases in \teff. This is due to a small amount of convective mixing of material from the edge of the He-core into the envelope for some models with particularly extreme core mass ratios. The mixing occurs only for some models with a core mass ratio of $\mcore/\mtotal \lesssim 0.15$ and it results in a slight decrease in \mcore and slight increase in \menv.

Our \snap models indicate that the maximum core mass ratio that a RSG can have increases with increasing He-core mass. We note that \citet{Eggleton:1998} finds a maximum allowed core mass ratio of $\approx 0.64$ for composite polytropic models of red giants, which is consistent with the trend as a function of \mcore that we observe in our models.

\subsubsection{Models with $\mcore > 7 \msun$ and $\yc = 0.50$} \label{sec:results:highmcore}
These models exhibit qualitatively different behaviour to models with $\mcore < 7 \msun$. Unlike models with lower \mcore, they do not easily develop an outer convective envelope. Small convective shells are formed in the envelope for intermediate $\menv \approx 6 - 20$. The location and mass of the convective shells in the envelope affect the radius $R$ and hence \teff. These cause the small `bumps' visible for models with $\mcore = 8.0$ and $8.4 \msun$ in Fig. \ref{fig:CC_envmass}c. Interestingly, models with $\yc = 0.01$ and $\mcore > 7 \msun$ do produce a convective envelope, suggesting that whether or not a star with a given core produces a RSG may depend on \yc.

The value of \menv at which models with $\mcore > 7 \msun$ form a convective H-burning shell with a radiative outer envelope is much lower than the models with $\mcore = 5.7$, 6.0 and 6.9. This is the opposite to the trend for lower core masses. The increase in \fhshell as a function of \menv is different for these core masses as well.

\subsubsection{Effect of Core Mass on H-Shell} \label{sec:results:hshell}
The grey lines in Fig. \ref{fig:CC_HRD}a indicate that the surface luminosity increases with \mcore for constant \menv and \yc. This is not unexpected, as a higher mass core is typically hotter, can generate more energy and produce a higher surface luminosity. Many authors have provided relationships between the luminosity and \mcore for core-He burning stars \citep[e.g.,][]{Eggleton:1989, Tout:1997, Hurley:2000}. 

Our grid of \snap models allows us to test the separate contributions from the core and H-shell to the luminosity. In Fig. \ref{fig:core_comp_lum}, we plot the surface luminosity, the luminosity of the He-core and the luminosity of the H-shell against \mcore for a set of models with $\menv = 3.0 \msun$ and $\yc = 0.50$ and \mcore from $2 - 9 \msun$. As we observed in Fig. \ref{fig:CC_HRD}, the surface luminosity increases with \mcore for constant \menv and \yc. However the increased luminosity does not all originate in the He-core. The energy produced in both the He-core and the H-shell increases with increasing \mcore. The value of \mcore modifies the temperature structure in the H-shell in two ways. Firstly, the temperature of the H-shell increases (by a small amount) with \mcore, because the higher mass He-cores are hotter and produce a higher temperature just above the core. Secondly, the mass of the region in which significant energy generation takes place increases. These two changes result in hotter, higher mass H-shells and higher nuclear energy generation rates in the H-shell.

\subsection{Effect of Core Composition}\label{sec:results:corecomp}
Our grid of \snap stellar structure models also allows us to isolate the effect of the core composition on the surface properties of the star, independent of the effects of the core mass and envelope mass. This analysis is difficult to accomplish with stellar evolution models, in which the three interior quantities listed above change simultaneously. Models with $\yc = 0.50$ and 0.01 exhibit qualitative differences in the internal and surface properties (Figs. \ref{fig:CC_envmass} and \ref{fig:CC_massratio}). In this section, we discuss the effect of the core composition.

\begin{figure*}
	\centering
	\includegraphics[width=\hsize]{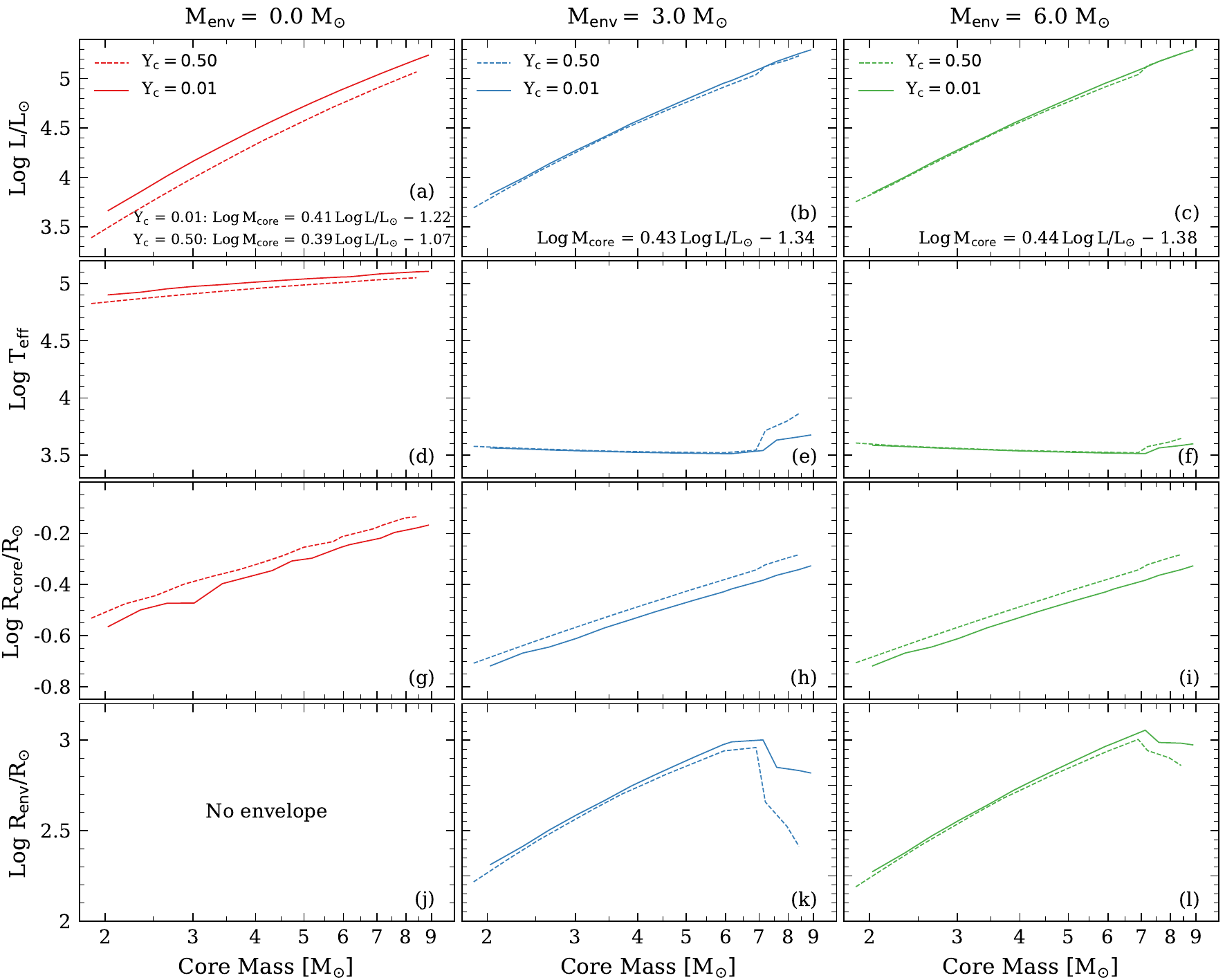}
	\caption{\textit{Left panels (a, d, g, and j)}: Surface luminosity, effective temperature, radius of the core and radius of the envelope as a function of core mass for models with no H-envelope. The dashed lines and solid lines indicate models with $\yc = 0.50$ and $\yc = 0.01$) respectively. \textit{Middle panels (b, e, h, and k)}: Same as left panels but for models with a constant envelope mass of $3 \msun$ and varying core mass. \textit{Right panels (c, f, i, and l)}: Same as left panels but for models with a constant envelope mass of $6 \msun$ and varying core mass.}
	\label{fig:ca_surface}
\end{figure*}

Figure \ref{fig:ca_surface}a, b, c compares the surface luminosity as function of \mcore at constant \menv for core compositions of $\yc = 0.50$ (dashed line) and $\yc = 0.01$ (solid line).  The models in panel A have $\menv = 0 \msun$. For these models, the surface luminosity is higher for models with $\yc = 0.01$ than for $\yc = 0.50$. This is due to the higher mean molecular weight of the core (\mucore) for models with $\yc = 0.01$ compared to 0.50. For the same core mass, a higher \mucore results in a higher central temperature, \tc, through the equation of state, and hence higher nuclear energy generation rates, \epsnuc, in the core. As these models have no H-envelope, the He-core is the only region of nuclear energy generation. This means that a higher \epsnuc in the core caused by with higher \mucore corresponds to a higher surface luminosity. We find mass-luminosity relationships of $\mathrm{L} \propto \mathrm{M}^{2.56}$ for $\yc = 0.50$ and $\mathrm{L} \propto \mathrm{M}^{2.43}$ for $\yc = 0.01$. These are consistent with the mass-luminosity relationships found by \citet{Langer:1989b}.

\begin{figure}
	\centering
	\includegraphics[width=\hsize]{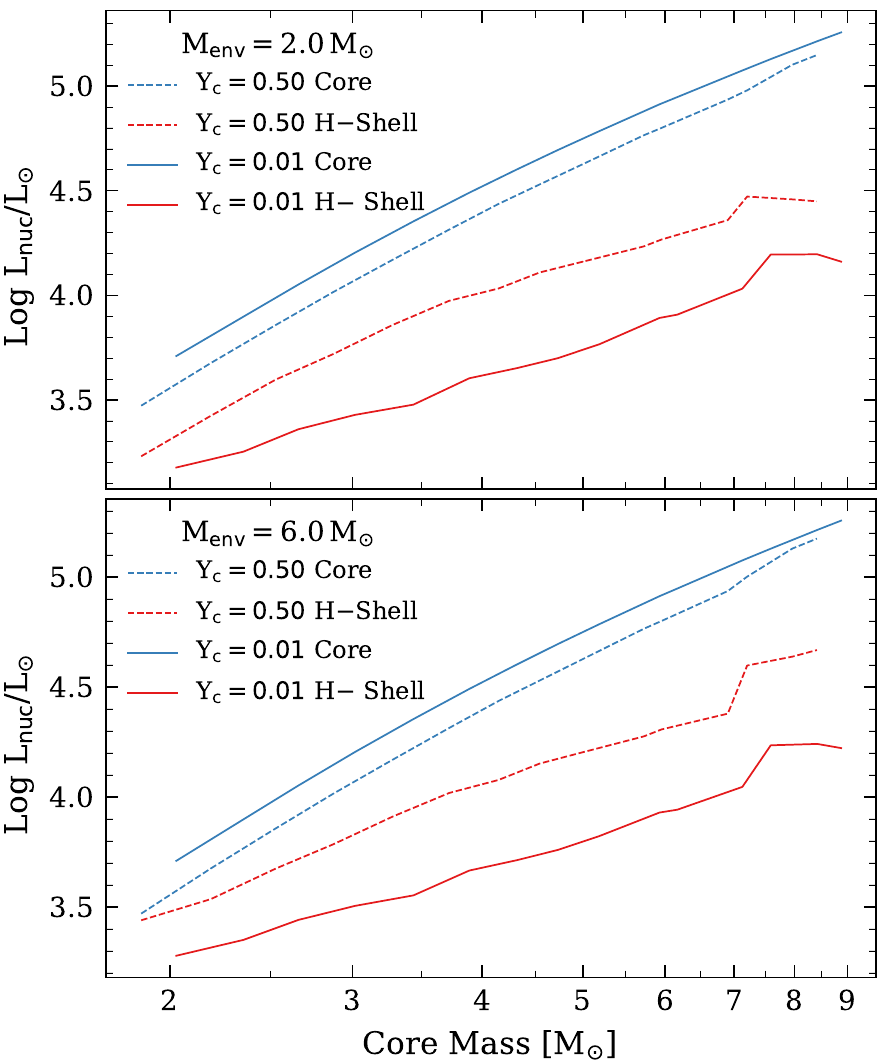}
	\caption{\textit{Upper panel}: Luminosity from the core (blue) and the H-shell (red) as a function of core mass for a constant envelope mass of $3 \msun$ and for $\yc = 0.50$ (dashed line) and $\yc = 0.01$ (solid line) -- the same models as panels b, e, h, and K in Fig. \ref{fig:ca_surface}. {\textit Lower panel:} Same as upper panel but for models with a constant envelope mass of $6 \msun$ -- the same models as panels c, f, i, and l in Fig. \ref{fig:ca_surface}}
	\label{fig:ca_eps_both}
\end{figure}

In contrast, for models with $\menv = 3 \msun$, the value of \yc (and thus \mucore) does not affect the surface luminosity (Fig. \ref{fig:ca_surface}b). For these models, there are two regions of nuclear energy generation, the He-core and the H-shell. The surface luminosity depends on the energy from both of these regions. While a lower \yc corresponds to a higher \mucore, \tc and \epsnuc in the core, it does not correspond to a higher surface luminosity. This is because the energy generation in the H-shell decreases with decreasing \yc. To illustrate this, we plot the total luminosity generated in the He-core, \lcore and the total luminosity generated in the H-shell, \lhshell (Fig. \ref{fig:ca_eps_both}) for models with $\menv = 3 \msun$. Models with $\yc = 0.50$ and 0.01 are indicated in dashed and solid lines respectively. For the same core mass, the \lcore is lower for $\yc = 0.50$ than for $\yc = 0.01$. In contrast, the \lhshell is lower for $\yc = 0.50$ than for $\yc = 0.01$. These effects nearly cancel out so that the total luminosity at the surface is similar for $\yc = 0.50$ and 0.01. For example, models with the same values of \mcore and \menv and \yc of 0.50 and 0.01, the luminosity differs by $\sim 0.02 \mathrm{dex}$ and \teff differs by $\sim 0.007 \,\mathrm{dex}$. The same effect is observed for $\menv = 6 \msun$ and for a wide range of envelope masses. This means that changes in the luminosity of a RSG as it evolves are due mostly to changes in \mcore and not due to changes in \yc or \menv.

We also study the trend in \teff as a function of \mcore for $\yc = 0.50$ and 0.01 (Fig. \ref{fig:ca_surface}d, e, f). For models with no H-envelope (Fig. \ref{fig:ca_surface}d), the luminosity and \teff are higher for models with $\yc = 0.01$ compared to $\yc = 0.50$, while the radius is smaller. For models with $\menv = 3$ and $6 \msun$ and a convective envelope, the value of \teff does not depend on \yc. This result holds for a wide range of envelope masses. In some cases for  $\mcore > 7 \msun$, \teff is slightly higher for $\yc = 0.50$ than 0.01. This is due to differences in the formation of convective shells in the envelope.

In Fig. \ref{fig:ca_surface}g, h, i, we plot the radius of the core as a function of the core mass. For a given core mass, models with $\yc = 0.50$ have cores with larger radii than models with $\yc = 0.01$. For the same \mcore, cores with lower values of \yc must be denser and hotter to produce the same amount of energy to compensate for the decreasing \yc. This results in a smaller core radius. We also plot the radius of the envelope against the core mass for the same \menv (Fig. \ref{fig:ca_surface}j, k, l). The envelope radius increases with \mcore at constant \menv and \yc. For the same \menv, the mass of the outer envelope that is unstable to convection decreases with increasing \mcore. At constant \menv, envelopes with a more massive outer convective region have lower radii.

% _______________________________________________________________

\section{Discussion}\label{sec:disc}

\subsection{Uncertain Masses of Red Supergiants} \label{sec:disc:rsgs}

For a RSG with a given luminosity and \teff, our models show that there is a large range of allowed total masses. This means that for all known RSGs in the Galaxy, it is impossible to know the current mass based on the luminosity and \teff alone. For instance, Betelgeuse and VY CMa may have quite different current masses to what we think they do. This degeneracy may be broken with an accurate measure for the surface gravity. However, this quantity is typically derived from evolutionary mass and either angular diameter (if the star can be resolved) or luminosity and \teff in spectroscopic analyses of RSGs, rather than derived based on diagnostics.

The core of a RSG is mostly unaffected of the presence of the envelope. The luminosity of the core is determined by \mcore and \yc, which is the classical result obtained for Helium stars \citep{Maeder:1987, Langer:1989}. However, the envelope is affected by the core. The energy produced in the H burning shell depends on the temperature profile at the base of the envelope, which in turn depends on \mcore and \yc. It contributes 10 -- 30 per cent of the total luminosity (for $0.50 < \yc < 0.01$ and $\mcore < 7 \msun$) in such a way that the role of \yc in the total energy production is almost eliminated. This means that the mass of the RSG envelope does not significantly impact the total luminosity.

Our finding about the uncertain current value of \menv of a RSG (and thus total current mass) has several implications for massive star evolution. First, two RSGs at a similar location in the HR diagram may have very different \menv and total mass. Second, not knowing the value of \menv makes it difficult to estimate the fraction of the envelope that will be lost as the star evolves during He core burning. This has effects on the duration of the plateau in the supernova lightcurve for those stars that are able to retain their H envelope. This is particularly relevant in light of the recent downward revision of the mass-loss rates of RSGs \citep[e.g.,][]{Beasor:2020}. Our \snap models open the possibility that RSGs could have much lower or much higher \menv than currently thought, however binary population synthesis models are needed to assess the distribution of \menv. The distribution of allowed \menv is not flat and based on our current knowledge of single and binary star evolution, some masses are preferred \citep{Zapartas:2019}.

This analysis can be extended to RSGs at the end of their lives, which is especially interesting for the fate of the star, supernova light curve properties and the compact remnant mass. \citet{Farrell2020} applied the \snap model approach to investigate RSGs at the end of their lives and found that it is not possible to determine the final mass of a red supergiant (RSG) at the pre-supernova (SN) stage from its luminosity L and effective temperature \teff alone. This result applies to RSG progenitors of core collapse supernovae, failed supernovae and direct collapse black holes.

We now turn our attention to the radius and \teff of RSGs. These quantities have significant impact on the morphological appearance of these stars and on the post-explosion properties, such as the early time lightcurve \citep[e.g.,][]{Dessart:2013, Gonzalez-Gaitan:2015,Morozova2015,Morozova2018,Hillier2019}. Because of their deep convective envelopes, RSGs are also ideal laboratories for studying the properties of convective mixing. For instance, it is well known that the \teff of RSGs in stellar models are strongly affected by the choice of mixing length parameter, \amlt \citep[e.g.,][]{Henyey:1965, Stothers:1995, Chun:2018}. In the models of \citet{Chun:2018}, the \teff of RSGs varies by up to $\sim$ 800\,K for different choices of \amlt. Our models show that, in addition to convective mixing, different envelope masses produce a variation of up to $\sim 400\,K$ for the same core mass.
This suggests that when calibrating the mixing length parameter \amlt by using the \teff of RSGs, it may be important to consider that the core and envelope masses may be substantially different to what is predicted by stellar evolution models. Possible processes that would modify \mcore and \menv include convective overshooting on the MS or mass loss during the RSG phase.

Our results about the behavior of the stellar radius as a function of envelope mass are also relevant for RSGs in binary systems, of which there are many observations \citep[e.g.,][]{Hansen:1944, McLaughlin:1950, Wright:1970, Stencel:1984, HagenBauer:2014, Harper:2016, Neugent:2018, Neugent:2019}.When a RSG undergoes RLOF, it often results in non-conservative mass-transfer. As the star loses mass via RLOF, the radius of the star increases faster than the Roche-Lobe increases \citep{Eggleton:2006}. 

Our models show that the maximum radius of RSGs occurs for a core mass ratio of $\approx 0.6$. This means that even in thermal equilibrium, as envelope mass is lost from the star, we expect that the envelope will continue to increase in radius until the core mass ratio increases above 0.6. For higher core mass ratios, the radius decreases with decreasing \menv and the mass transfer episode may finish.

\subsection{Stripped stars} \label{sec:disc:strip}

\begin{figure} 
	\centering
	\includegraphics[width=\hsize]{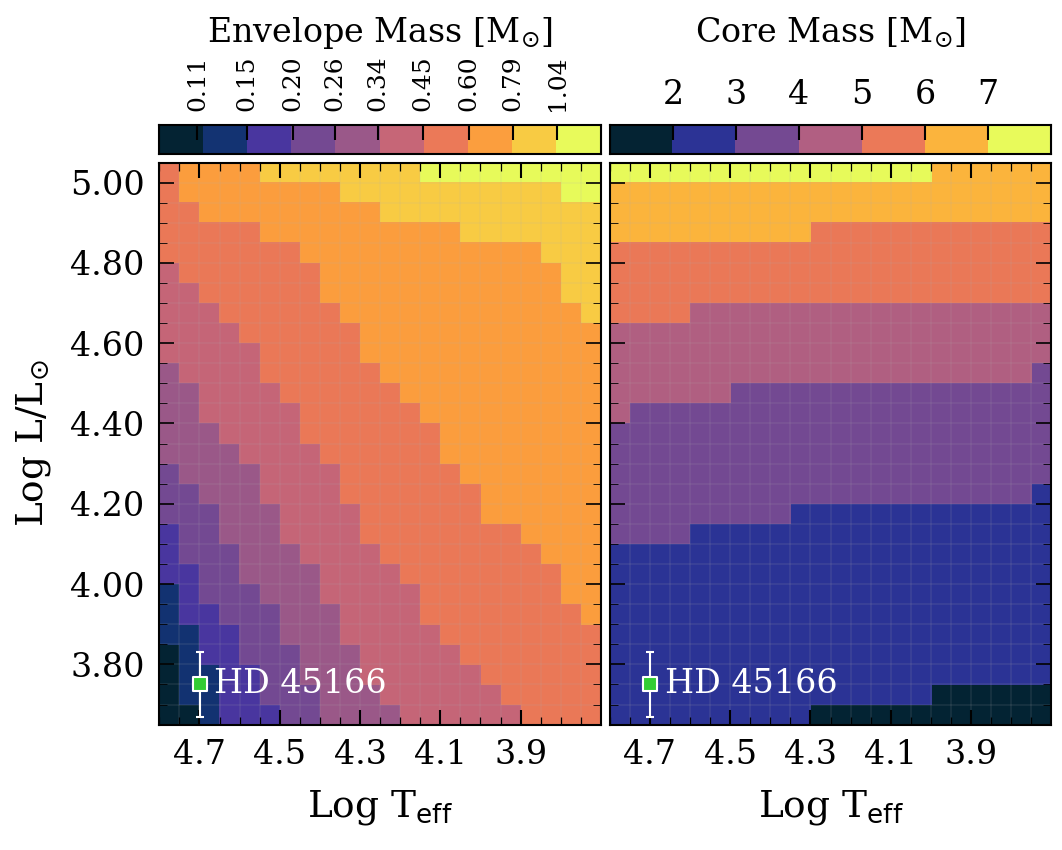}
	\caption{\textit{Left}: Envelope Mass \menv that we derive by interpolating between our \snap models of stripped stars (with $\yc = 0.50$) for a range of luminosities and \teff. \textit{Right}: Same as \textit{left} but for the Core Mass \mcore.}
	\label{fig:stripped_stars}
\end{figure}

Based on the number of stars in binary systems, we expect a large number of stars to exist that are stripped of their hydrogen-rich envelope \citep{Sana:2012}. However, only a small number have actually been observed \citep[e.g.,][]{Gies:1998, Groh:2008, Peters:2013, wang:2017, Chojnowski:2018}, This may be due to the presence of companion stars that are brighter at visual wavelengths \citep{Gotberg:2018} or as a result of biases and selection effects \citep{deMink:2014, Schootemeijer:2018}. 

Our results have implications for the detectability of stripped stars. Our results show that the surface properties of stripped stars, in particular the effective temperature, strongly depend on the mass of the envelope left after binary interaction. This will change the flux distribution in different filters, and impact the completeness limit of future observing surveys that will aim to detect those objects.

By interpolating between our \snap models, it is possible to determine the allowed values of \mcore and \menv for a given observed luminosity and \teff of a stripped star. Knowledge of the core and envelope masses of stripped stars can provide constraints on the physics of RLOF in binary models. When combined with a mass-loss rate, the values of \mcore and \menv can be used to infer their final fates. Depending on the mass of hydrogen left in the envelope at the end of their evolution, these stripped stars are likely to produce SNe IIb or Ib \citep{Podsiadlowski:1993, Woosley:1994, Filippenko:1997, Stancliffe:2009, Gal-Yam:2012, Gotberg:2017}.

Most stripped stars detected so far have low mass, with HD 45166 being the most massive and only detected system within the mass range that can be compared to our models. Based on orbital dynamics, the primary star has a current mass of 4.2~ \msun \citep{Steiner:2005}. Spectroscopic analysis using CMFGEN radiative transfer models derived a luminosity of $\logllsun = 3.75$, \teff of $50\,000$K and mass-loss rate of $2.2\times10^{-7}\msun$/yr \citep{Groh:2008}. Using our models, and assuming $\yc = 0.50$ (the middle of core-He burning), we derive $\mcore = 2.30 \pm 0.15 \msun$ and $\menv = 0.15 \pm 0.02 \msun$. Taking into account the range of possible values of \yc from 0.98 to 0.00, we obtain $\mcore = 2.30^{+0.35}_{-0.23} \msun$ and $\menv = 0.15^{+0.11}_{-0.08} \msun$. 

This mass is consistent with detailed binary models \citep[e.g.,][]{Gotberg:2018}. However it is lower than the mass of $4.2 \pm 0.7 \msun$ obtained by \citet{Steiner:2005}. There are several possibilities for this discrepancy. First, it may be due to the assumptions for the secondary star to HD 45166, which is assumed to be $4.8 \msun$ based on its B7V spectral type \citep{Steiner:2005}. For example, the secondary star may be out of thermal equilibrium. Assuming our primary mass is correct, the secondary star has a mass of $2.70 \msun$ based on the mass ratio derived by \citet{Steiner:2005}. Secondly, it is possible that the primary star is out of thermal and/or hydrostatic equilibrium, in which case our models would not be applicable. It is also possible, but less likely, that HD45166 is a post core-He burning star, but in this case the derived value of \mcore and total mass would be even lower. Regardless of the exact mass of HD 45166, its He core is massive enough to explode as a CCSN. We expect that this mass discrepancy is much larger than  potential uncertainties in our \snap models (see Sec. \ref{sec:caveats}.)

The energy generation of core-He burning stripped stars is dominated by \mcore. For stripped stars with high \teff, a given luminosity and \teff could correspond to a high value of \mcore and high \yc or a lower value of \mcore and low \yc. For a given luminosity and \teff, most of the uncertainty in the value of \mcore is due to the degeneracy between \mcore and \yc, rather than the observational uncertainty in the luminosity and \teff. For a given \teff, the value of \mcore and \menv increases with increasing luminosity while for a given luminosity, the value of \mcore increases and \menv decreases with increasing \teff. This behaviour is illustrated in Fig. \ref{fig:stripped_stars}, where we show the different values allowed for \menv and \mcore for a given position in the HR diagram.

We will provide an online tool to derive the values of \mcore, \menv of stripped stars and the associated uncertainties based on the luminosity and \teff which we hope will be of great benefit to interpret future observations of stripped stars. When such further observations are available, by deriving the values of \mcore and \menv of a population of stripped stars, and with knowledge of the lifetimes of these burning stages, it may be possible to estimate a mass loss rate for hot, stripped stars.

\subsection{Additional Caveats} \label{sec:caveats}
One of the advantages of studying \snap stellar structure models is that they are independent of many of the usual sources of uncertainty that affect stellar evolution models. These include mass loss by stellar winds, mass-exchange during a binary interaction, the evolutionary effects of rotation and of convective overshooting. However, our approach is still subject to uncertainties of physical inputs to the models. We discuss these below.

% Opacities
We use Type I and Type II opacities from OPAL \citep{Iglesias:1993, Iglesias:1996} at high temperatures and opacities from \citet{Ferguson:2005} at low temperatures. The opacities for H, He, C, N and O are quite well known. Most of the uncertainty arises in the Fe opacities \citep{Bailey:2015} and these can have quite substantial effects on the structure of the star. Changes to the opacities could in some cases systematically shift the \teff of our models.

% EOS
The equation of state (EOS) used in \mesa is from OPAL \citep{Rogers:2002}, SCVH \citep{Saumon:1995} and the HELM EOS \citep{Timmes:2000}. For stars during core-H and core-He burning, the EOS is expected to be relatively accurate \citep{Timmes:2000}. Therefore, it is likely that any uncertainties associated with the EOS do not have large effects on the models in this grid.

% Nuclear Reaction Rates
The nuclear reaction rates are a third potential source of uncertainty. The uncertainties in the reactions during the CNO-cycle and in the triple alpha reaction are relatively small, however there is some uncertainty in the rate of $^{12}C(\alpha,\gamma)^{16}O$ \citep{deBoer:2017}. The effect of any uncertainties associated with these nuclear reaction rates on the \snap structure models in this work is likely small.

% Outer Boundary
We use \mesa's \texttt{simple\_photosphere atmosphere} boundary condition which applies a simple grey atmosphere. This treatment of the outer boundary is likely appropriate for our models as they are still far from the Eddington limit. Stellar evolution models have shown that the treatment of the outer boundary will have a larger effect on the radius for high mass stars close to the Eddington limit \citep{Langer:1989b, Schaerer:1996, Schaerer:1996a, Grafener:2012, Groh:2014}.

% Rotation
All of our models are non-rotating. Massive stars exhibit a range of rotational velocities \citep{Hunter:2008, Huang:2010, Ramirez-Agudelo:2013, Ramirez-Agudelo:2015, Dufton:2019} and rotation can have an important impact on their evolution \citep{Maeder:1987, Meynet:2000, Heger:2005, Brott:2011a}. However as our models are not evolving, only the hydrostatic effect of rotation could potentially affect the results in this paper. This effect is small, except for fast rotating stars \citep{Maeder:2000}.

% Magnetic Fields
Our models do not take into account the effect of internal or surface magnetic fields. However, as we consider only structure models in this paper, the evolutionary effects of magnetic fields on mass-loss rates and angular momentum loss rates \citep[e.g.,][]{Meynet:2011, Keszthelyi:2019} likely do not have significant effects on our analysis. The hydrostatic effect of a magnetic field may have a small impact on the structure and energy transport inside the star, but further work is needed to properly address this effect.

% Mixing length theory
We adopt a mixing length parameter of $\amlt = 1.82$ for the models in this paper. This value was calibrated based on solar observations by \citet{Choi:2016}. The choice of \amlt can significantly affect the \teff of RSGs. A higher value of \amlt produces RSGs with lower values of \teff. \citet{Chun:2018} compared observations of RSGs to stellar evolution models and found that models $\amlt = 2$ or 2.5 best reproduced the observations. 

% Abundance structure in the star
In this paper, we select 3 key structural properties, i.e. the \mcore, \menv and core composition in terms of \xc and \yc, and draw connections with the surface properties. However, these structural properties clearly do not fully describe the interior of a star. Other structural properties, such as different abundance profiles in the envelope due to mixing, mass-accretion from a companion or mergers, may have important effects on the surface properties. For example, \citet{Schootemeijer:2019a} studied the properties and lifetimes of red and blue supergiants in terms of the H-gradient outside the He-core.

% Must be in thermal equilibrium
Most stars are in hydrostatic and in thermal equilibrium, that is the radial acceleration is zero and that the luminosity emitted at the surface is equal to the rate of energy production by nuclear reactions in the interior. The method described in this paper to construct \snap stellar structure models is only appropriate for stars that are in hydrostatic and thermal equilibrium. It is not possible to construct stellar structure model after the end of core-He burning or close to core-collapse as the envelope is out of thermal equilibrium during these stages. The latest stage in the evolution of a star at which is is possible to consistently construct \snap structure models is the end of core-He burning.

% _______________________________________________________________

\subsection{Directions for Future Work}
Our \snap model approach provides the foundation for several possible directions for future work. We briefly outline some of them below.

In this work, we studied models at solar metallicity. Our analysis can be extended for stars at lower metallicity. Several papers have studied the evolution of stars at different metallicities \citep[e.g.,][]{Brott:2011, Yoon:2012, Szecsi:2015, Choi:2016, Groh:2019a}. \snap models may help us to understand the effects of metallicity on stellar structure, independent of the evolutionary effects of mass loss, rotation and binaries.

While we cannot compute stellar structure models in hydrostatic and thermal equilibrium at the end of central carbon burning (due to the fact that the envelopes are out of thermal equilibrium at this point), the structure models at the end of core-He burning could be evolved to the end of core-C burning to study the surface properties of supernova progenitors. Furthermore, these models could be evolved to core-collapse and exploded in 1-D explosion models, similar to the approach of \citet{Ugliano:2012}, \citet{Sukhbold:2016} and others. This would allow us to systematically study the appearance of supernovae as a function of progenitor structure.

\snap stellar structure models may help to shed light on what affects the \teff of post-main sequence massive stars. As well as insight into stellar evolution, understanding which stars live as BSGs and RSGs has implications for the number of ionising photons emitted by stars, the kinetic energy feedback to a galaxy and chemical yields from massive stars.

Our \snap models will benefit greatly from current and future astero-seismological results for massive stars, such as from the TESS and PLATO space missions. In particular, improvements in our understanding of stellar structure from astero-seismological studies \citep{Buldgen:2015, Eggenberger:2017, Aerts:2019a} could be implemented in our models in the future. \snap stellar structure models could also be applied to low mass stars which benefit from larger observational samples.

Gravitational wave observations of double neutron star mergers \citep{Abbott:2017, Abbott:2017a} and double black hole mergers \citep{Abbott:2016} have opened a new frontier in astrophysics. These observations have provided new insights into compact objects and the endpoints of massive stars. Our \snap stellar structure models could be used to study the mass function of compact remnants and the boundaries between white dwarfs and neutron stars, and neutron stars and black holes.

%_______________________________________________________________

\section{Conclusions} \label{sec:conclusions}
In this paper, we introduced our \snap technique to construct stellar structure models in hydrostatic and thermal equilibrium. We then applied our approach to study the surface properties of core-H and core-He burning stars with a range of core and envelope masses and core compositions.

\begin{enumerate}
    \item We find that there is a limited range of core masses, envelope masses and core compositions that can form core-H burning structures in hydrostatic and thermal equilibrium. We quantified the relationship between the convective core mass and the total mass for different central H mass fractions.
    
    \item Over a wide range of He-core masses ($\mcore \approx 2 - 9 \msun$), core-He burning stars show similar trends in luminosity and \teff as a function of the core mass ratio (\mcore/\mtotal).
    
    \item Our models with core mass ratios of $\mcore/\mtotal > 0.8$ correspond to stripped stars produced as a consequence of significant mass loss or binary interaction. They show that \teff has a strong dependence on \menv (due to the increased effect of opacity from the H-rich envelope), \mcore and the core composition. When a large observational sample of stripped stars becomes available, our results can be used to constrain their \mcore, \menv, mass-loss rates and the physics of binary interaction. Our models also show that the surface luminosity of these stars increases slightly with increasing envelope mass due to increased energy generation in the H-shell, in which $0 - 25$ per cent of the total nuclear energy is generated.
    
    \item Stars with $\mcore/\mtotal$ from 0.2 to 0.8 have convective outer envelopes, low \teff and  will appear as RSGs. They exhibit a small variation in luminosity (0.02 dex) and \teff ($\sim 400 \mathrm{K}$), over a wide range of envelope masses ($\sim 2 - 17 \msun$). This means that given current uncertainties in the physics driving stellar evolution, it is not possible to derive red supergiant masses from luminosities and \teff alone. In these stars, we find that 10 to 35 per cent of the nuclear energy generation occurs in the H-shell, depending on the core mass (\mcore) and the central He mass fraction (\yc). We derive the following relationship between \mcore and the total luminosity of a red supergiant during core He burning: $\log \mcore \simeq 0.44\logllsun - 1.38$.
    
    \item At $\mcore/\mtotal \approx 0.2$, our models exhibit a bi-stability in the solution of stellar structure equations. The solution of the stellar structure equations switches from a convective outer envelope with a radiative H-burning shell to a radiative outer envelope with a convective H-burning shell. This switch is accompanied by a large increase in luminosity and \teff.
    
    \item Stars with greater than 80 per cent of the mass in the H-envelope correspond to mass gainers and merger products. The luminosity and \teff of these stars are dominated by properties of the envelope. More than 70 per cent of their energy generation comes from the H-shell. Some of these stars may resemble OB-type stars and others may resemble blue supergiants.
    
    \item For a constant envelope mass and He-core composition, the total energy produced in the H-shell increases with increasing core mass. This is because higher mass cores produce a larger, hotter H-burning shell which increases CNO burning in the shell. Despite this, the fraction of energy produced in the H-shell decreases with increasing core mass. This is because higher mass cores are hotter and produce more energy which means the stars requires less energy generation in the H-shell to support an envelope of a given mass.
    
    \item For core-He burning stars with the same core mass (\mcore) and envelope mass (\menv), the luminosity of the He-core increases with decreasing \yc, due to the effect of the mean molecular weight of the core (\mucore). For stars with envelope masses of $\menv \lesssim 1 \msun$, the increased luminosity of the He-core results in an increased surface luminosity. However, for stars with $\menv \gtrsim 1 \msun$, the increased luminosity in the He-core is nearly cancelled out by a corresponding decrease in the luminosity of the H-shell. As a result, in these stars the luminosity at the surface is not strongly affected by \mucore, and is set by \mcore and \menv.
    
\end{enumerate}

\section*{Acknowledgements}
We thank the referee, Prof. Alex de Koter, for his detailed review and comments that improved our manuscript. We would like to thank the \mesa (Modules for Experiments in Stellar Astrophysics) collaboration for making their software package freely available to the astrophysics community \citep{Paxton:2011, Paxton:2013, Paxton:2015, Paxton:2018, Paxton:2019}.  E.F. would like to thank GG for helping to create Fig. \ref{fig:method_summary} and acknowledges funding from IRC Project 208026, Award 15330. GM, SE, and CG have received funding from the European Research Council (ERC) under the European Union's Horizon 2020 research and innovation program (grant agreement No 833925, project STAREX).

%_______________________________________________________________

\bibliographystyle{mnras}
\bibliography{main}

\begin{thebibliography}{}
\makeatletter
\relax
\def\mn@urlcharsother{\let\do\@makeother \do\$\do\&\do\#\do\^\do\_\do\%\do\~}
\def\mn@doi{\begingroup\mn@urlcharsother \@ifnextchar [ {\mn@doi@}
  {\mn@doi@[]}}
\def\mn@doi@[#1]#2{\def\@tempa{#1}\ifx\@tempa\@empty \href
  {http://dx.doi.org/#2} {doi:#2}\else \href {http://dx.doi.org/#2} {#1}\fi
  \endgroup}
\def\mn@eprint#1#2{\mn@eprint@#1:#2::\@nil}
\def\mn@eprint@arXiv#1{\href {http://arxiv.org/abs/#1} {{\tt arXiv:#1}}}
\def\mn@eprint@dblp#1{\href {http://dblp.uni-trier.de/rec/bibtex/#1.xml}
  {dblp:#1}}
\def\mn@eprint@#1:#2:#3:#4\@nil{\def\@tempa {#1}\def\@tempb {#2}\def\@tempc
  {#3}\ifx \@tempc \@empty \let \@tempc \@tempb \let \@tempb \@tempa \fi \ifx
  \@tempb \@empty \def\@tempb {arXiv}\fi \@ifundefined
  {mn@eprint@\@tempb}{\@tempb:\@tempc}{\expandafter \expandafter \csname
  mn@eprint@\@tempb\endcsname \expandafter{\@tempc}}}

\bibitem[\protect\citeauthoryear{Abbott et~al.,}{Abbott
  et~al.}{2016}]{Abbott:2016}
Abbott B.~P.,  et~al., 2016, \mn@doi [\prl] {10.1103/PhysRevLett.116.061102},
  116, 061102

\bibitem[\protect\citeauthoryear{Abbott et~al.,}{Abbott
  et~al.}{2017a}]{Abbott:2017}
Abbott B.~P.,  et~al., 2017a, \mn@doi [\prl] {10.1103/PhysRevLett.119.161101},
  119, 161101

\bibitem[\protect\citeauthoryear{Abbott et~al.,}{Abbott
  et~al.}{2017b}]{Abbott:2017a}
Abbott B.~P.,  et~al., 2017b, \mn@doi [\apjl] {10.3847/2041-8213/aa91c9}, 848,
  L12

\bibitem[\protect\citeauthoryear{Aerts et~al.,}{Aerts
  et~al.}{2019}]{Aerts:2019a}
Aerts C.,  et~al., 2019, \mn@doi [\aap] {10.1051/0004-6361/201834762}, 624, A75

\bibitem[\protect\citeauthoryear{{Aguilera-Dena}, Langer, Moriya  \&
  Schootemeijer}{{Aguilera-Dena} et~al.}{2018}]{Aguilera-Dena:2018}
{Aguilera-Dena} D.~R.,  Langer N.,  Moriya T.~J.,   Schootemeijer A.,  2018,
  \mn@doi [\apjl] {10.3847/1538-4357/aabfc1}, 858, 115

\bibitem[\protect\citeauthoryear{Almeida et~al.,}{Almeida
  et~al.}{2017}]{Almeida:2017}
Almeida L.~A.,  et~al., 2017, \mn@doi [\aap] {10.1051/0004-6361/201629844},
  598, A84

\bibitem[\protect\citeauthoryear{Bailey et~al.,}{Bailey
  et~al.}{2015}]{Bailey:2015}
Bailey J.~E.,  et~al., 2015, \mn@doi [\nat] {10.1038/nature14048}, 517, 56

\bibitem[\protect\citeauthoryear{Beasor \& Davies}{Beasor \&
  Davies}{2018}]{Beasor:2018}
Beasor E.~R.,  Davies B.,  2018, \mn@doi [\mnras] {10.1093/mnras/stx3174}, 475,
  55

\bibitem[\protect\citeauthoryear{Beasor, Davies, Smith, {van Loon}, Gehrz  \&
  Figer}{Beasor et~al.}{2020}]{Beasor:2020}
Beasor E.~R.,  Davies B.,  Smith N.,  {van Loon} J.~T.,  Gehrz R.~D.,   Figer
  D.~F.,  2020, arXiv:2001.07222 [astro-ph]

\bibitem[\protect\citeauthoryear{Belczynski, Kalogera  \& Bulik}{Belczynski
  et~al.}{2002}]{Belczynski:2002}
Belczynski K.,  Kalogera V.,   Bulik T.,  2002, \mn@doi [\apjl]
  {10.1086/340304}, 572, 407

\bibitem[\protect\citeauthoryear{Belczynski, Kalogera, Rasio, Taam, Zezas,
  Bulik, Maccarone  \& Ivanova}{Belczynski et~al.}{2008}]{Belczynski:2008}
Belczynski K.,  Kalogera V.,  Rasio F.~A.,  Taam R.~E.,  Zezas A.,  Bulik T.,
  Maccarone T.~J.,   Ivanova N.,  2008, \mn@doi [\apjl Supplement Series]
  {10.1086/521026}, 174, 223

\bibitem[\protect\citeauthoryear{Boian \& Groh}{Boian \&
  Groh}{2018}]{Boian:2018}
Boian I.,  Groh J.~H.,  2018, \mn@doi [\aap] {10.1051/0004-6361/201731794},
  617, A115

\bibitem[\protect\citeauthoryear{Braun \& Langer}{Braun \&
  Langer}{1995}]{Braun:1995}
Braun H.,  Langer N.,  1995, \aap, 297, 483

\bibitem[\protect\citeauthoryear{Brott et~al.,}{Brott
  et~al.}{2011a}]{Brott:2011}
Brott I.,  et~al., 2011a, \mn@doi [\aap] {10.1051/0004-6361/201016113}, 530,
  A115

\bibitem[\protect\citeauthoryear{Brott et~al.,}{Brott
  et~al.}{2011b}]{Brott:2011a}
Brott I.,  et~al., 2011b, \mn@doi [\aap] {10.1051/0004-6361/201016114}, 530,
  A116

\bibitem[\protect\citeauthoryear{Buldgen, Reese  \& Dupret}{Buldgen
  et~al.}{2015}]{Buldgen:2015}
Buldgen G.,  Reese D.~R.,   Dupret M.~A.,  2015, \mn@doi [\aap]
  {10.1051/0004-6361/201526390}, 583, A62

\bibitem[\protect\citeauthoryear{Castro, Fossati, Langer, {Sim{\'o}n-D{\'i}az},
  Schneider  \& Izzard}{Castro et~al.}{2014}]{Castro:2014a}
Castro N.,  Fossati L.,  Langer N.,  {Sim{\'o}n-D{\'i}az} S.,  Schneider F.
  R.~N.,   Izzard R.~G.,  2014, \mn@doi [\aap] {10.1051/0004-6361/201425028},
  570, L13

\bibitem[\protect\citeauthoryear{Chieffi \& Limongi}{Chieffi \&
  Limongi}{2013}]{Chieffi:2013}
Chieffi A.,  Limongi M.,  2013, \mn@doi [\apjl] {10.1088/0004-637X/764/1/21},
  764, 21

\bibitem[\protect\citeauthoryear{Chini, Hoffmeister, Nasseri, Stahl  \&
  Zinnecker}{Chini et~al.}{2012}]{Chini:2012}
Chini R.,  Hoffmeister V.~H.,  Nasseri A.,  Stahl O.,   Zinnecker H.,  2012,
  \mn@doi [\mnras] {10.1111/j.1365-2966.2012.21317.x}, 424, 1925

\bibitem[\protect\citeauthoryear{Chiosi \& Maeder}{Chiosi \&
  Maeder}{1986}]{Chiosi:1986}
Chiosi C.,  Maeder A.,  1986, \mn@doi [\araa]
  {10.1146/annurev.aa.24.090186.001553}, 24, 329

\bibitem[\protect\citeauthoryear{Choi, Dotter, Conroy, Cantiello, Paxton  \&
  Johnson}{Choi et~al.}{2016}]{Choi:2016}
Choi J.,  Dotter A.,  Conroy C.,  Cantiello M.,  Paxton B.,   Johnson B.~D.,
  2016, \mn@doi [\apjl] {10.3847/0004-637X/823/2/102}, 823, 102

\bibitem[\protect\citeauthoryear{Chojnowski et~al.,}{Chojnowski
  et~al.}{2018}]{Chojnowski:2018}
Chojnowski S.~D.,  et~al., 2018, \mn@doi [\apjl] {10.3847/1538-4357/aad964},
  865, 76

\bibitem[\protect\citeauthoryear{Chun, Yoon, Jung, Kim  \& Kim}{Chun
  et~al.}{2018}]{Chun:2018}
Chun S.-H.,  Yoon S.-C.,  Jung M.-K.,  Kim D.~U.,   Kim J.,  2018, \mn@doi
  [\apjl] {10.3847/1538-4357/aa9a37}, 853, 79

\bibitem[\protect\citeauthoryear{Cox \& Salpeter}{Cox \&
  Salpeter}{1961}]{Cox:1961}
Cox J.~P.,  Salpeter E.~E.,  1961, \mn@doi [\apjl] {10.1086/147082}, 133, 764

\bibitem[\protect\citeauthoryear{De~Beck, Decin, {de Koter}, Justtanont,
  Verhoelst, Kemper  \& Menten}{De~Beck et~al.}{2010}]{DeBeck:2010}
De~Beck E.,  Decin L.,  {de Koter} A.,  Justtanont K.,  Verhoelst T.,  Kemper
  F.,   Menten K.~M.,  2010, \mn@doi [\aap] {10.1051/0004-6361/200913771}, 523,
  A18

\bibitem[\protect\citeauthoryear{Dessart, Hillier, Waldman  \& Livne}{Dessart
  et~al.}{2013}]{Dessart:2013}
Dessart L.,  Hillier D.~J.,  Waldman R.,   Livne E.,  2013, \mn@doi [\mnras]
  {10.1093/mnras/stt861}, 433, 1745

\bibitem[\protect\citeauthoryear{Dominik, Belczynski, Fryer, Holz, Berti,
  Bulik, Mandel  \& O'Shaughnessy}{Dominik et~al.}{2012}]{Dominik:2012}
Dominik M.,  Belczynski K.,  Fryer C.,  Holz D.~E.,  Berti E.,  Bulik T.,
  Mandel I.,   O'Shaughnessy R.,  2012, \mn@doi [\apjl]
  {10.1088/0004-637X/759/1/52}, 759, 52

\bibitem[\protect\citeauthoryear{Dufton, Evans, Hunter, Lennon  \&
  Schneider}{Dufton et~al.}{2019}]{Dufton:2019}
Dufton P.~L.,  Evans C.~J.,  Hunter I.,  Lennon D.~J.,   Schneider F. R.~N.,
  2019, \mn@doi [\aap] {10.1051/0004-6361/201935415}, 626, A50

\bibitem[\protect\citeauthoryear{Dunstall et~al.,}{Dunstall
  et~al.}{2015}]{Dunstall:2015}
Dunstall P.~R.,  et~al., 2015, \mn@doi [\aap] {10.1051/0004-6361/201526192},
  580, A93

\bibitem[\protect\citeauthoryear{Eggenberger et~al.,}{Eggenberger
  et~al.}{2017}]{Eggenberger:2017}
Eggenberger P.,  et~al., 2017, \mn@doi [\aap] {10.1051/0004-6361/201629459},
  599, A18

\bibitem[\protect\citeauthoryear{Eggleton}{Eggleton}{2006}]{Eggleton:2006}
Eggleton P.,  2006, Evolutionary {{Processes}} in {{Binary}} and {{Multiple
  Stars}}.
{Cambridge University Press}

\bibitem[\protect\citeauthoryear{Eggleton, Fitchett  \& Tout}{Eggleton
  et~al.}{1989}]{Eggleton:1989}
Eggleton P.~P.,  Fitchett M.~J.,   Tout C.~A.,  1989, \mn@doi [\apjl]
  {10.1086/168190}, 347, 998

\bibitem[\protect\citeauthoryear{Eggleton, Faulkner  \& Cannon}{Eggleton
  et~al.}{1998}]{Eggleton:1998}
Eggleton P.~P.,  Faulkner J.,   Cannon R.~C.,  1998, \mn@doi [\mnras]
  {10.1046/j.1365-8711.1998.01655.x}, 298, 831

\bibitem[\protect\citeauthoryear{Ekstr{\"o}m et~al.,}{Ekstr{\"o}m
  et~al.}{2012}]{Ekstrom:2012}
Ekstr{\"o}m S.,  et~al., 2012, \mn@doi [\aap] {10.1051/0004-6361/201117751},
  537, A146

\bibitem[\protect\citeauthoryear{Eldridge \& Stanway}{Eldridge \&
  Stanway}{2009}]{Eldridge:2009}
Eldridge J.~J.,  Stanway E.~R.,  2009, \mn@doi [\mnras]
  {10.1111/j.1365-2966.2009.15514.x}, 400, 1019

\bibitem[\protect\citeauthoryear{Eldridge \& Tout}{Eldridge \&
  Tout}{2004}]{Eldridge:2004}
Eldridge J.~J.,  Tout C.~A.,  2004, \mn@doi [\mnras]
  {10.1111/j.1365-2966.2004.08041.x}, 353, 87

\bibitem[\protect\citeauthoryear{Eldridge, Izzard  \& Tout}{Eldridge
  et~al.}{2008}]{Eldridge:2008}
Eldridge J.~J.,  Izzard R.~G.,   Tout C.~A.,  2008, \mn@doi [\mnras]
  {10.1111/j.1365-2966.2007.12738.x}, 384, 1109

\bibitem[\protect\citeauthoryear{Eldridge, Stanway, Xiao, McClelland, Taylor,
  Ng, Greis  \& Bray}{Eldridge et~al.}{2017}]{Eldridge:2017}
Eldridge J.~J.,  Stanway E.~R.,  Xiao L.,  McClelland L. A.~S.,  Taylor G.,  Ng
  M.,  Greis S. M.~L.,   Bray J.~C.,  2017, \mn@doi [\pasa]
  {10.1017/pasa.2017.51}, 34, e058

\bibitem[\protect\citeauthoryear{{Farrell}, {Groh}, {Meynet}  \&
  {Eldridge}}{{Farrell} et~al.}{2020}]{Farrell2020}
{Farrell} E.~J.,  {Groh} J.~H.,  {Meynet} G.,   {Eldridge} J.~J.,  2020,
  \mn@doi [\mnras] {10.1093/mnrasl/slaa035}, \href
  {https://ui.adsabs.harvard.edu/abs/2020MNRAS.tmpL..33F} {}

\bibitem[\protect\citeauthoryear{Ferguson, Alexander, Allard, Barman, Bodnarik,
  Hauschildt, {Heffner-Wong}  \& Tamanai}{Ferguson
  et~al.}{2005}]{Ferguson:2005}
Ferguson J.~W.,  Alexander D.~R.,  Allard F.,  Barman T.,  Bodnarik J.~G.,
  Hauschildt P.~H.,  {Heffner-Wong} A.,   Tamanai A.,  2005, \mn@doi [\apjl]
  {10.1086/428642}, 623, 585

\bibitem[\protect\citeauthoryear{Filippenko}{Filippenko}{1997}]{Filippenko:1997}
Filippenko A.~V.,  1997, \mn@doi [\araa] {10.1146/annurev.astro.35.1.309}, 35,
  309

\bibitem[\protect\citeauthoryear{Fragos, Andrews, {Ramirez-Ruiz}, Meynet,
  Kalogera, Taam  \& Zezas}{Fragos et~al.}{2019}]{Fragos:2019}
Fragos T.,  Andrews J.~J.,  {Ramirez-Ruiz} E.,  Meynet G.,  Kalogera V.,  Taam
  R.~E.,   Zezas A.,  2019, \mn@doi [\apjl] {10.3847/2041-8213/ab40d1}, 883,
  L45

\bibitem[\protect\citeauthoryear{Fraser et~al.,}{Fraser
  et~al.}{2013}]{Fraser:2013}
Fraser M.,  et~al., 2013, \mn@doi [\mnras] {10.1093/mnras/stt813}, 433, 1312

\bibitem[\protect\citeauthoryear{Fuller}{Fuller}{2017}]{Fuller:2017}
Fuller J.,  2017, \mn@doi [\mnras] {10.1093/mnras/stx1314}, 470, 1642

\bibitem[\protect\citeauthoryear{{Gal-Yam}}{{Gal-Yam}}{2012}]{Gal-Yam:2012}
{Gal-Yam} A.,  2012, \mn@doi [\sci] {10.1126/science.1203601}, 337, 927

\bibitem[\protect\citeauthoryear{{Gal-Yam} \& Leonard}{{Gal-Yam} \&
  Leonard}{2009}]{Gal-Yam:2009}
{Gal-Yam} A.,  Leonard D.~C.,  2009, \mn@doi [\nat] {10.1038/nature07934}, 458,
  865

\bibitem[\protect\citeauthoryear{{Gal-Yam} et~al.,}{{Gal-Yam}
  et~al.}{2014}]{Gal-Yam:2014}
{Gal-Yam} A.,  et~al., 2014, \mn@doi [\nat] {10.1038/nature13304}, 509, 471

\bibitem[\protect\citeauthoryear{Garmany, Conti  \& Massey}{Garmany
  et~al.}{1980}]{Garmany:1980}
Garmany C.~D.,  Conti P.~S.,   Massey P.,  1980, \mn@doi [\apjl]
  {10.1086/158537}, 242, 1063

\bibitem[\protect\citeauthoryear{Georgy}{Georgy}{2012}]{Georgy:2012}
Georgy C.,  2012, \mn@doi [\aap] {10.1051/0004-6361/201118372}, 538, L8

\bibitem[\protect\citeauthoryear{Giannone}{Giannone}{1967}]{Giannone:1967a}
Giannone P.,  1967, \zap, 65, 226

\bibitem[\protect\citeauthoryear{Giannone, Kohl  \& Weigert}{Giannone
  et~al.}{1968}]{Giannone:1968}
Giannone P.,  Kohl K.,   Weigert A.,  1968, \zap, 68, 107

\bibitem[\protect\citeauthoryear{Gies, Bagnuolo, Ferrara, Kaye, Thaller, Penny
  \& Peters}{Gies et~al.}{1998}]{Gies:1998}
Gies D.~R.,  Bagnuolo W.~G.,  Ferrara E.~C.,  Kaye A.~B.,  Thaller M.~L.,
  Penny L.~R.,   Peters G.~J.,  1998, \mn@doi [\apjl] {10.1086/305113}, 493,
  440

\bibitem[\protect\citeauthoryear{{Gonz{\'a}lez-Gait{\'a}n}
  et~al.,}{{Gonz{\'a}lez-Gait{\'a}n} et~al.}{2015}]{Gonzalez-Gaitan:2015}
{Gonz{\'a}lez-Gait{\'a}n} S.,  et~al., 2015, \mn@doi [\mnras]
  {10.1093/mnras/stv1097}, 451, 2212

\bibitem[\protect\citeauthoryear{G{\"o}tberg, {de Mink}  \& Groh}{G{\"o}tberg
  et~al.}{2017}]{Gotberg:2017}
G{\"o}tberg Y.,  {de Mink} S.~E.,   Groh J.~H.,  2017, \mn@doi [\aap]
  {10.1051/0004-6361/201730472}, 608, A11

\bibitem[\protect\citeauthoryear{G{\"o}tberg, {de Mink}, Groh, Kupfer,
  Crowther, Zapartas  \& Renzo}{G{\"o}tberg et~al.}{2018}]{Gotberg:2018}
G{\"o}tberg Y.,  {de Mink} S.~E.,  Groh J.~H.,  Kupfer T.,  Crowther P.~A.,
  Zapartas E.,   Renzo M.,  2018, \mn@doi [\aap] {10.1051/0004-6361/201732274},
  615, A78

\bibitem[\protect\citeauthoryear{Gr{\"a}fener, Owocki  \& Vink}{Gr{\"a}fener
  et~al.}{2012}]{Grafener:2012}
Gr{\"a}fener G.,  Owocki S.~P.,   Vink J.~S.,  2012, \mn@doi [\aap]
  {10.1051/0004-6361/201117497}, 538, A40

\bibitem[\protect\citeauthoryear{Grevesse \& Sauval}{Grevesse \&
  Sauval}{1998}]{Grevesse:1998}
Grevesse N.,  Sauval A.~J.,  1998, \mn@doi [Space \sci Reviews]
  {10.1023/A:1005161325181}, 85, 161

\bibitem[\protect\citeauthoryear{Groh}{Groh}{2014}]{Groh:2014a}
Groh J.~H.,  2014, \mn@doi [\aap] {10.1051/0004-6361/201424852}, 572, L11

\bibitem[\protect\citeauthoryear{Groh, Oliveira  \& Steiner}{Groh
  et~al.}{2008}]{Groh:2008}
Groh J.~H.,  Oliveira A.~S.,   Steiner J.~E.,  2008, \mn@doi [\aap]
  {10.1051/0004-6361:200809511}, 485, 245

\bibitem[\protect\citeauthoryear{Groh, Georgy  \& Ekstr{\"o}m}{Groh
  et~al.}{2013}]{Groh:2013a}
Groh J.~H.,  Georgy C.,   Ekstr{\"o}m S.,  2013, \mn@doi [\aap]
  {10.1051/0004-6361/201322369}, 558, L1

\bibitem[\protect\citeauthoryear{Groh, Meynet, Ekstr{\"o}m  \& Georgy}{Groh
  et~al.}{2014}]{Groh:2014}
Groh J.~H.,  Meynet G.,  Ekstr{\"o}m S.,   Georgy C.,  2014, \mn@doi [\aap]
  {10.1051/0004-6361/201322573}, 564, A30

\bibitem[\protect\citeauthoryear{Groh et~al.,}{Groh et~al.}{2019}]{Groh:2019a}
Groh J.~H.,  et~al., 2019, \mn@doi [\aap] {10.1051/0004-6361/201833720}, 627,
  A24

\bibitem[\protect\citeauthoryear{Hagen~Bauer \& Bennett}{Hagen~Bauer \&
  Bennett}{2014}]{HagenBauer:2014}
Hagen~Bauer W.,  Bennett P.~D.,  2014, \mn@doi [\apjl Supplement Series]
  {10.1088/0067-0049/211/2/27}, 211, 27

\bibitem[\protect\citeauthoryear{Hansen \& M}{Hansen \& M}{1944}]{Hansen:1944}
Hansen V.,  M J.,  1944, \mn@doi [\apjl] {10.1086/144632}, 100, 8

\bibitem[\protect\citeauthoryear{Harper, Griffin, Bennett  \& O'Riain}{Harper
  et~al.}{2016}]{Harper:2016}
Harper G.~M.,  Griffin R. E.~M.,  Bennett P.~D.,   O'Riain N.,  2016, \mn@doi
  [\mnras] {10.1093/mnras/stv2668}, 456, 1346

\bibitem[\protect\citeauthoryear{Heger, Langer  \& Woosley}{Heger
  et~al.}{2000}]{Heger:2000}
Heger A.,  Langer N.,   Woosley S.~E.,  2000, \mn@doi [\apjl] {10.1086/308158},
  528, 368

\bibitem[\protect\citeauthoryear{Heger, Woosley  \& Spruit}{Heger
  et~al.}{2005}]{Heger:2005}
Heger A.,  Woosley S.~E.,   Spruit H.~C.,  2005, \mn@doi [\apjl]
  {10.1086/429868}, 626, 350

\bibitem[\protect\citeauthoryear{Hellings}{Hellings}{1983}]{Hellings:1983}
Hellings P.,  1983, \mn@doi [\apss] {10.1007/BF00661941}, 96, 37

\bibitem[\protect\citeauthoryear{Hellings}{Hellings}{1984}]{Hellings:1984}
Hellings P.,  1984, \mn@doi [\apss] {10.1007/BF00653994}, 104, 83

\bibitem[\protect\citeauthoryear{Henyey, Lelevier  \& Levee}{Henyey
  et~al.}{1959}]{Henyey:1959}
Henyey L.~G.,  Lelevier R.,   Levee R.~D.,  1959, \mn@doi [\apjl]
  {10.1086/146590}, 129, 2

\bibitem[\protect\citeauthoryear{Henyey, Vardya  \& Bodenheimer}{Henyey
  et~al.}{1965}]{Henyey:1965}
Henyey L.,  Vardya M.~S.,   Bodenheimer P.,  1965, \mn@doi [\apjl]
  {10.1086/148357}, 142, 841

\bibitem[\protect\citeauthoryear{Herwig}{Herwig}{2000}]{Herwig:2000}
Herwig F.,  2000, \aap, p.~17

\bibitem[\protect\citeauthoryear{Higgins \& Vink}{Higgins \&
  Vink}{2019}]{Higgins:2019}
Higgins E.~R.,  Vink J.~S.,  2019, \mn@doi [\aap]
  {10.1051/0004-6361/201834123}, 622, A50

\bibitem[\protect\citeauthoryear{{Hillier} \& {Dessart}}{{Hillier} \&
  {Dessart}}{2019}]{Hillier2019}
{Hillier} D.~J.,  {Dessart} L.,  2019, \mn@doi [\aap]
  {10.1051/0004-6361/201935100}, \href
  {https://ui.adsabs.harvard.edu/abs/2019A&A...631A...8H} {631, A8}

\bibitem[\protect\citeauthoryear{Huang, Gies  \& McSwain}{Huang
  et~al.}{2010}]{Huang:2010}
Huang W.,  Gies D.~R.,   McSwain M.~V.,  2010, \mn@doi [\apjl]
  {10.1088/0004-637X/722/1/605}, 722, 605

\bibitem[\protect\citeauthoryear{Hunter et~al.,}{Hunter
  et~al.}{2008}]{Hunter:2008}
Hunter I.,  et~al., 2008, \mn@doi [\apjl] {10.1086/587436}, 676, L29

\bibitem[\protect\citeauthoryear{Hurley, Pols  \& Tout}{Hurley
  et~al.}{2000}]{Hurley:2000}
Hurley J.~R.,  Pols O.~R.,   Tout C.~A.,  2000, \mn@doi [\mnras]
  {10.1046/j.1365-8711.2000.03426.x}, 315, 543

\bibitem[\protect\citeauthoryear{Hurley, Tout  \& Pols}{Hurley
  et~al.}{2002}]{Hurley:2002}
Hurley J.~R.,  Tout C.~A.,   Pols O.~R.,  2002, \mn@doi [\mnras]
  {10.1046/j.1365-8711.2002.05038.x}, 329, 897

\bibitem[\protect\citeauthoryear{Iglesias \& Rogers}{Iglesias \&
  Rogers}{1993}]{Iglesias:1993}
Iglesias C.~A.,  Rogers F.~J.,  1993, \mn@doi [\apjl] {10.1086/172958}, 412,
  752

\bibitem[\protect\citeauthoryear{Iglesias \& Rogers}{Iglesias \&
  Rogers}{1996}]{Iglesias:1996}
Iglesias C.~A.,  Rogers F.~J.,  1996, \mn@doi [\apjl] {10.1086/177381}, 464,
  943

\bibitem[\protect\citeauthoryear{Ivanova}{Ivanova}{2018}]{Ivanova:2018}
Ivanova N.,  2018, \mn@doi [\apjl] {10.3847/2041-8213/aac101}, 858, L24

\bibitem[\protect\citeauthoryear{Izzard, Dray, Karakas, Lugaro  \& Tout}{Izzard
  et~al.}{2006}]{Izzard:2006}
Izzard R.~G.,  Dray L.~M.,  Karakas A.~I.,  Lugaro M.,   Tout C.~A.,  2006,
  \mn@doi [\aap] {10.1051/0004-6361:20066129}, 460, 565

\bibitem[\protect\citeauthoryear{{Justham}, {Podsiadlowski}  \&
  {Vink}}{{Justham} et~al.}{2014}]{Justham2014}
{Justham} S.,  {Podsiadlowski} P.,   {Vink} J.~S.,  2014, \mn@doi [\apj]
  {10.1088/0004-637X/796/2/121}, \href
  {https://ui.adsabs.harvard.edu/abs/2014ApJ...796..121J} {796, 121}

\bibitem[\protect\citeauthoryear{Keszthelyi, Meynet, Georgy, Wade, Petit  \&
  {David-Uraz}}{Keszthelyi et~al.}{2019}]{Keszthelyi:2019}
Keszthelyi Z.,  Meynet G.,  Georgy C.,  Wade G.~A.,  Petit V.,   {David-Uraz}
  A.,  2019, \mn@doi [\mnras] {10.1093/mnras/stz772}, 485, 5843

\bibitem[\protect\citeauthoryear{Kobulnicky \& Fryer}{Kobulnicky \&
  Fryer}{2007}]{Kobulnicky:2007}
Kobulnicky H.~A.,  Fryer C.~L.,  2007, \mn@doi [\apjl] {10.1086/522073}, 670,
  747

\bibitem[\protect\citeauthoryear{Kobulnicky et~al.,}{Kobulnicky
  et~al.}{2014}]{Kobulnicky:2014}
Kobulnicky H.~A.,  et~al., 2014, \mn@doi [\apjl Supplement Series]
  {10.1088/0067-0049/213/2/34}, 213, 34

\bibitem[\protect\citeauthoryear{Kotak \& Vink}{Kotak \&
  Vink}{2006}]{Kotak:2006}
Kotak R.,  Vink J.~S.,  2006, \mn@doi [\aap] {10.1051/0004-6361:20065800}, 460,
  L5

\bibitem[\protect\citeauthoryear{Langer}{Langer}{1989}]{Langer:1989b}
Langer N.,  1989, \aap, 210, 93

\bibitem[\protect\citeauthoryear{Langer, El~Eid  \& Baraffe}{Langer
  et~al.}{1989}]{Langer:1989}
Langer N.,  El~Eid M.~F.,   Baraffe I.,  1989, \aap, 224, L17

\bibitem[\protect\citeauthoryear{Lauterborn, Refsdal  \& Weigert}{Lauterborn
  et~al.}{1971a}]{Lauterborn:1971a}
Lauterborn D.,  Refsdal S.,   Weigert A.,  1971a, \aap, 10, 97

\bibitem[\protect\citeauthoryear{Lauterborn, Refsdal  \& Roth}{Lauterborn
  et~al.}{1971b}]{Lauterborn:1971}
Lauterborn D.,  Refsdal S.,   Roth M.~L.,  1971b, \aap, 13, 119

\bibitem[\protect\citeauthoryear{Maeder}{Maeder}{1980}]{Maeder:1980}
Maeder A.,  1980, \aap, 92, 101

\bibitem[\protect\citeauthoryear{Maeder}{Maeder}{2009}]{Maeder:2009}
Maeder A.,  2009, Physics, {{Formation}} and {{Evolution}} of {{Rotating
  Stars}}.
{Springer Berlin Heidelberg}, \mn@doi{10.1007/978-3-540-76949-1}

\bibitem[\protect\citeauthoryear{Maeder \& Mermilliod}{Maeder \&
  Mermilliod}{1981}]{Maeder:1981}
Maeder A.,  Mermilliod J.~C.,  1981, \aap, 93, 136

\bibitem[\protect\citeauthoryear{Maeder \& Meynet}{Maeder \&
  Meynet}{1987}]{Maeder:1987}
Maeder A.,  Meynet G.,  1987, \aap, 182, 243

\bibitem[\protect\citeauthoryear{Maeder \& Meynet}{Maeder \&
  Meynet}{1988}]{Maeder:1988}
Maeder A.,  Meynet G.,  1988, \aap Supplement Series, 76, 411

\bibitem[\protect\citeauthoryear{Maeder \& Meynet}{Maeder \&
  Meynet}{1989}]{Maeder:1989}
Maeder A.,  Meynet G.,  1989, \aap, 210, 155

\bibitem[\protect\citeauthoryear{Maeder \& Meynet}{Maeder \&
  Meynet}{2000}]{Maeder:2000}
Maeder A.,  Meynet G.,  2000, \mn@doi [\araa] {10.1146/annurev.astro.38.1.143},
  38, 143

\bibitem[\protect\citeauthoryear{Maeder \& Meynet}{Maeder \&
  Meynet}{2012}]{Maeder:2012}
Maeder A.,  Meynet G.,  2012, \mn@doi [Reviews of Modern Physics]
  {10.1103/RevModPhys.84.25}, 84, 25

\bibitem[\protect\citeauthoryear{Martins \& Palacios}{Martins \&
  Palacios}{2013}]{Martins:2013}
Martins F.,  Palacios A.,  2013, \mn@doi [\aap] {10.1051/0004-6361/201322480},
  560, A16

\bibitem[\protect\citeauthoryear{Mauron \& Josselin}{Mauron \&
  Josselin}{2011}]{Mauron:2011}
Mauron N.,  Josselin E.,  2011, \mn@doi [\aap] {10.1051/0004-6361/201013993},
  526, A156

\bibitem[\protect\citeauthoryear{McLaughlin}{McLaughlin}{1950}]{McLaughlin:1950}
McLaughlin D.~B.,  1950, \mn@doi [\apjl] {10.1086/145288}, 111, 449

\bibitem[\protect\citeauthoryear{Meynet \& Maeder}{Meynet \&
  Maeder}{2000}]{Meynet:2000}
Meynet G.,  Maeder A.,  2000, \aap, p.~20

\bibitem[\protect\citeauthoryear{Meynet, Maeder, Schaller, Schaerer  \&
  Charbonnel}{Meynet et~al.}{1994}]{Meynet:1994}
Meynet G.,  Maeder A.,  Schaller G.,  Schaerer D.,   Charbonnel C.,  1994, \aap
  Supplement Series, 103, 97

\bibitem[\protect\citeauthoryear{Meynet, Eggenberger  \& Maeder}{Meynet
  et~al.}{2011}]{Meynet:2011}
Meynet G.,  Eggenberger P.,   Maeder A.,  2011, \mn@doi [\aap]
  {10.1051/0004-6361/201016017}, 525, L11

\bibitem[\protect\citeauthoryear{Meynet et~al.,}{Meynet
  et~al.}{2015}]{Meynet:2015}
Meynet G.,  et~al., 2015, \mn@doi [\aap] {10.1051/0004-6361/201424671}, 575,
  A60

\bibitem[\protect\citeauthoryear{Moe \& Di~Stefano}{Moe \&
  Di~Stefano}{2017}]{Moe:2017}
Moe M.,  Di~Stefano R.,  2017, \mn@doi [\apjl Supplement Series]
  {10.3847/1538-4365/aa6fb6}, 230, 15

\bibitem[\protect\citeauthoryear{{Morozova}, {Piro}, {Renzo}, {Ott}, {Clausen},
  {Couch}, {Ellis}  \& {Roberts}}{{Morozova} et~al.}{2015}]{Morozova2015}
{Morozova} V.,  {Piro} A.~L.,  {Renzo} M.,  {Ott} C.~D.,  {Clausen} D.,
  {Couch} S.~M.,  {Ellis} J.,   {Roberts} L.~F.,  2015, \mn@doi [\apj]
  {10.1088/0004-637X/814/1/63}, \href
  {https://ui.adsabs.harvard.edu/abs/2015ApJ...814...63M} {814, 63}

\bibitem[\protect\citeauthoryear{{Morozova}, {Piro}  \& {Valenti}}{{Morozova}
  et~al.}{2018}]{Morozova2018}
{Morozova} V.,  {Piro} A.~L.,   {Valenti} S.,  2018, \mn@doi [\apj]
  {10.3847/1538-4357/aab9a6}, \href
  {https://ui.adsabs.harvard.edu/abs/2018ApJ...858...15M} {858, 15}

\bibitem[\protect\citeauthoryear{Neugent, Levesque  \& Massey}{Neugent
  et~al.}{2018}]{Neugent:2018}
Neugent K.~F.,  Levesque E.~M.,   Massey P.,  2018, \mn@doi [\apj]
  {10.3847/1538-3881/aae4e0}, 156, 225

\bibitem[\protect\citeauthoryear{Neugent, Levesque, Massey  \& Morrell}{Neugent
  et~al.}{2019}]{Neugent:2019}
Neugent K.~F.,  Levesque E.~M.,  Massey P.,   Morrell N.~I.,  2019, \mn@doi
  [\apjl] {10.3847/1538-4357/ab1012}, 875, 124

\bibitem[\protect\citeauthoryear{Nugis}{Nugis}{2000}]{Nugis:2000}
Nugis T.,  2000, \aap, p.~18

\bibitem[\protect\citeauthoryear{Paczy{\'n}ski}{Paczy{\'n}ski}{1971}]{Paczynski:1971}
Paczy{\'n}ski B.,  1971, \actaa, 21, 417

\bibitem[\protect\citeauthoryear{Pastorello et~al.,}{Pastorello
  et~al.}{2007}]{Pastorello:2007}
Pastorello A.,  et~al., 2007, \mn@doi [\nat] {10.1038/nature05825}, 447, 829

\bibitem[\protect\citeauthoryear{Paxton, Bildsten, Dotter, Herwig, Lesaffre  \&
  Timmes}{Paxton et~al.}{2011}]{Paxton:2011}
Paxton B.,  Bildsten L.,  Dotter A.,  Herwig F.,  Lesaffre P.,   Timmes F.,
  2011, \mn@doi [\apjl Supplement Series] {10.1088/0067-0049/192/1/3}, 192, 3

\bibitem[\protect\citeauthoryear{Paxton et~al.,}{Paxton
  et~al.}{2013}]{Paxton:2013}
Paxton B.,  et~al., 2013, \mn@doi [\apjl Supplement Series]
  {10.1088/0067-0049/208/1/4}, 208, 4

\bibitem[\protect\citeauthoryear{Paxton et~al.,}{Paxton
  et~al.}{2015}]{Paxton:2015}
Paxton B.,  et~al., 2015, \mn@doi [\apjl Supplement Series]
  {10.1088/0067-0049/220/1/15}, 220, 15

\bibitem[\protect\citeauthoryear{Paxton et~al.,}{Paxton
  et~al.}{2018}]{Paxton:2018}
Paxton B.,  et~al., 2018, \mn@doi [\apjl Supplement Series]
  {10.3847/1538-4365/aaa5a8}, 234, 34

\bibitem[\protect\citeauthoryear{Paxton et~al.,}{Paxton
  et~al.}{2019}]{Paxton:2019}
Paxton B.,  et~al., 2019, \mn@doi [\apjl Supplement Series]
  {10.3847/1538-4365/ab2241}, 243, 10

\bibitem[\protect\citeauthoryear{Peters, Pewett, Gies, Touhami  \&
  Grundstrom}{Peters et~al.}{2013}]{Peters:2013}
Peters G.~J.,  Pewett T.~D.,  Gies D.~R.,  Touhami Y.~N.,   Grundstrom E.~D.,
  2013, \mn@doi [\apjl] {10.1088/0004-637X/765/1/2}, 765, 2

\bibitem[\protect\citeauthoryear{Podsiadlowski, Joss  \& Hsu}{Podsiadlowski
  et~al.}{1992}]{Podsiadlowski:1992}
Podsiadlowski P.,  Joss P.~C.,   Hsu J. J.~L.,  1992, \mn@doi [\apjl]
  {10.1086/171341}, 391, 246

\bibitem[\protect\citeauthoryear{Podsiadlowski, Hsu, Joss  \&
  Ross}{Podsiadlowski et~al.}{1993}]{Podsiadlowski:1993}
Podsiadlowski P.,  Hsu J. J.~L.,  Joss P.~C.,   Ross R.~R.,  1993, \mn@doi
  [\nat] {10.1038/364509a0}, 364, 509

\bibitem[\protect\citeauthoryear{Pols, Tout, Eggleton  \& Han}{Pols
  et~al.}{1995}]{Pols:1995}
Pols O.~R.,  Tout C.~A.,  Eggleton P.~P.,   Han Z.,  1995, \mn@doi [\mnras]
  {10.1093/mnras/274.3.964}, 274, 964

\bibitem[\protect\citeauthoryear{Pols, Schr{\"o}der, Hurley, Tout  \&
  Eggleton}{Pols et~al.}{1998}]{Pols:1998}
Pols O.~R.,  Schr{\"o}der K.-P.,  Hurley J.~R.,  Tout C.~A.,   Eggleton P.~P.,
  1998, \mn@doi [\mnras] {10.1046/j.1365-8711.1998.01658.x}, 298, 525

\bibitem[\protect\citeauthoryear{{Ram{\'i}rez-Agudelo}
  et~al.,}{{Ram{\'i}rez-Agudelo} et~al.}{2013}]{Ramirez-Agudelo:2013}
{Ram{\'i}rez-Agudelo} O.~H.,  et~al., 2013, \mn@doi [\aap]
  {10.1051/0004-6361/201321986}, 560, A29

\bibitem[\protect\citeauthoryear{{Ram{\'i}rez-Agudelo}
  et~al.,}{{Ram{\'i}rez-Agudelo} et~al.}{2015}]{Ramirez-Agudelo:2015}
{Ram{\'i}rez-Agudelo} O.~H.,  et~al., 2015, \mn@doi [\aap]
  {10.1051/0004-6361/201425424}, 580, A92

\bibitem[\protect\citeauthoryear{Renzo, Ott, Shore  \& {de Mink}}{Renzo
  et~al.}{2017}]{Renzo:2017}
Renzo M.,  Ott C.~D.,  Shore S.~N.,   {de Mink} S.~E.,  2017, \mn@doi [\aap]
  {10.1051/0004-6361/201730698}, 603, A118

\bibitem[\protect\citeauthoryear{Rogers \& Nayfonov}{Rogers \&
  Nayfonov}{2002}]{Rogers:2002}
Rogers F.~J.,  Nayfonov A.,  2002, \mn@doi [\apjl] {10.1086/341894}, 576, 1064

\bibitem[\protect\citeauthoryear{Sana et~al.,}{Sana et~al.}{2012}]{Sana:2012}
Sana H.,  et~al., 2012, \mn@doi [\sci] {10.1126/science.1223344}, 337, 444

\bibitem[\protect\citeauthoryear{Sana et~al.,}{Sana et~al.}{2013}]{Sana:2013}
Sana H.,  et~al., 2013, \mn@doi [\aap] {10.1051/0004-6361/201219621}, 550, A107

\bibitem[\protect\citeauthoryear{Saumon, Chabrier  \& {van Horn}}{Saumon
  et~al.}{1995}]{Saumon:1995}
Saumon D.,  Chabrier G.,   {van Horn} H.~M.,  1995, \mn@doi [\apjl Supplement
  Series] {10.1086/192204}, 99, 713

\bibitem[\protect\citeauthoryear{Schaerer}{Schaerer}{1996}]{Schaerer:1996a}
Schaerer D.,  1996, \aap, 309, 129

\bibitem[\protect\citeauthoryear{Schaerer, {de Koter}, Schmutz  \&
  Maeder}{Schaerer et~al.}{1996}]{Schaerer:1996}
Schaerer D.,  {de Koter} A.,  Schmutz W.,   Maeder A.,  1996, \aap, 310, 837

\bibitem[\protect\citeauthoryear{{Schneider} et~al.,}{{Schneider}
  et~al.}{2014}]{Schneider:2014}
{Schneider} F.~R.~N.,  et~al., 2014, \mn@doi [\apj]
  {10.1088/0004-637X/780/2/117}, \href
  {https://ui.adsabs.harvard.edu/abs/2014ApJ...780..117S} {780, 117}

\bibitem[\protect\citeauthoryear{Schootemeijer \& Langer}{Schootemeijer \&
  Langer}{2018}]{Schootemeijer:2018}
Schootemeijer A.,  Langer N.,  2018, \mn@doi [\aap]
  {10.1051/0004-6361/201731895}, 611, A75

\bibitem[\protect\citeauthoryear{Schootemeijer, Langer, Grin  \&
  Wang}{Schootemeijer et~al.}{2019}]{Schootemeijer:2019a}
Schootemeijer A.,  Langer N.,  Grin N.~J.,   Wang C.,  2019, \mn@doi [\aap]
  {10.1051/0004-6361/201935046}, 625, A132

\bibitem[\protect\citeauthoryear{Schwarzschild}{Schwarzschild}{1961}]{Schwarzschild:1961}
Schwarzschild M.,  1961, \mn@doi [\apjl] {10.1086/147121}, 134, 1

\bibitem[\protect\citeauthoryear{Schwarzschild \& H{\"a}rm}{Schwarzschild \&
  H{\"a}rm}{1958}]{Schwarzschild:1958}
Schwarzschild M.,  H{\"a}rm R.,  1958, \mn@doi [\apjl] {10.1086/146548}, 128,
  348

\bibitem[\protect\citeauthoryear{Smith}{Smith}{2014}]{Smith:2014}
Smith N.,  2014, \mn@doi [\araa] {10.1146/annurev-astro-081913-040025}, 52, 487

\bibitem[\protect\citeauthoryear{Smith et~al.,}{Smith
  et~al.}{2007}]{Smith:2007}
Smith N.,  et~al., 2007, \mn@doi [\apjl] {10.1086/519949}, 666, 1116

\bibitem[\protect\citeauthoryear{Sota, Apell{\'a}niz, Morrell, Barb{\'a},
  Walborn, Gamen, Arias  \& Alfaro}{Sota et~al.}{2014}]{Sota:2014}
Sota A.,  Apell{\'a}niz J.~M.,  Morrell N.~I.,  Barb{\'a} R.~H.,  Walborn
  N.~R.,  Gamen R.~C.,  Arias J.~I.,   Alfaro E.~J.,  2014, \mn@doi [\apjl
  Supplement Series] {10.1088/0067-0049/211/1/10}, 211, 10

\bibitem[\protect\citeauthoryear{Stancliffe \& Eldridge}{Stancliffe \&
  Eldridge}{2009}]{Stancliffe:2009}
Stancliffe R.~J.,  Eldridge J.~J.,  2009, \mn@doi [\mnras]
  {10.1111/j.1365-2966.2009.14849.x}, 396, 1699

\bibitem[\protect\citeauthoryear{Steiner \& Oliveira}{Steiner \&
  Oliveira}{2005}]{Steiner:2005}
Steiner J.~E.,  Oliveira A.~S.,  2005, \mn@doi [\aap]
  {10.1051/0004-6361:20052782}, 444, 895

\bibitem[\protect\citeauthoryear{Stencel, Hopkins, Hagen, Fried, Schmidtke,
  Kondo  \& Chapman}{Stencel et~al.}{1984}]{Stencel:1984}
Stencel R.~E.,  Hopkins J.~L.,  Hagen W.,  Fried R.,  Schmidtke P.~C.,  Kondo
  Y.,   Chapman R.~D.,  1984, \mn@doi [\apjl] {10.1086/162153}, 281, 751

\bibitem[\protect\citeauthoryear{Stothers}{Stothers}{1970}]{Stothers:1970}
Stothers R.,  1970, \mn@doi [\mnras] {10.1093/mnras/151.1.65}, 151, 65

\bibitem[\protect\citeauthoryear{Stothers}{Stothers}{1974}]{Stothers:1974}
Stothers R.,  1974, \mn@doi [\apjl] {10.1086/153291}, 194, 699

\bibitem[\protect\citeauthoryear{Stothers \& Chin}{Stothers \&
  Chin}{1985}]{Stothers:1985}
Stothers R.~B.,  Chin C.-W.,  1985, \mn@doi [\apjl] {10.1086/163150}, 292, 222

\bibitem[\protect\citeauthoryear{Stothers \& Chin}{Stothers \&
  Chin}{1995}]{Stothers:1995}
Stothers R.~B.,  Chin C.-W.,  1995, \mn@doi [\apjl] {10.1086/175270}, 440, 297

\bibitem[\protect\citeauthoryear{Sukhbold, Ertl, Woosley, Brown  \&
  Janka}{Sukhbold et~al.}{2016}]{Sukhbold:2016}
Sukhbold T.,  Ertl T.,  Woosley S.~E.,  Brown J.~M.,   Janka H.-T.,  2016,
  \mn@doi [\apjl] {10.3847/0004-637X/821/1/38}, 821, 38

\bibitem[\protect\citeauthoryear{Sz{\'e}csi, Langer, Yoon, Sanyal, {de Mink},
  Evans  \& Dermine}{Sz{\'e}csi et~al.}{2015}]{Szecsi:2015}
Sz{\'e}csi D.,  Langer N.,  Yoon S.-C.,  Sanyal D.,  {de Mink} S.,  Evans
  C.~J.,   Dermine T.,  2015, \mn@doi [\aap] {10.1051/0004-6361/201526617},
  581, A15

\bibitem[\protect\citeauthoryear{Timmes \& Swesty}{Timmes \&
  Swesty}{2000}]{Timmes:2000}
Timmes F.~X.,  Swesty F.~D.,  2000, \mn@doi [\apjl Supplement Series]
  {10.1086/313304}, 126, 501

\bibitem[\protect\citeauthoryear{Tout, Pols, Eggleton  \& Han}{Tout
  et~al.}{1996}]{Tout:1996}
Tout C.~A.,  Pols O.~R.,  Eggleton P.~P.,   Han Z.,  1996, \mn@doi [\mnras]
  {10.1093/mnras/281.1.257}, 281, 257

\bibitem[\protect\citeauthoryear{Tout, Aarseth, Pols  \& Eggleton}{Tout
  et~al.}{1997}]{Tout:1997}
Tout C.~A.,  Aarseth S.~J.,  Pols O.~R.,   Eggleton P.~P.,  1997, \mn@doi
  [\mnras] {10.1093/mnras/291.4.732}, 291, 732

\bibitem[\protect\citeauthoryear{Ugliano, Janka, Marek  \& Arcones}{Ugliano
  et~al.}{2012}]{Ugliano:2012}
Ugliano M.,  Janka H.-T.,  Marek A.,   Arcones A.,  2012, \mn@doi [\apjl]
  {10.1088/0004-637X/757/1/69}, 757, 69

\bibitem[\protect\citeauthoryear{Vanbeveren, Van~Bever  \& Belkus}{Vanbeveren
  et~al.}{2007}]{Vanbeveren:2007}
Vanbeveren D.,  Van~Bever J.,   Belkus H.,  2007, \mn@doi [\apjl]
  {10.1086/519454}, 662, L107

\bibitem[\protect\citeauthoryear{Vink, {de Koter}  \& Lamers}{Vink
  et~al.}{2001}]{Vink:2001}
Vink J.~S.,  {de Koter} A.,   Lamers H. J. G. L.~M.,  2001, \mn@doi [\aap]
  {10.1051/0004-6361:20010127}, 369, 574

\bibitem[\protect\citeauthoryear{Wang, Gies  \& Peters}{Wang
  et~al.}{2017}]{wang:2017}
Wang L.,  Gies D.~R.,   Peters G.~J.,  2017, \mn@doi [\apjl]
  {10.3847/1538-4357/aa740a}, 843, 60

\bibitem[\protect\citeauthoryear{Woosley \& Weaver}{Woosley \&
  Weaver}{1994}]{Woosley:1994}
Woosley S.~E.,  Weaver T.~A.,  1994, \mn@doi [\apjl] {10.1086/173813}, 423, 371

\bibitem[\protect\citeauthoryear{Wright}{Wright}{1970}]{Wright:1970}
Wright K.~O.,  1970, \mn@doi [Vistas in Astronomy]
  {10.1016/0083-6656(70)90038-3}, 12, 147

\bibitem[\protect\citeauthoryear{Yaron et~al.,}{Yaron
  et~al.}{2017}]{Yaron:2017}
Yaron O.,  et~al., 2017, \mn@doi [\nat Physics] {10.1038/nphys4025}, 13, 510

\bibitem[\protect\citeauthoryear{Yoon}{Yoon}{2015}]{Yoon:2015}
Yoon S.-C.,  2015, \mn@doi [\pasa] {10.1017/pasa.2015.16}, 32, e015

\bibitem[\protect\citeauthoryear{Yoon, Dierks  \& Langer}{Yoon
  et~al.}{2012}]{Yoon:2012}
Yoon S.-C.,  Dierks A.,   Langer N.,  2012, \mn@doi [\aap]
  {10.1051/0004-6361/201117769}, 542, A113

\bibitem[\protect\citeauthoryear{Zapartas et~al.,}{Zapartas
  et~al.}{2019}]{Zapartas:2019}
Zapartas E.,  et~al., 2019, \mn@doi [\aap] {10.1051/0004-6361/201935854}, 631,
  A5

\bibitem[\protect\citeauthoryear{{de Jager}, Nieuwenhuijzen  \& {van der
  Hucht}}{{de Jager} et~al.}{1988}]{deJager:1988}
{de Jager} C.,  Nieuwenhuijzen H.,   {van der Hucht} K.~A.,  1988, \aap
  Supplement Series, 72, 259

\bibitem[\protect\citeauthoryear{{de Mink} \& Belczynski}{{de Mink} \&
  Belczynski}{2015}]{deMink:2015}
{de Mink} S.~E.,  Belczynski K.,  2015, \mn@doi [\apjl]
  {10.1088/0004-637X/814/1/58}, 814, 58

\bibitem[\protect\citeauthoryear{{de Mink}, Langer, Izzard, Sana  \& {de
  Koter}}{{de Mink} et~al.}{2013}]{deMink:2013}
{de Mink} S.~E.,  Langer N.,  Izzard R.~G.,  Sana H.,   {de Koter} A.,  2013,
  \mn@doi [\apjl] {10.1088/0004-637X/764/2/166}, 764, 166

\bibitem[\protect\citeauthoryear{{de Mink}, Sana, Langer, Izzard  \&
  Schneider}{{de Mink} et~al.}{2014}]{deMink:2014}
{de Mink} S.~E.,  Sana H.,  Langer N.,  Izzard R.~G.,   Schneider F. R.~N.,
  2014, \mn@doi [\apjl] {10.1088/0004-637X/782/1/7}, 782, 7

\bibitem[\protect\citeauthoryear{{deBoer} et~al.,}{{deBoer}
  et~al.}{2017}]{deBoer:2017}
{deBoer} R.~J.,  et~al., 2017, \mn@doi [Reviews of Modern Physics]
  {10.1103/RevModPhys.89.035007}, 89, 035007

\bibitem[\protect\citeauthoryear{{van Bever} \& Vanbeveren}{{van Bever} \&
  Vanbeveren}{1998}]{vanBever:1998}
{van Bever} J.,  Vanbeveren D.,  1998, \aap, 334, 21

\makeatother
\end{thebibliography}

\appendix

\section{Core and Envelope Masses for Core-H Burning Models} \label{app:parameter_space_h}
\begin{figure*}
	\centering
	\includegraphics[width=\hsize]{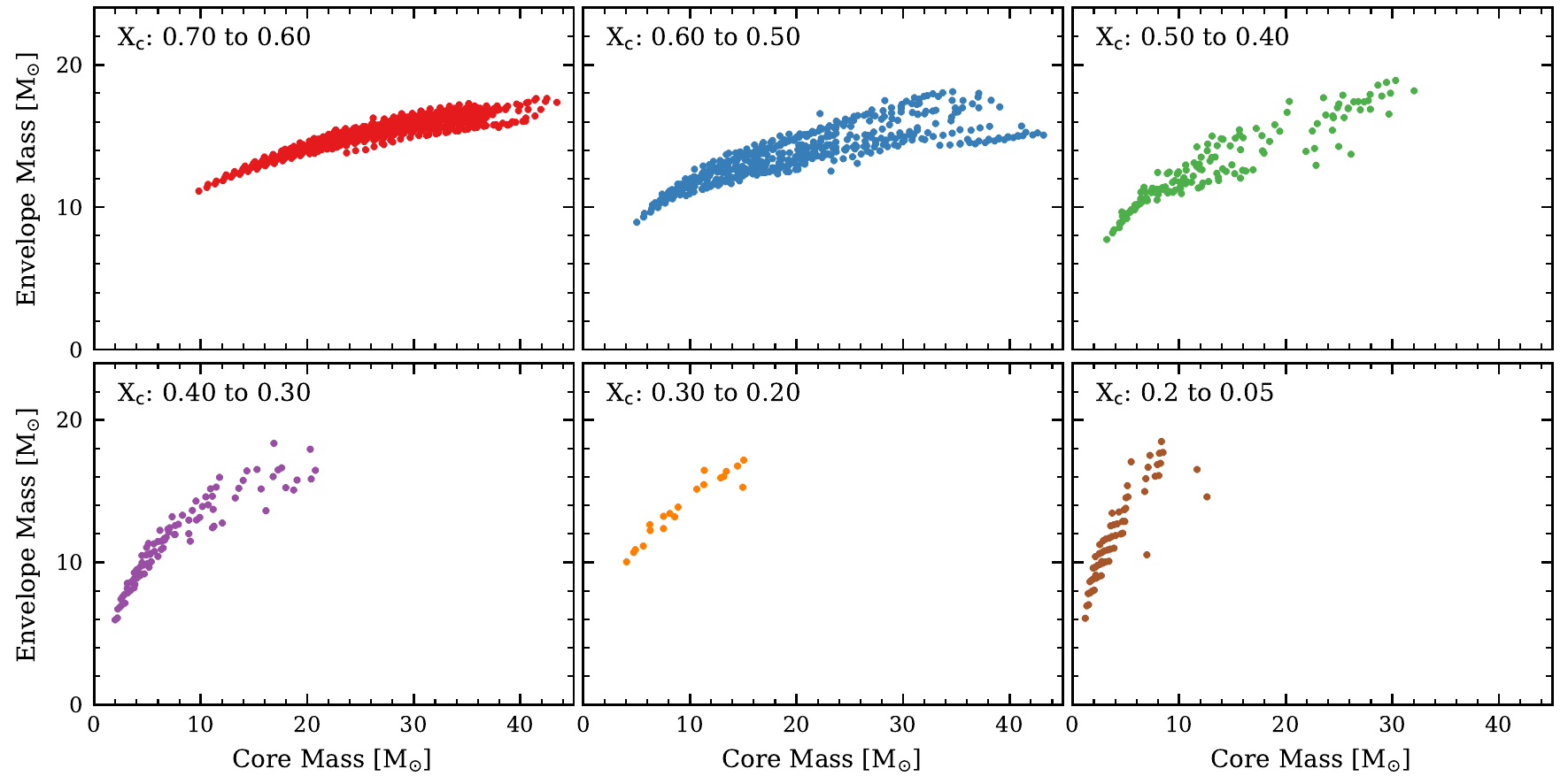}
	\caption{Summary of the 3 key structural parameters (convective core mass, envelope mass and core composition) for each of our core-H burning structure models. Each dot represents an individual core-H burning stellar structure model with the given core mass and envelope mass in \msun. The models are divided up into bins by their central H mass fraction (\xc) as indicated in each plot.}
	\label{fig:parameter_space_ms1}
\end{figure*}

\section{Core and Envelope Masses for Core-He Burning Models} \label{app:parameter_space_he}
\begin{figure}
	\centering
	\includegraphics[width=\hsize]{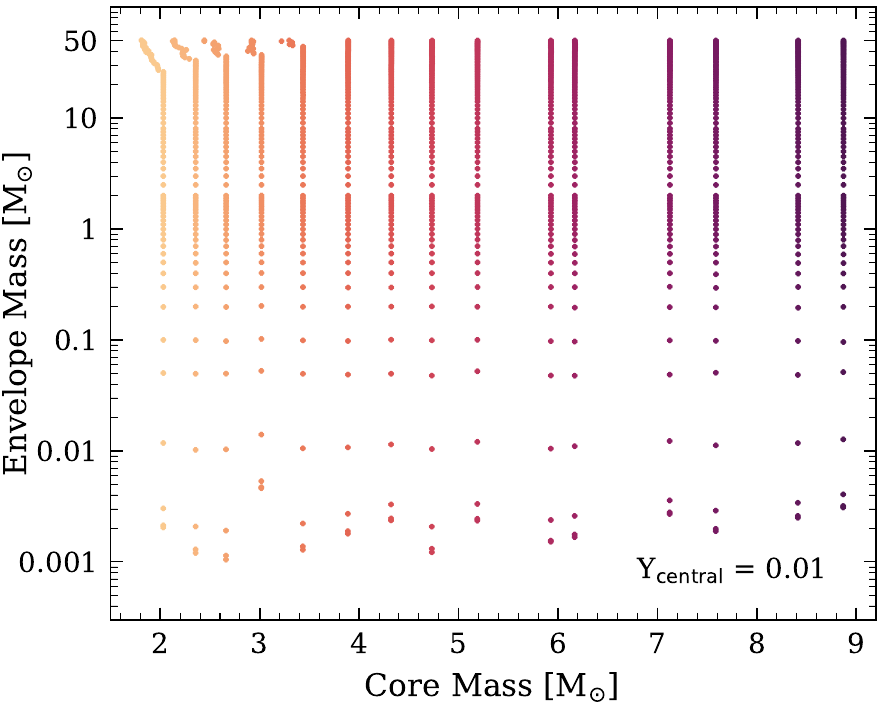}
	\caption{Each dot represents a core-He burning structure, similar to Fig. \ref{fig:parameter_he}. but with with $\yc = 0.01$}
	\label{fig:parameter_he_001}
\end{figure}

\section{Central Temperature vs. Core mass ratio} \label{app:logtc_massratio}
\begin{figure}
	\centering
	\includegraphics[width=\hsize]{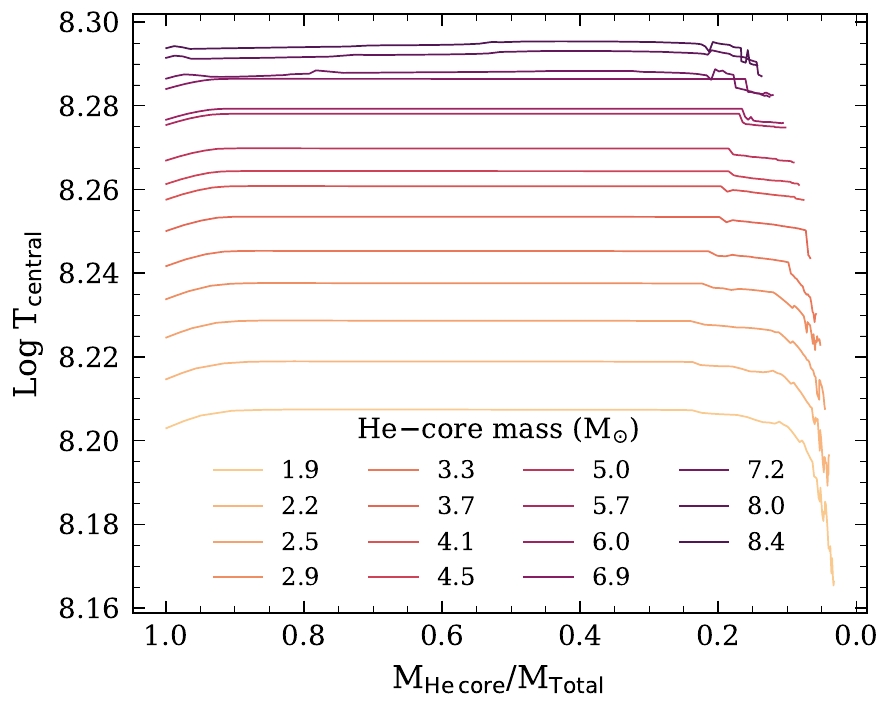}
	\caption{The central temperature as a function of envelope mass for core-He burning models of constant core mass. For most combinations of core and envelope masses, the central temperature of the core is not affected by the mass of the envelope.}
	\label{fig:logtc_massratio}
\end{figure}

\section{Surface properties for core-He burning models} \label{sec:app:tables_surface}
\begin{table*}
	\centering
	\caption{Summary of surface properties for core-He burning models with a He-core mass of $4.1 \msun$, a central Helium abundance of 0.50 and envelope masses from $10^{-3} - 25 \msun$. \fhshell refers to the fraction of the total nuclear energy generated in the H-Shell. The same table for envelope masses from $25 - 50 \msun$ can be found below in Table \ref{tab:c4d1_app_surface_props2}.}
	\label{tab:c4d1_app_surface_props}
	\begin{tabular}{rrrrrrrr}
		\hline
        \mcore [\msun] & \menv [\msun] & $\logteff$ [K] & \teff [K] &  \logllsun & \logg & \logrrsun & \fhshell \\
        4.1 & 2$\times10^{-3}$ & 4.96 & 91700 & 4.37 & 5.48 & -0.21 & 0.00 \\
        4.1 & 4$\times10^{-3}$ & 4.96 & 91700 & 4.37 & 5.48 & -0.21 & 0.00 \\
        4.1 & 0.01 & 4.96 & 91750 & 4.38 & 5.48 & -0.21 & 0.00 \\
        4.1 & 0.05 & 4.96 & 91680 & 4.39 & 5.48 & -0.21 & 0.00 \\
        4.1 & 0.1 & 4.95 & 88160 & 4.40 & 5.40 & -0.17 & 0.00 \\
        4.1 & 0.2 & 4.88 & 75700 & 4.43 & 5.11 & -0.02 & 0.00 \\
        4.1 & 0.3 & 4.84 & 68800 & 4.46 & 4.93 & 0.08 & 0.03 \\
        4.1 & 0.4 & 4.62 & 41630 & 4.50 & 4.03 & 0.53 & 0.10 \\
        4.1 & 0.5 & 4.45 & 28000 & 4.53 & 3.32 & 0.89 & 0.17 \\
        4.1 & 0.6 & 4.26 & 18010 & 4.55 & 2.54 & 1.29 & 0.22 \\
        4.1 & 0.7 & 3.96 & 9200 & 4.56 & 1.37 & 1.88 & 0.24 \\
        4.1 & 0.8 & 3.62 & 4150 & 4.57 & -0.01 & 2.57 & 0.26 \\
        4.1 & 0.9 & 3.57 & 3690 & 4.57 & -0.21 & 2.67 & 0.26 \\
        4.1 & 1.0 & 3.55 & 3580 & 4.57 & -0.26 & 2.70 & 0.26 \\
        4.1 & 1.1 & 3.55 & 3520 & 4.57 & -0.28 & 2.72 & 0.27 \\
        4.1 & 1.2 & 3.54 & 3480 & 4.57 & -0.29 & 2.73 & 0.27 \\
        4.1 & 1.3 & 3.54 & 3460 & 4.58 & -0.29 & 2.73 & 0.27 \\
        4.1 & 1.4 & 3.54 & 3440 & 4.58 & -0.30 & 2.74 & 0.27 \\
        4.1 & 1.5 & 3.53 & 3420 & 4.58 & -0.30 & 2.74 & 0.28 \\
        4.1 & 1.6 & 3.53 & 3410 & 4.58 & -0.30 & 2.75 & 0.28 \\
        4.1 & 1.7 & 3.53 & 3400 & 4.58 & -0.29 & 2.75 & 0.28 \\
        4.1 & 1.8 & 3.53 & 3400 & 4.58 & -0.29 & 2.75 & 0.28 \\
        4.1 & 1.9 & 3.53 & 3390 & 4.58 & -0.29 & 2.75 & 0.28 \\
        4.1 & 2.0 & 3.53 & 3390 & 4.58 & -0.28 & 2.76 & 0.28 \\
        4.1 & 2.5 & 3.53 & 3380 & 4.58 & -0.26 & 2.76 & 0.29 \\
        4.1 & 3.0 & 3.53 & 3380 & 4.59 & -0.23 & 2.76 & 0.29 \\
        4.1 & 3.5 & 3.53 & 3390 & 4.59 & -0.19 & 2.76 & 0.29 \\
        4.1 & 4.0 & 3.53 & 3400 & 4.59 & -0.16 & 2.76 & 0.30 \\
        4.1 & 4.5 & 3.53 & 3410 & 4.59 & -0.13 & 2.75 & 0.30 \\
        4.1 & 5.0 & 3.53 & 3430 & 4.59 & -0.10 & 2.75 & 0.30 \\
        4.1 & 5.5 & 3.54 & 3440 & 4.59 & -0.07 & 2.75 & 0.30 \\
        4.1 & 6.0 & 3.54 & 3460 & 4.59 & -0.04 & 2.74 & 0.30 \\
        4.1 & 6.5 & 3.54 & 3470 & 4.59 & -0.01 & 2.74 & 0.31 \\
        4.1 & 7.0 & 3.54 & 3490 & 4.59 & 0.02 & 2.73 & 0.31 \\
        4.1 & 7.5 & 3.54 & 3510 & 4.60 & 0.04 & 2.73 & 0.31 \\
        4.1 & 8.0 & 3.55 & 3520 & 4.60 & 0.07 & 2.73 & 0.31 \\
        4.1 & 9.0 & 3.55 & 3550 & 4.60 & 0.12 & 2.72 & 0.31 \\
        4.1 & 10.0 & 3.55 & 3590 & 4.60 & 0.16 & 2.71 & 0.32 \\
        4.1 & 11.0 & 3.56 & 3620 & 4.60 & 0.21 & 2.71 & 0.32 \\
        4.1 & 12.0 & 3.56 & 3650 & 4.60 & 0.25 & 2.70 & 0.32 \\
        4.1 & 13.0 & 3.57 & 3680 & 4.60 & 0.29 & 2.69 & 0.32 \\
        4.1 & 14.0 & 3.57 & 3700 & 4.60 & 0.32 & 2.69 & 0.33 \\
        4.1 & 15.0 & 3.57 & 3730 & 4.61 & 0.36 & 2.68 & 0.33 \\
        4.1 & 16.0 & 3.58 & 3760 & 4.61 & 0.39 & 2.68 & 0.33 \\
        4.1 & 17.0 & 3.58 & 3800 & 4.61 & 0.42 & 2.67 & 0.34 \\
        4.1 & 18.0 & 4.29 & 19620 & 5.04 & 2.87 & 1.46 & 0.75 \\
        4.1 & 19.0 & 4.32 & 20770 & 5.09 & 2.94 & 1.43 & 0.77 \\
        4.1 & 20.0 & 4.33 & 21490 & 5.12 & 2.98 & 1.42 & 0.79 \\
        4.1 & 21.0 & 4.34 & 22080 & 5.15 & 3.02 & 1.41 & 0.80 \\
        4.1 & 22.0 & 4.36 & 22710 & 5.18 & 3.05 & 1.40 & 0.82 \\
        4.1 & 23.0 & 4.37 & 23270 & 5.21 & 3.08 & 1.39 & 0.84 \\
        4.1 & 24.0 & 4.38 & 23740 & 5.24 & 3.11 & 1.39 & 0.84 \\
        4.1 & 25.0 & 4.38 & 24260 & 5.26 & 3.13 & 1.39 & 0.85 \\
    	\hline
	\end{tabular}
\end{table*}

\begin{table*}
	\centering
	\caption{Same as Table \ref{tab:c4d1_app_surface_props} but for envelope masses from $25 - 50 \msun$.}
	\label{tab:c4d1_app_surface_props2}
	\begin{tabular}{rrrrrrrr}
		\hline
        \mcore [\msun] & \menv [\msun] & \logteff [K] & \teff [K] &  \logllsun & \logg & \logrrsun & \fhshell \\
        4.1 & 26.0 & 4.39 & 24760 & 5.29 & 3.15 & 1.38 & 0.86 \\
        4.1 & 27.0 & 4.40 & 25230 & 5.32 & 3.18 & 1.38 & 0.87 \\
        4.1 & 28.0 & 4.41 & 25690 & 5.34 & 3.20 & 1.37 & 0.88 \\
        4.1 & 29.0 & 4.42 & 26120 & 5.36 & 3.22 & 1.37 & 0.89 \\
        4.1 & 30.0 & 4.42 & 26530 & 5.39 & 3.23 & 1.37 & 0.89 \\
        4.1 & 31.0 & 4.43 & 26940 & 5.41 & 3.25 & 1.37 & 0.90 \\
        4.1 & 32.0 & 4.44 & 27320 & 5.43 & 3.27 & 1.37 & 0.90 \\
        4.1 & 33.0 & 4.44 & 27690 & 5.45 & 3.28 & 1.36 & 0.91 \\
        4.1 & 34.0 & 4.45 & 28050 & 5.47 & 3.29 & 1.36 & 0.91 \\
        4.1 & 35.0 & 4.45 & 28400 & 5.49 & 3.31 & 1.36 & 0.92 \\
        4.1 & 36.0 & 4.46 & 28730 & 5.51 & 3.32 & 1.36 & 0.92 \\
        4.1 & 37.0 & 4.46 & 29060 & 5.53 & 3.33 & 1.36 & 0.92 \\
        4.1 & 38.0 & 4.47 & 29380 & 5.55 & 3.34 & 1.36 & 0.93 \\
        4.1 & 39.0 & 4.47 & 29680 & 5.57 & 3.35 & 1.36 & 0.93 \\
        4.1 & 40.0 & 4.48 & 29970 & 5.58 & 3.36 & 1.36 & 0.93 \\
        4.1 & 41.0 & 4.48 & 30250 & 5.60 & 3.37 & 1.36 & 0.93 \\
        4.1 & 42.0 & 4.48 & 30520 & 5.62 & 3.38 & 1.36 & 0.94 \\
        4.1 & 43.0 & 4.49 & 30940 & 5.63 & 3.40 & 1.36 & 0.94 \\
        4.1 & 44.0 & 4.49 & 31200 & 5.65 & 3.40 & 1.36 & 0.94 \\
        4.1 & 45.0 & 4.50 & 31450 & 5.66 & 3.41 & 1.36 & 0.95 \\
        4.1 & 46.0 & 4.50 & 31690 & 5.68 & 3.42 & 1.36 & 0.95 \\
        4.1 & 47.0 & 4.50 & 31930 & 5.69 & 3.42 & 1.36 & 0.95 \\
        4.1 & 48.0 & 4.51 & 32150 & 5.71 & 3.43 & 1.36 & 0.95 \\
        4.1 & 49.0 & 4.51 & 32370 & 5.72 & 3.44 & 1.36 & 0.95 \\
        4.1 & 50.0 & 4.51 & 32590 & 5.73 & 3.44 & 1.36 & 0.95 \\
    	\hline
	\end{tabular}
\end{table*}

 \section{Interior Graphs}

\begin{figure*}
  \centering
  \includegraphics[width=\hsize]{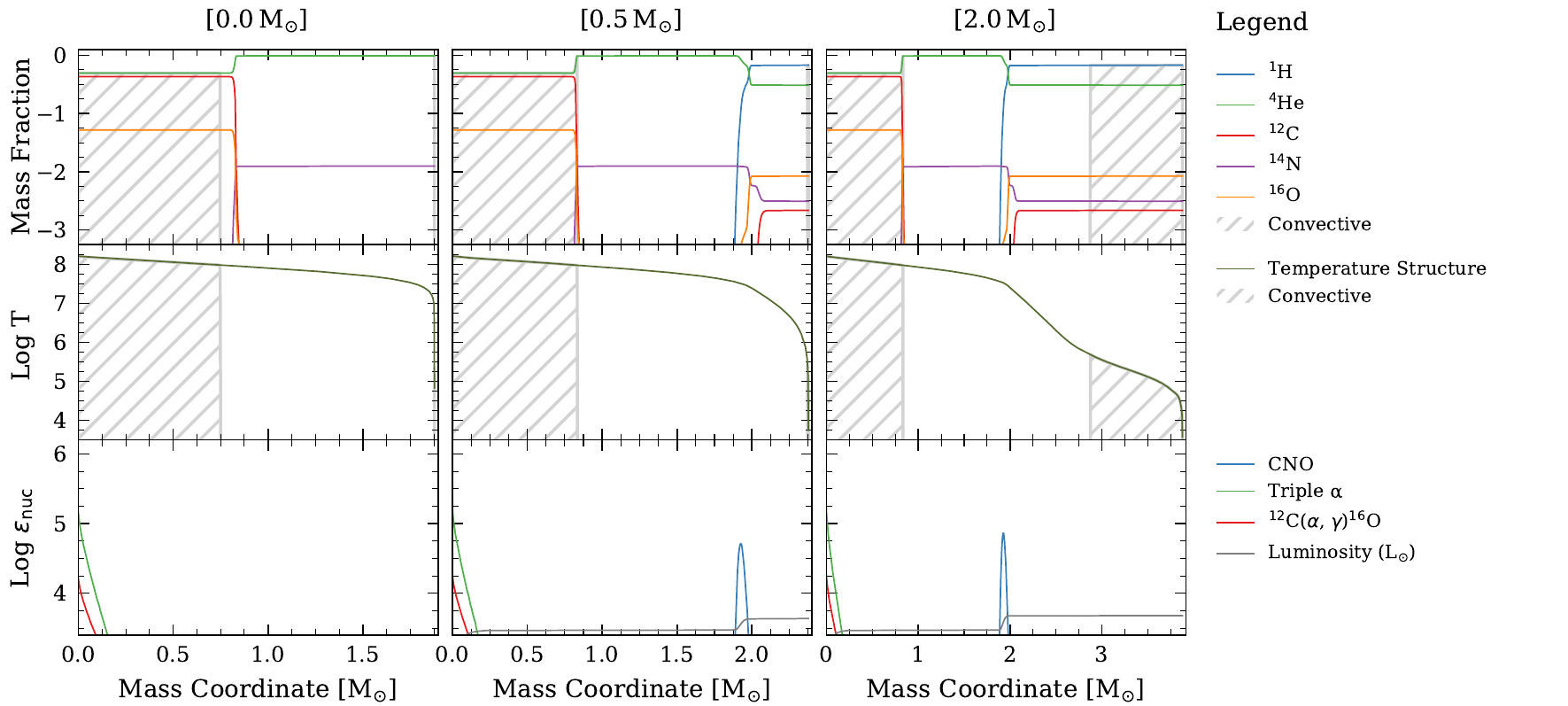}
  \includegraphics[width=\hsize]{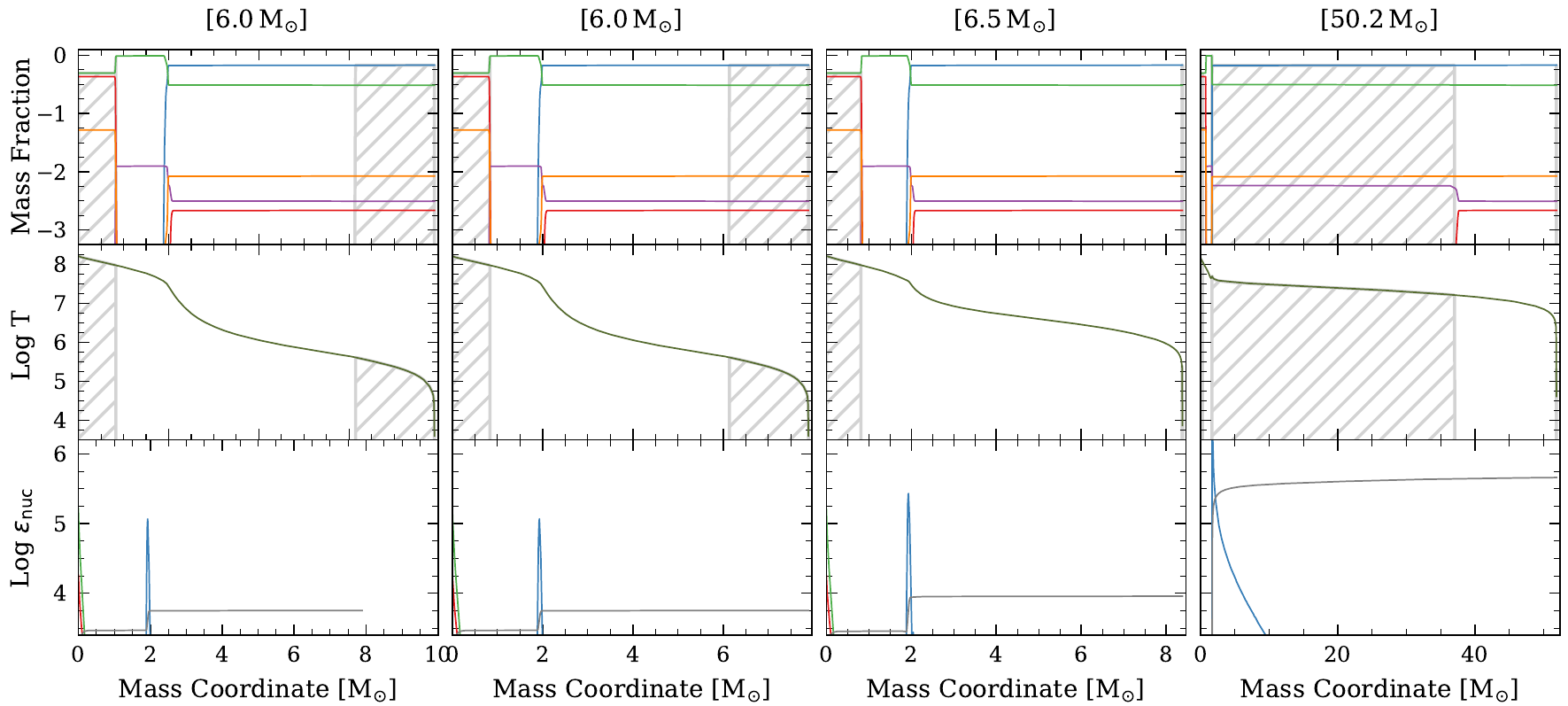}
  \caption{Internal structure profiles of core--He burning models with $\mcore = 1.9\msun$, $\yc = 0.50$, and a range of envelope masses (indicated at top of each panel) selected to represent the qualitative behaviour of the models. See caption of Fig. \ref{fig:interior_example_C4D1} for further details.}
 \label{fig:appendix:mass8}
 \end{figure*}

 \begin{figure*}
  \centering
  \includegraphics[width=\hsize]{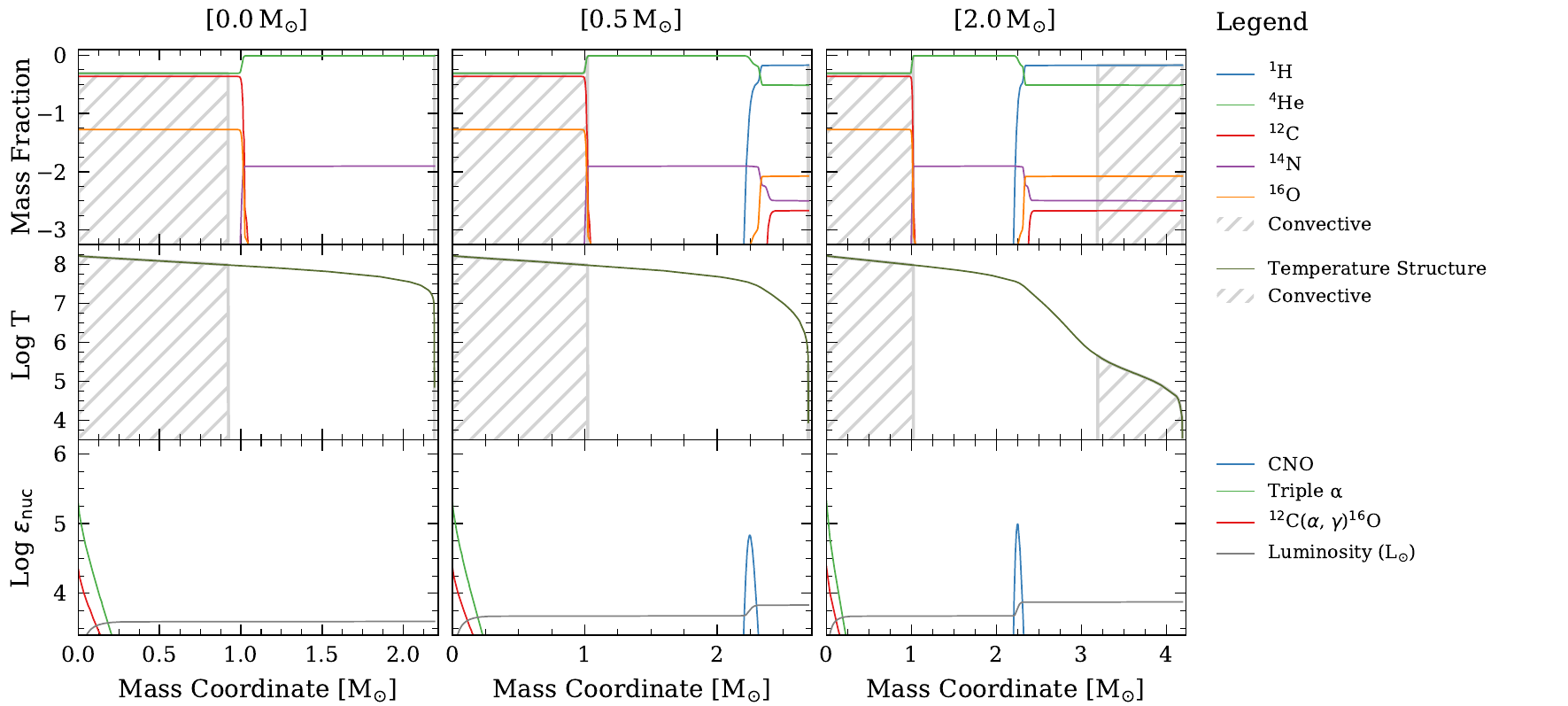}
  \includegraphics[width=\hsize]{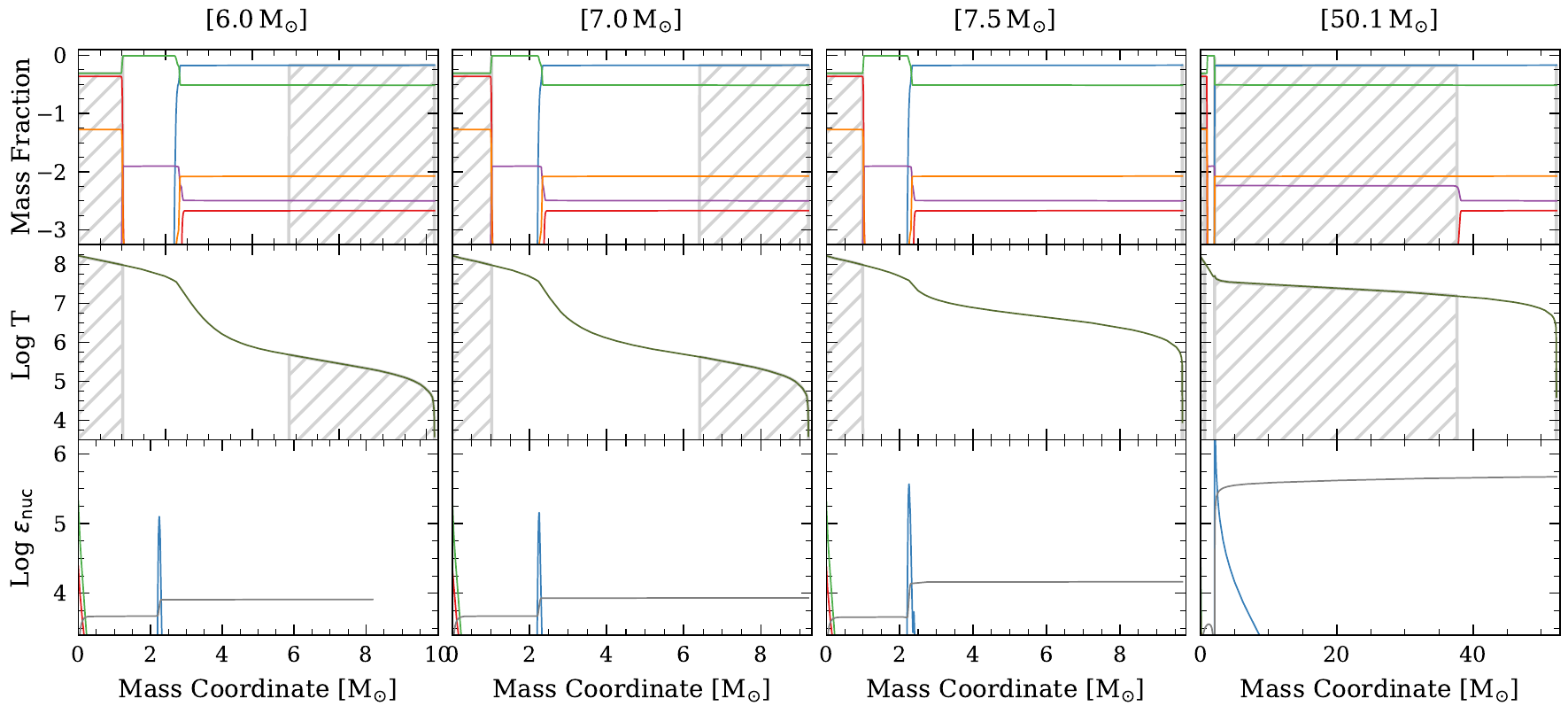}
  \caption{Internal structure profiles of core--He burning models with $\mcore = 2.2\msun$, $\yc = 0.50$, and a range of envelope masses (indicated at top of each panel) selected to represent the qualitative behaviour of the models. See caption of Fig. \ref{fig:interior_example_C4D1} for further details.}
 \label{fig:appendix:mass9}
 \end{figure*}

 \begin{figure*}
  \centering
  \includegraphics[width=\hsize]{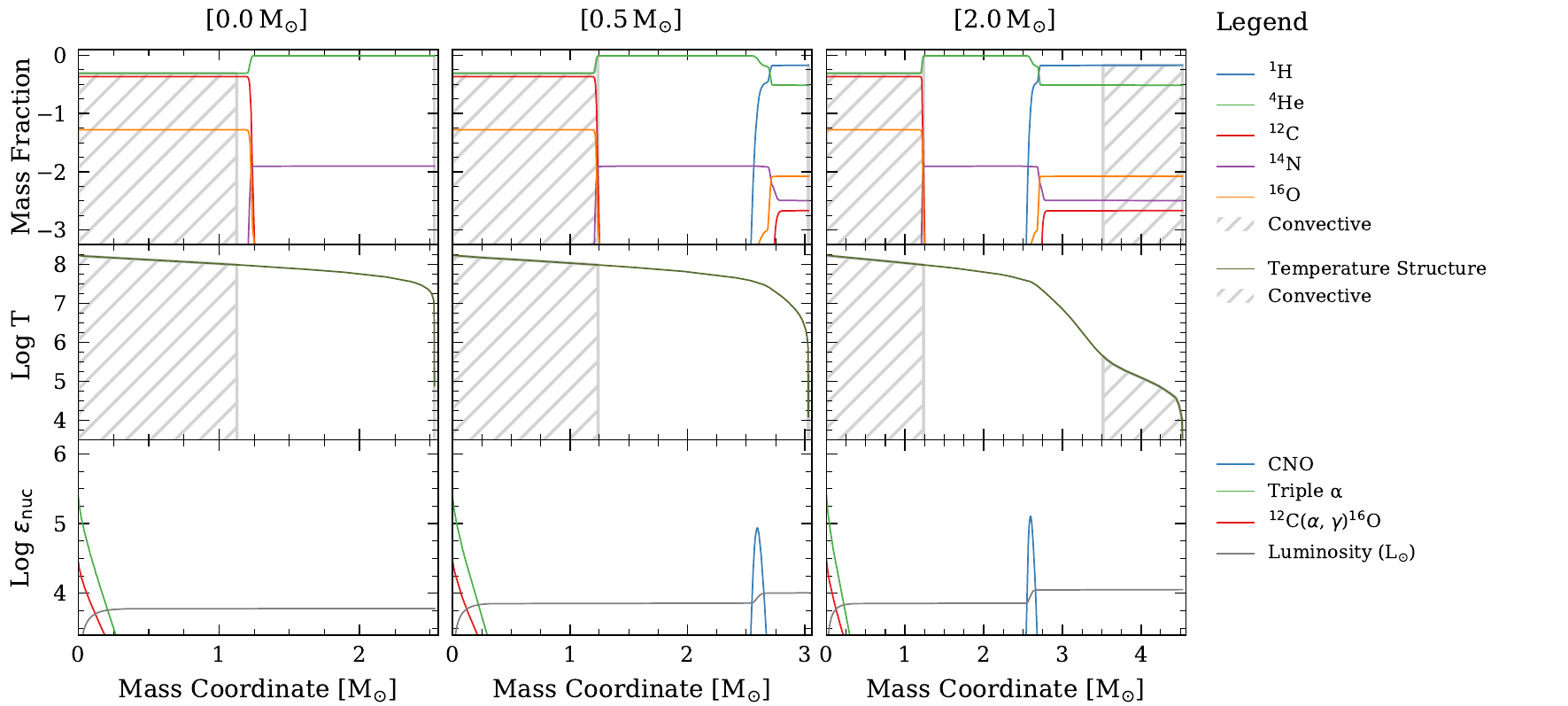}
  \includegraphics[width=\hsize]{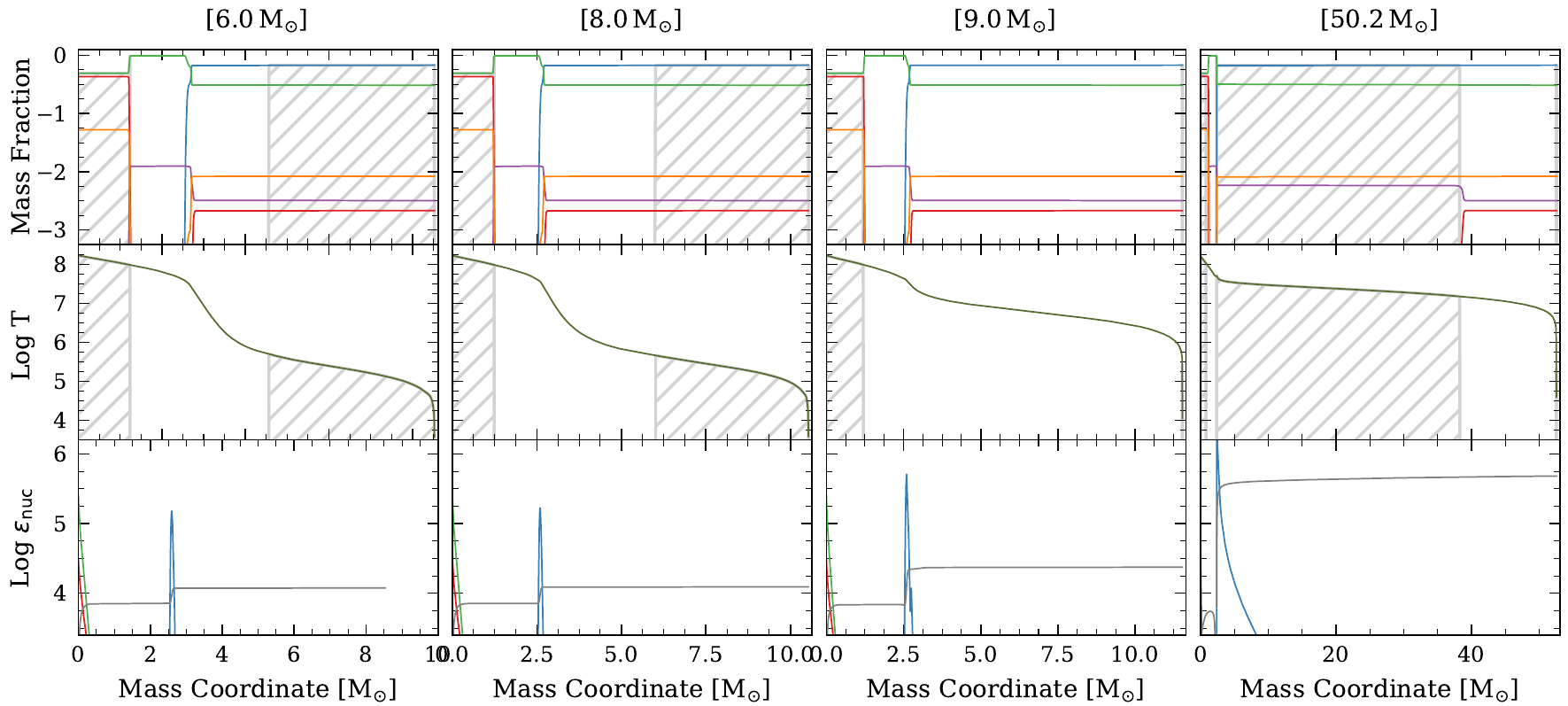}
  \caption{Internal structure profiles of core--He burning models with $\mcore = 2.5\msun$, $\yc = 0.50$, and a range of envelope masses (indicated at top of each panel) selected to represent the qualitative behaviour of the models. See caption of Fig. \ref{fig:interior_example_C4D1} for further details.}
 \label{fig:appendix:mass10}
 \end{figure*}

 \begin{figure*}
  \centering
  \includegraphics[width=\hsize]{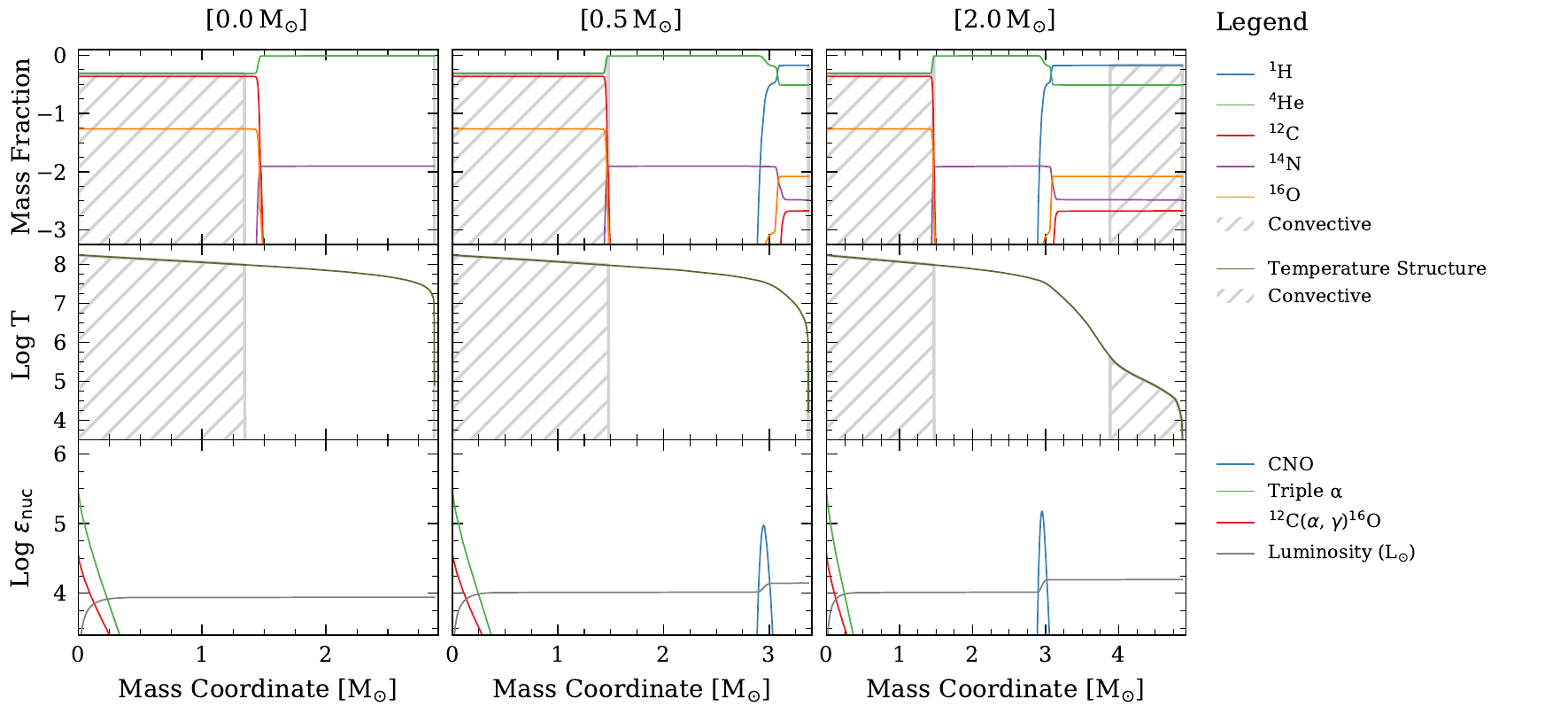}
  \includegraphics[width=\hsize]{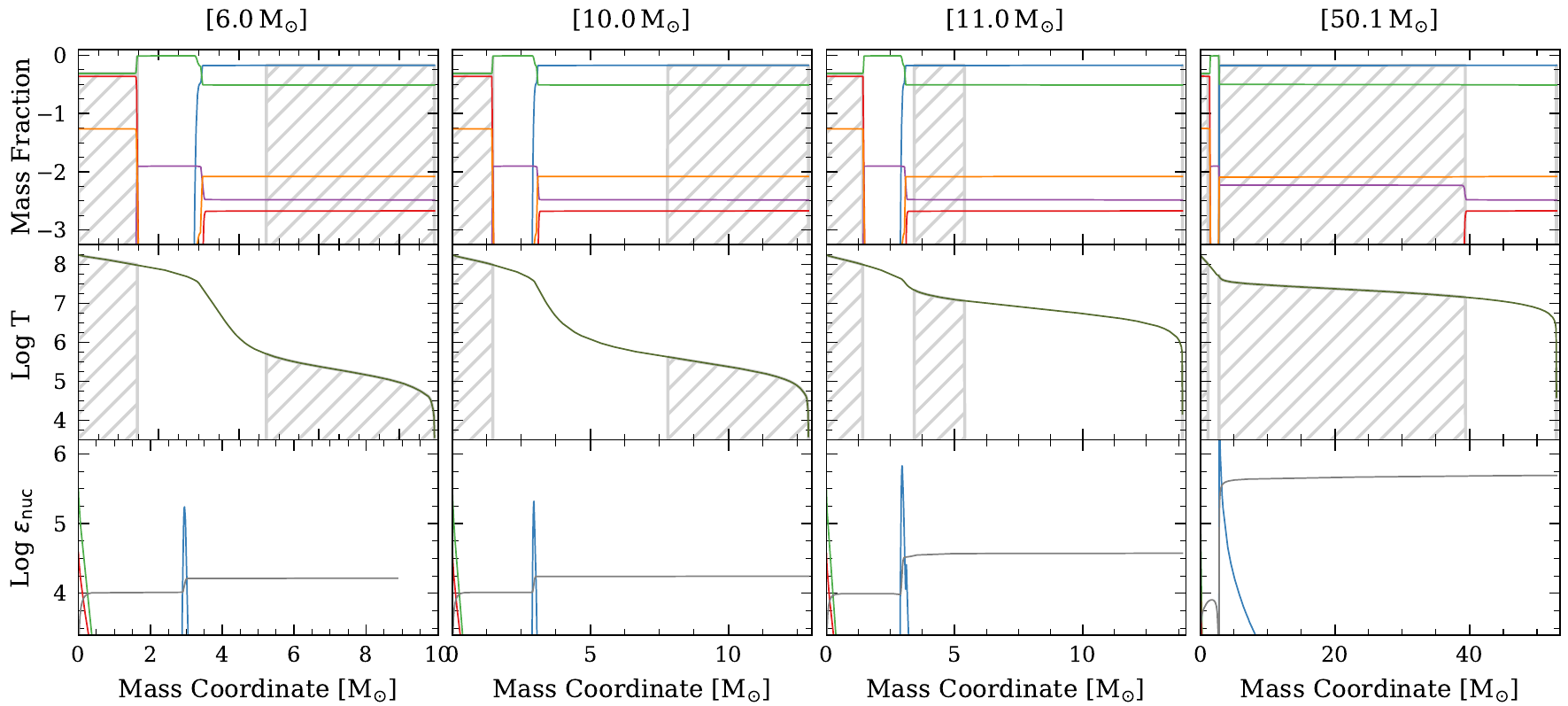}
  \caption{Internal structure profiles of core--He burning models with $\mcore = 2.9\msun$, $\yc = 0.50$, and a range of envelope masses (indicated at top of each panel) selected to represent the qualitative behaviour of the models. See caption of Fig. \ref{fig:interior_example_C4D1} for further details.}
 \label{fig:appendix:mass11}
 \end{figure*}

 \begin{figure*}
  \centering
  \includegraphics[width=\hsize]{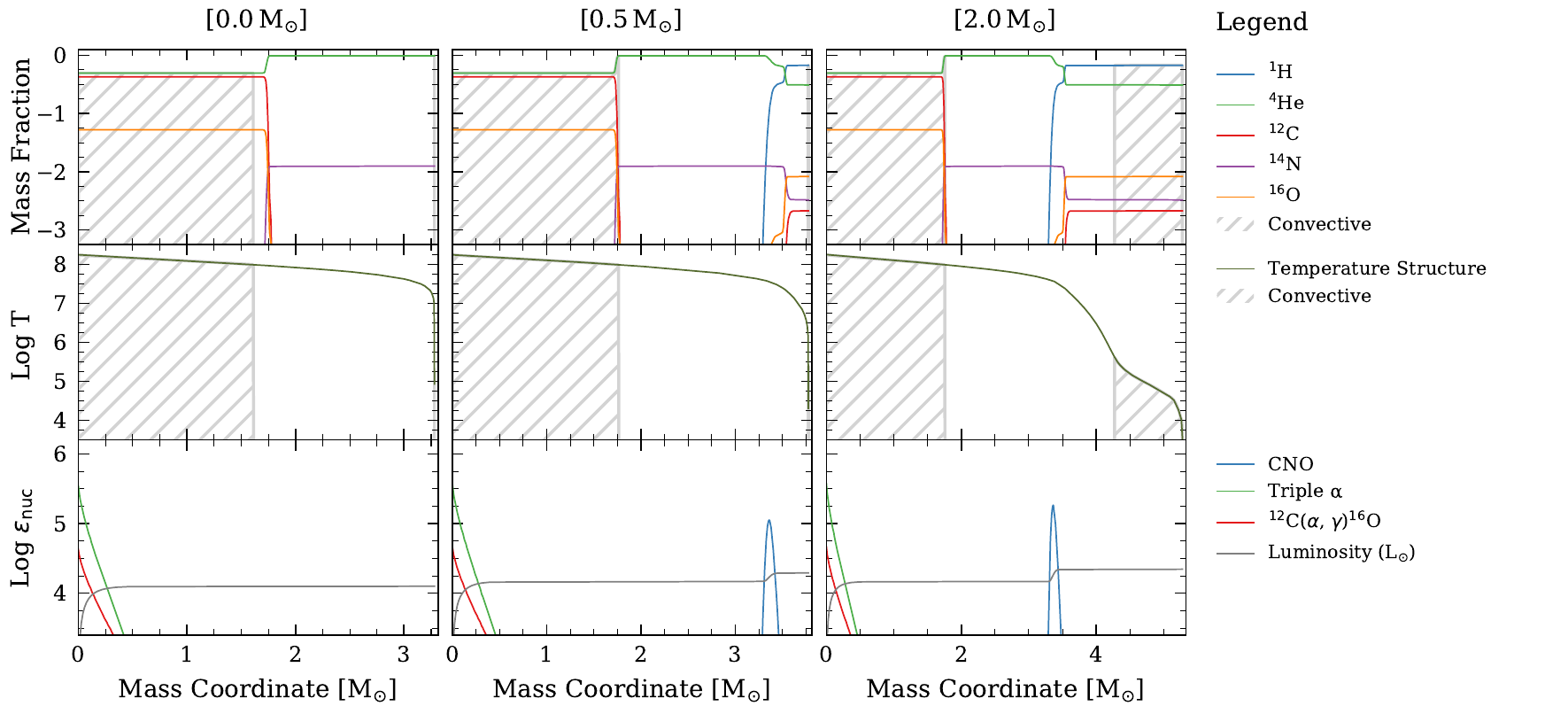}
  \includegraphics[width=\hsize]{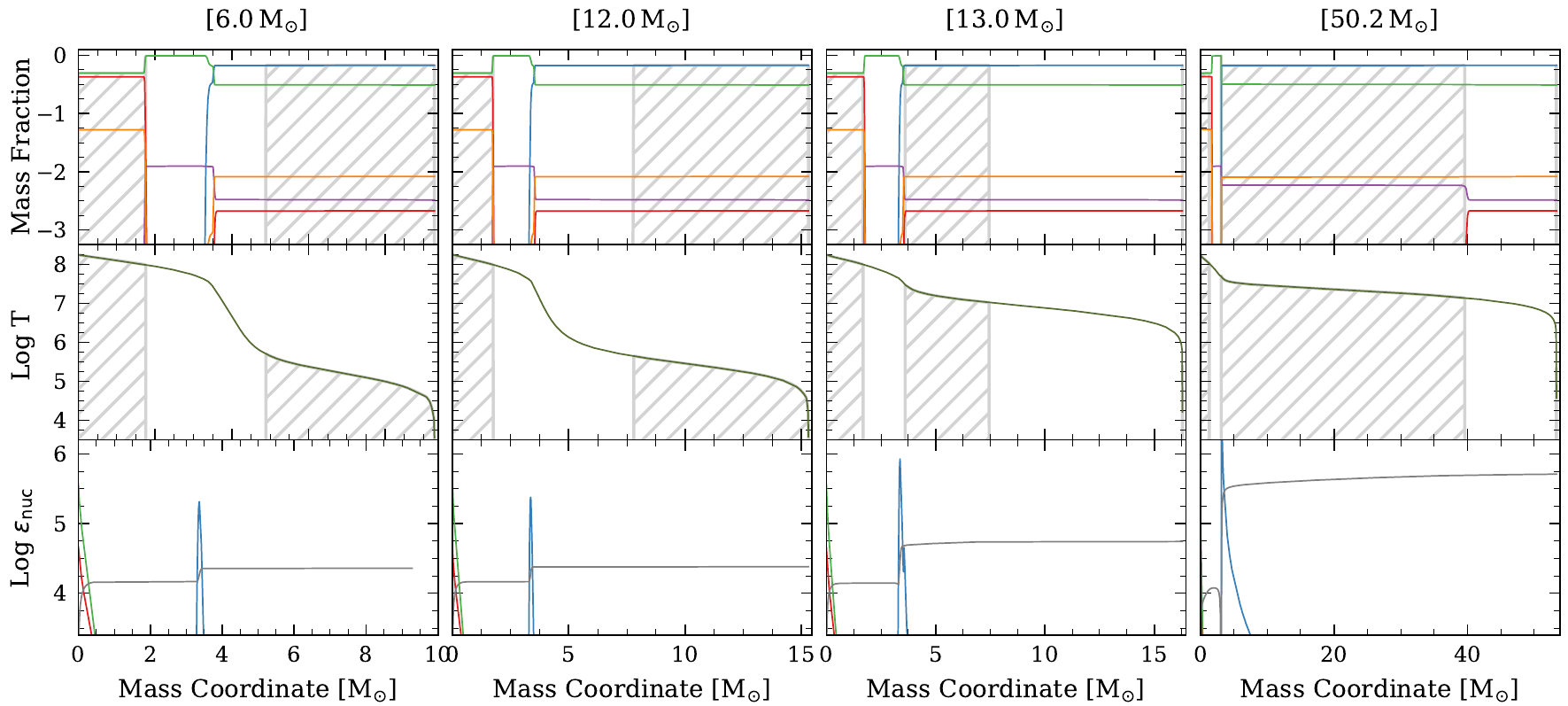}
  \caption{Internal structure profiles of core--He burning models with $\mcore = 3.3\msun$, $\yc = 0.50$, and a range of envelope masses (indicated at top of each panel) selected to represent the qualitative behaviour of the models. See caption of Fig. \ref{fig:interior_example_C4D1} for further details.}
 \label{fig:appendix:mass12}
 \end{figure*}

 \begin{figure*}
  \centering
  \includegraphics[width=\hsize]{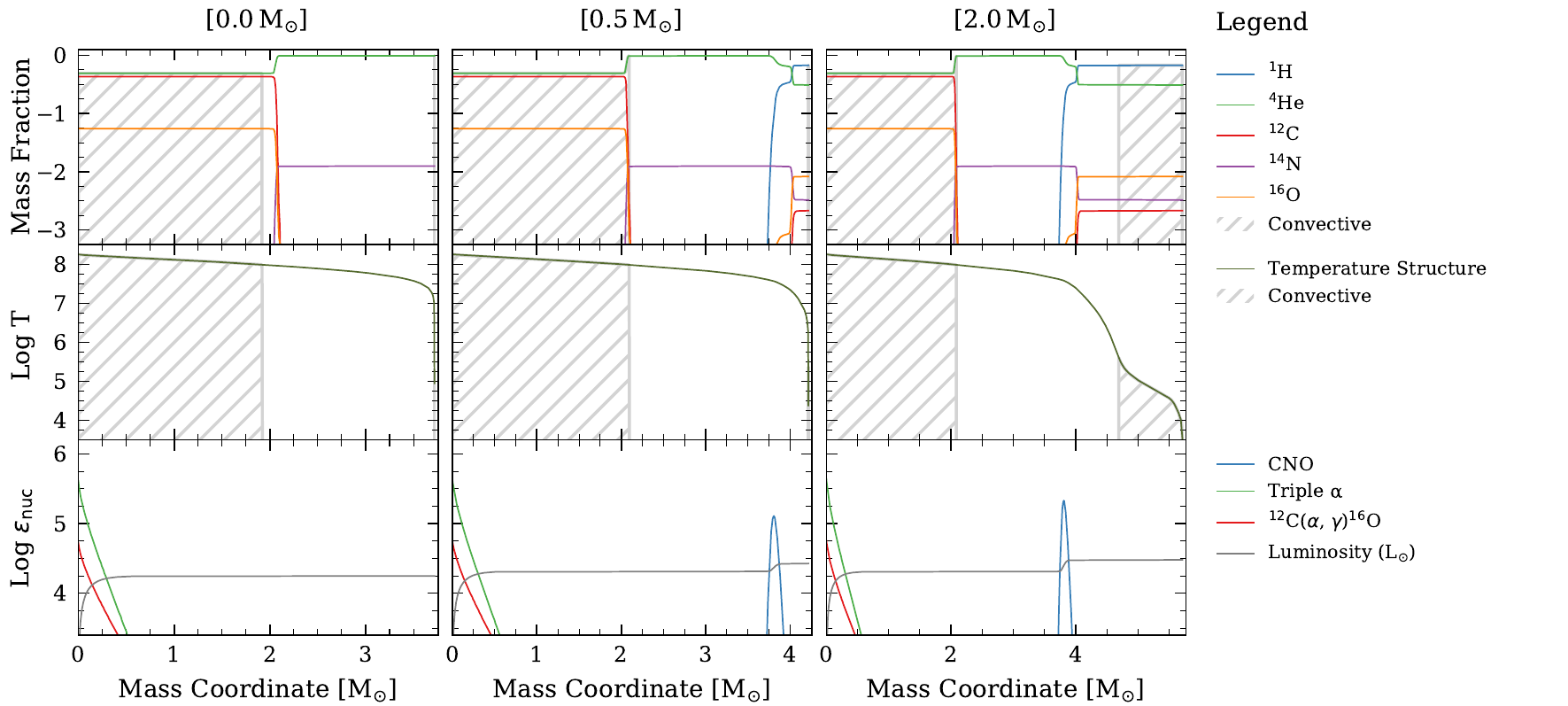}
  \includegraphics[width=\hsize]{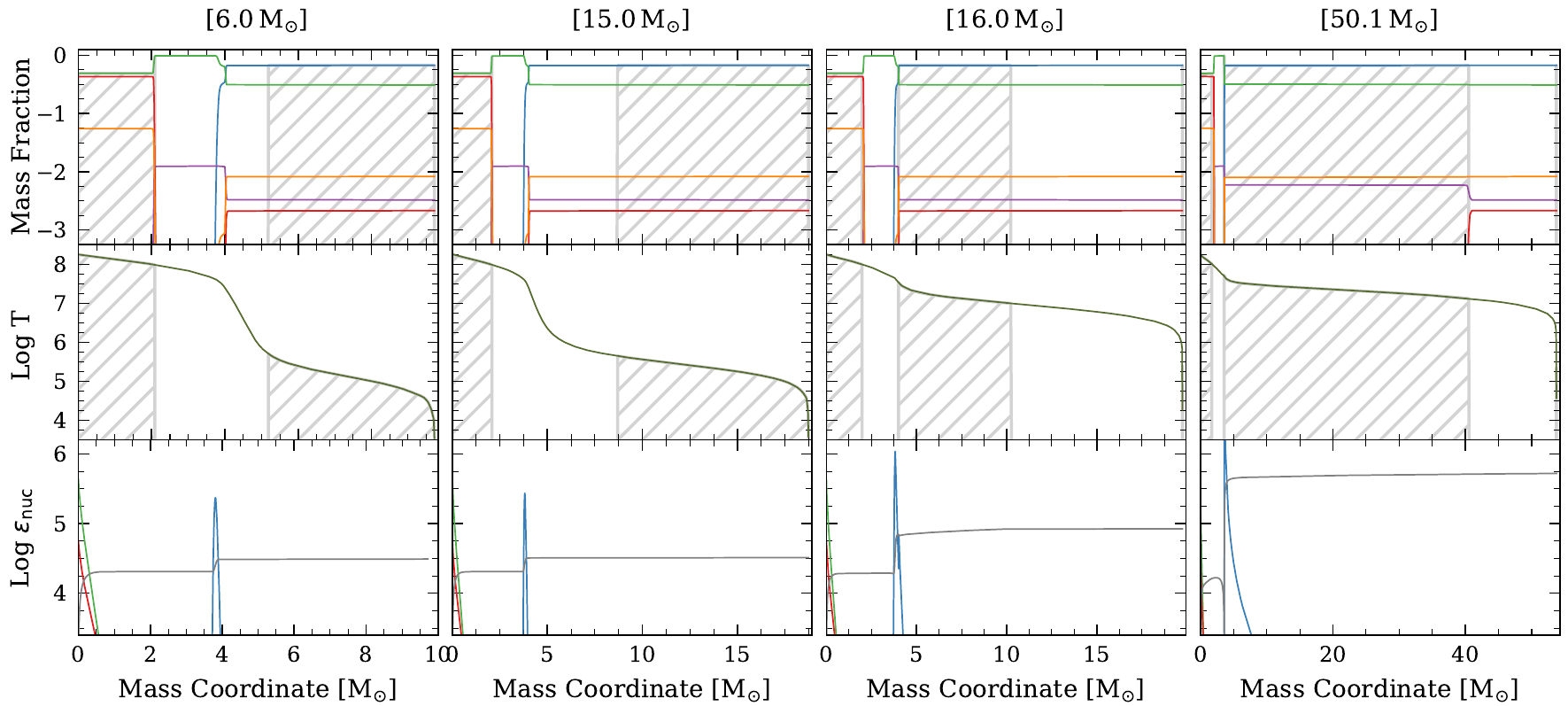}
  \caption{Internal structure profiles of core--He burning models with $\mcore = 3.7\msun$, $\yc = 0.50$, and a range of envelope masses (indicated at top of each panel) selected to represent the qualitative behaviour of the models. See caption of Fig. \ref{fig:interior_example_C4D1} for further details.}
 \label{fig:appendix:mass13}
 \end{figure*}

 \begin{figure*}
  \centering
  \includegraphics[width=\hsize]{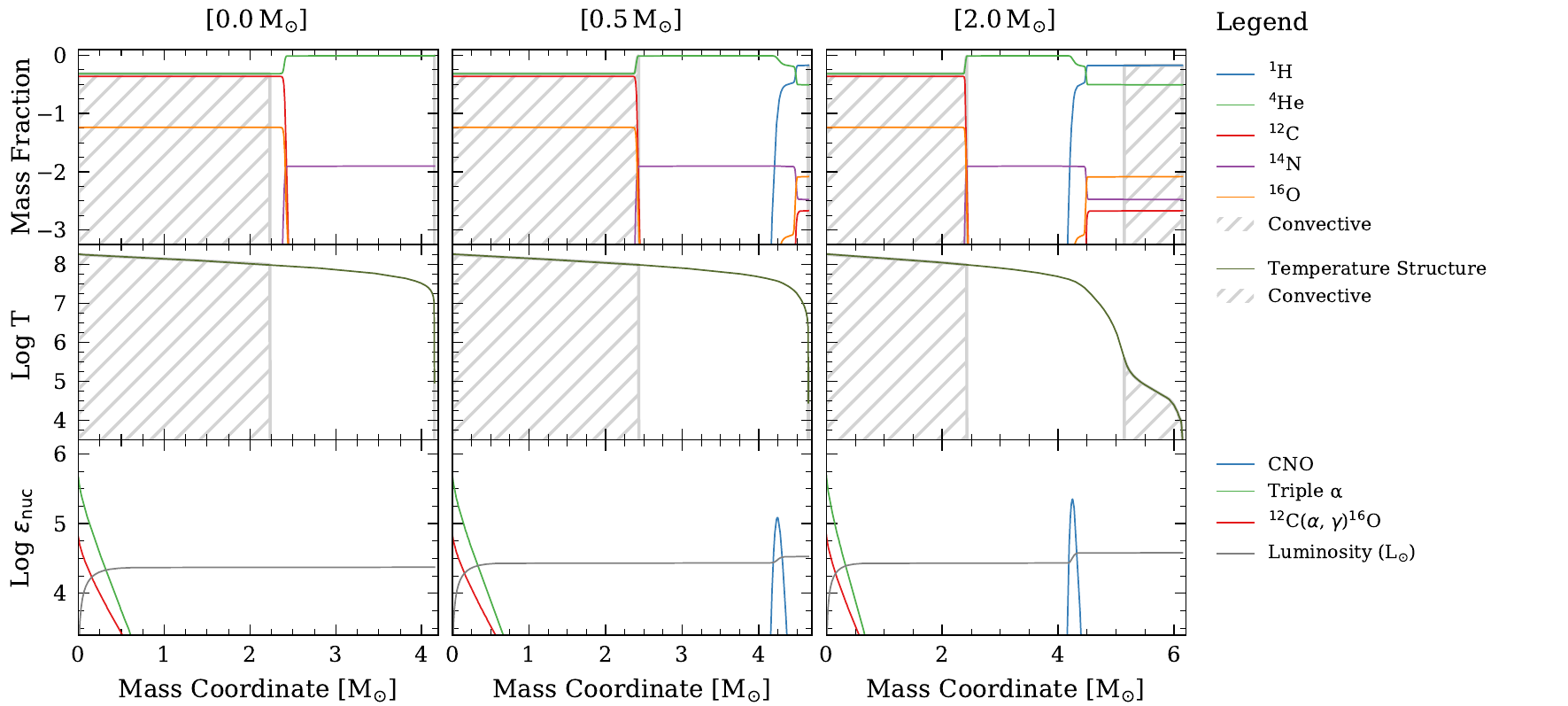}
  \includegraphics[width=\hsize]{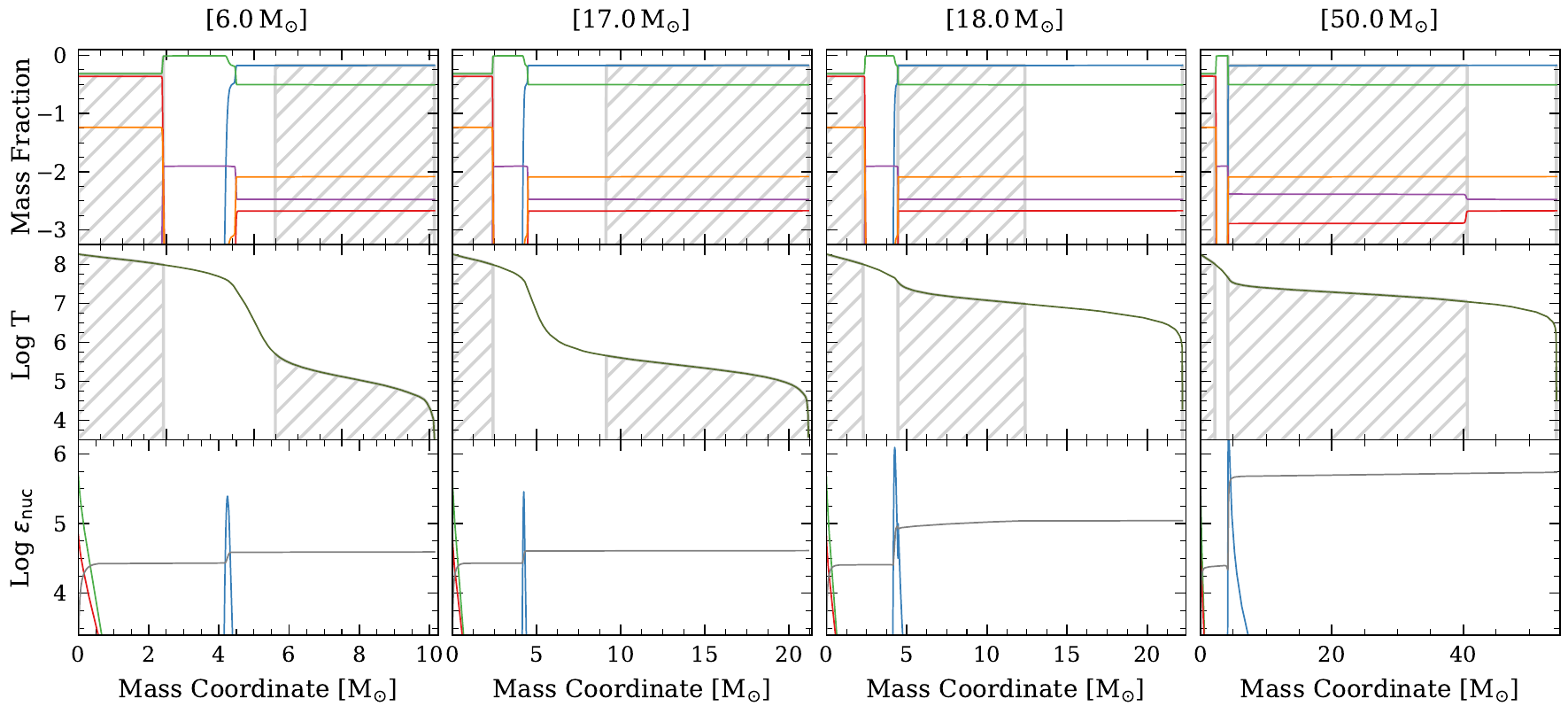}
  \caption{Internal structure profiles of core--He burning models with $\mcore = 4.1\msun$, $\yc = 0.50$, and a range of envelope masses (indicated at top of each panel) selected to represent the qualitative behaviour of the models. See caption of Fig. \ref{fig:interior_example_C4D1} for further details.}
 \label{fig:appendix:mass14}
 \end{figure*}

 \begin{figure*}
  \centering
  \includegraphics[width=\hsize]{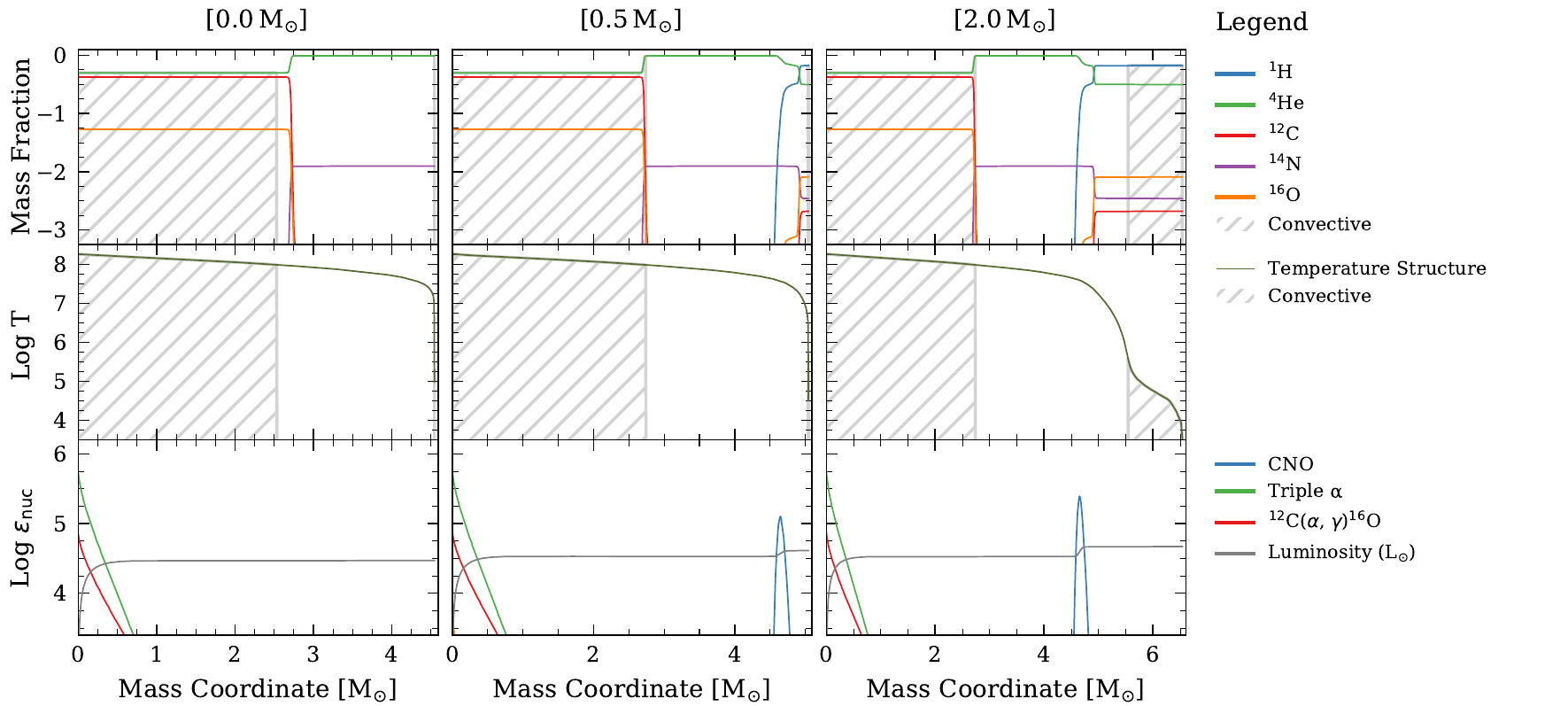}
  \includegraphics[width=\hsize]{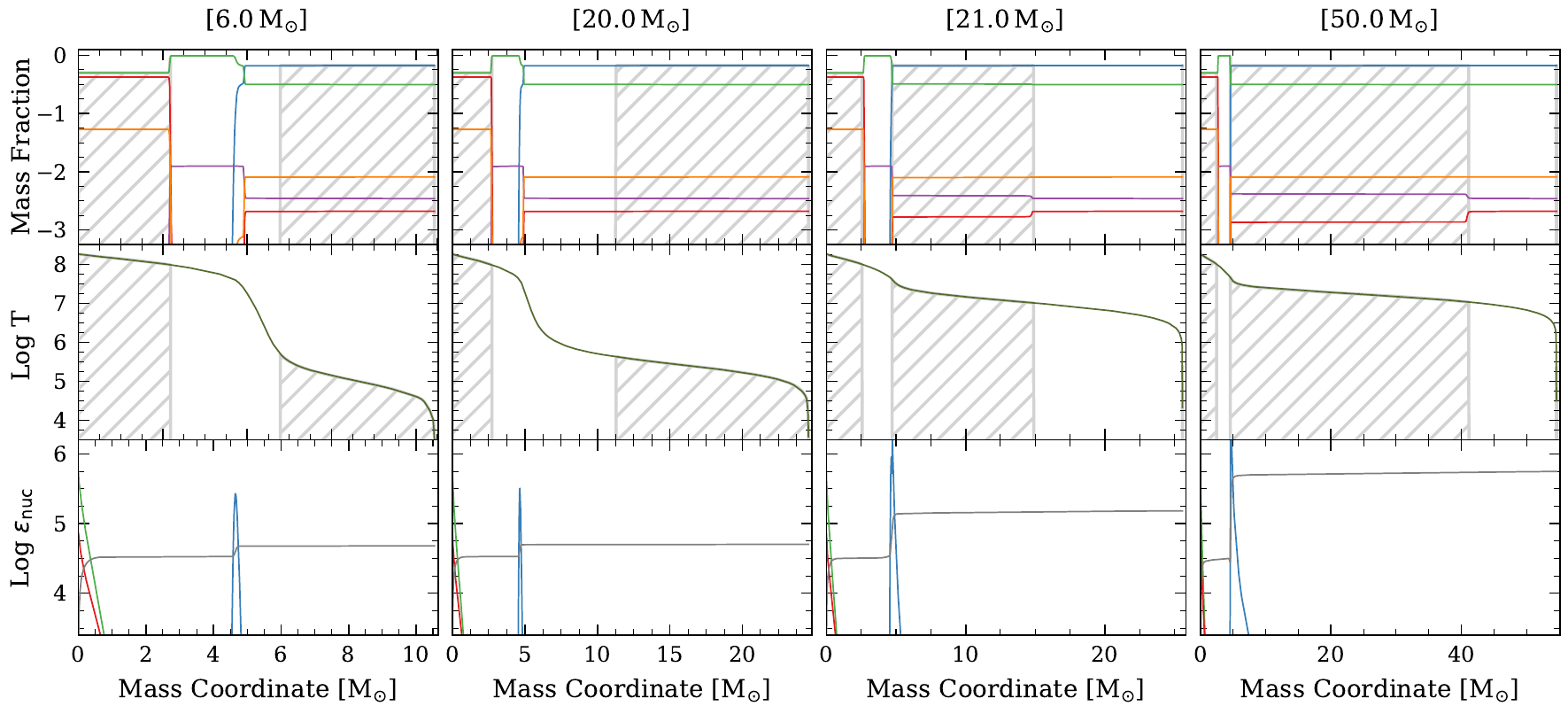}
  \caption{Internal structure profiles of core--He burning models with $\mcore = 4.5\msun$, $\yc = 0.50$, and a range of envelope masses (indicated at top of each panel) selected to represent the qualitative behaviour of the models. See caption of Fig. \ref{fig:interior_example_C4D1} for further details.}
 \label{fig:appendix:mass15}
 \end{figure*}

 \begin{figure*}
  \centering
  \includegraphics[width=\hsize]{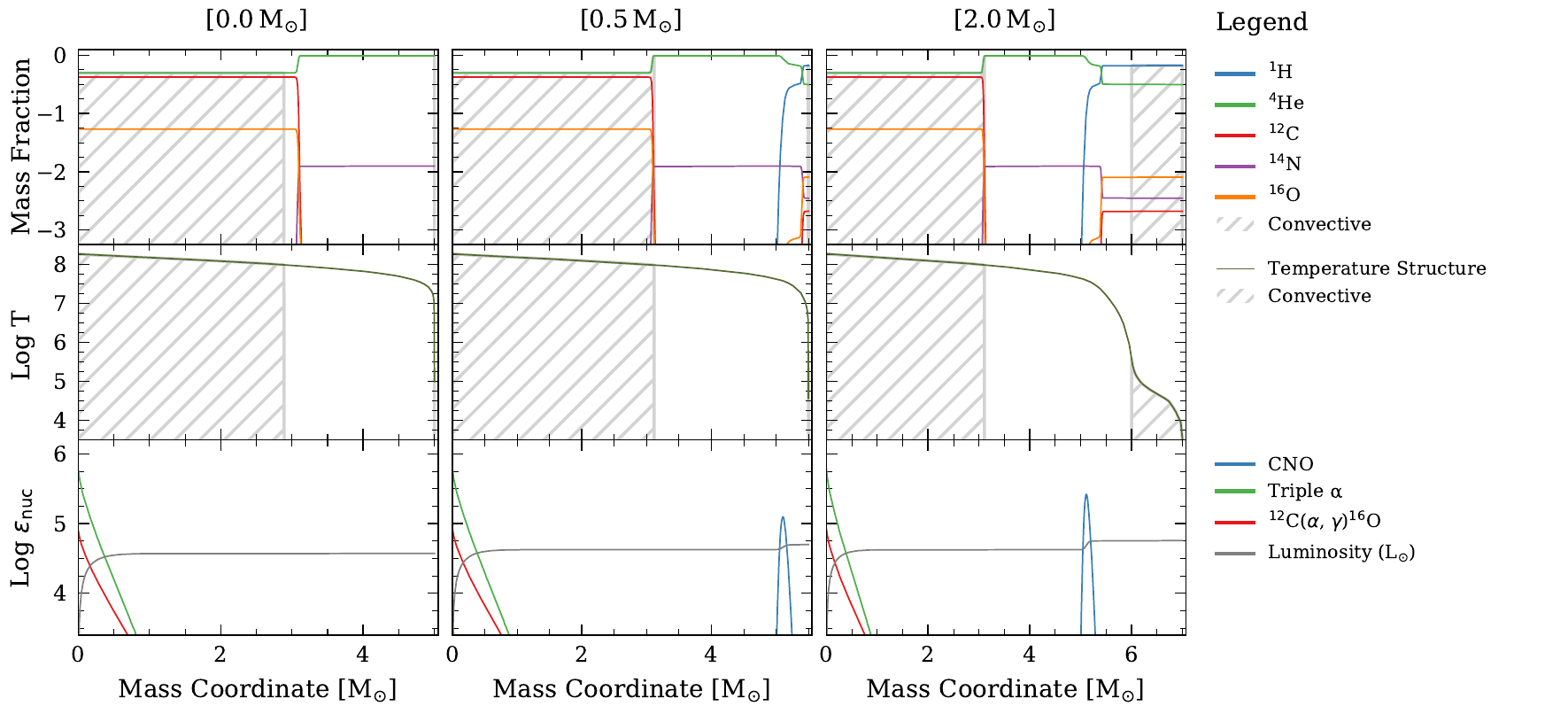}
  \includegraphics[width=\hsize]{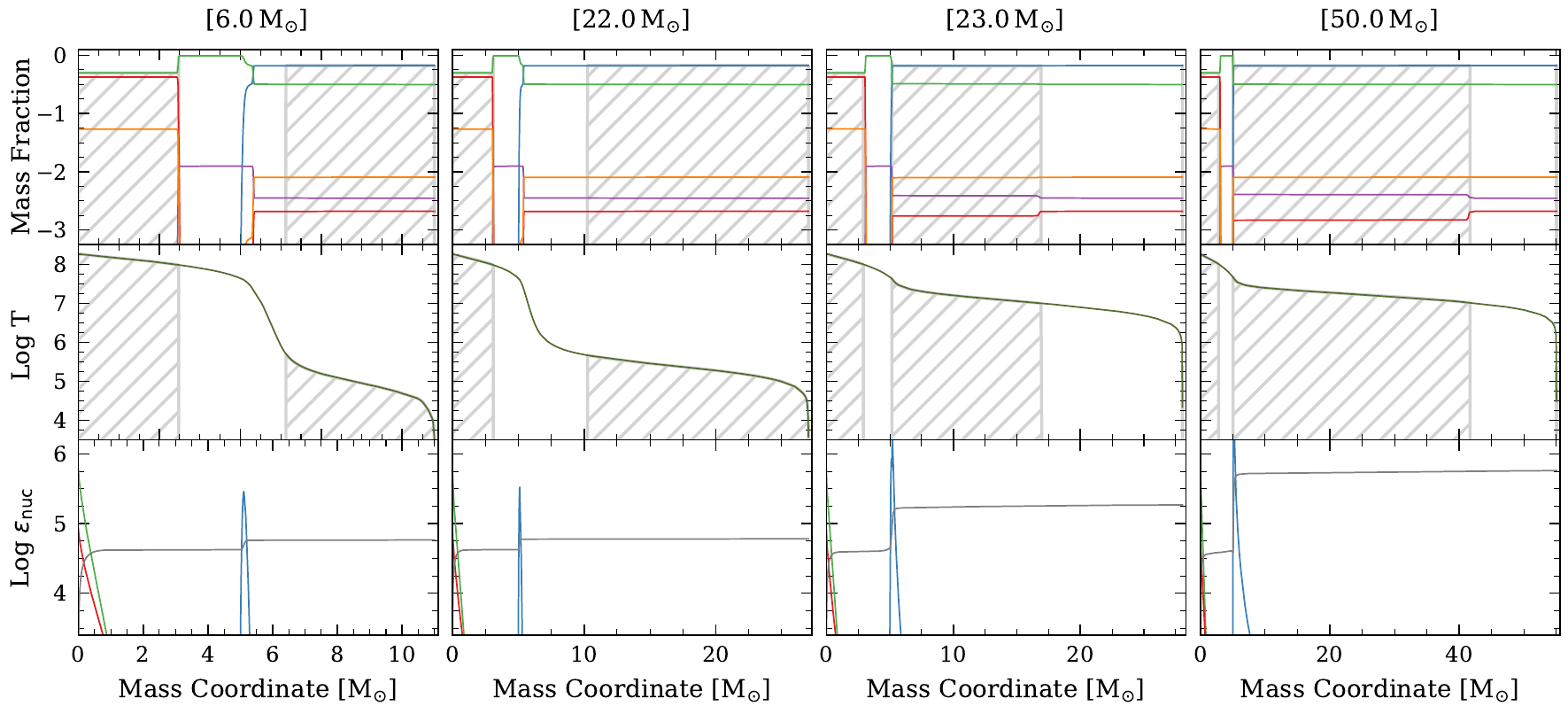}
  \caption{Internal structure profiles of core--He burning models with $\mcore = 5.0\msun$, $\yc = 0.50$, and a range of envelope masses (indicated at top of each panel) selected to represent the qualitative behaviour of the models. See caption of Fig. \ref{fig:interior_example_C4D1} for further details.}
 \label{fig:appendix:mass16}
 \end{figure*}

 \begin{figure*}
  \centering
  \includegraphics[width=\hsize]{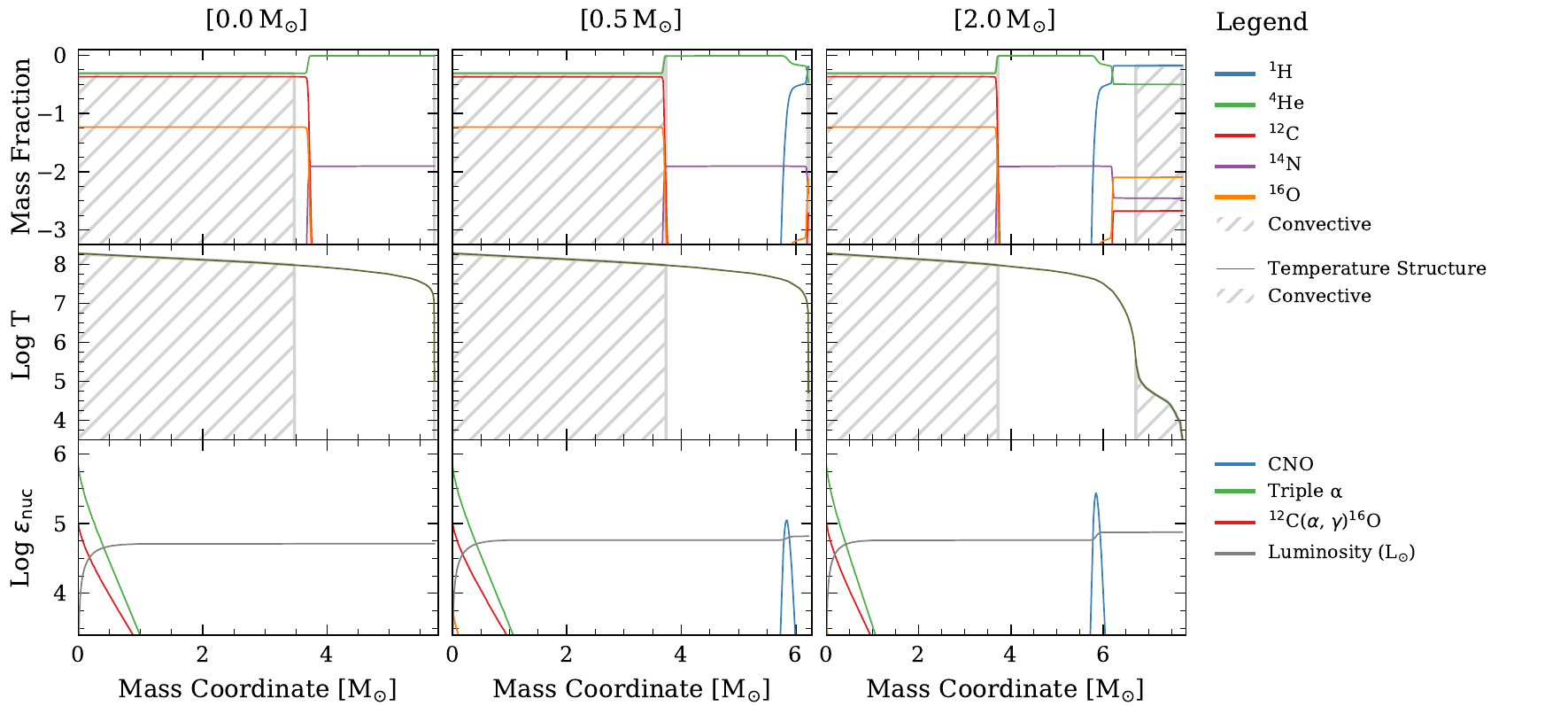}
  \includegraphics[width=\hsize]{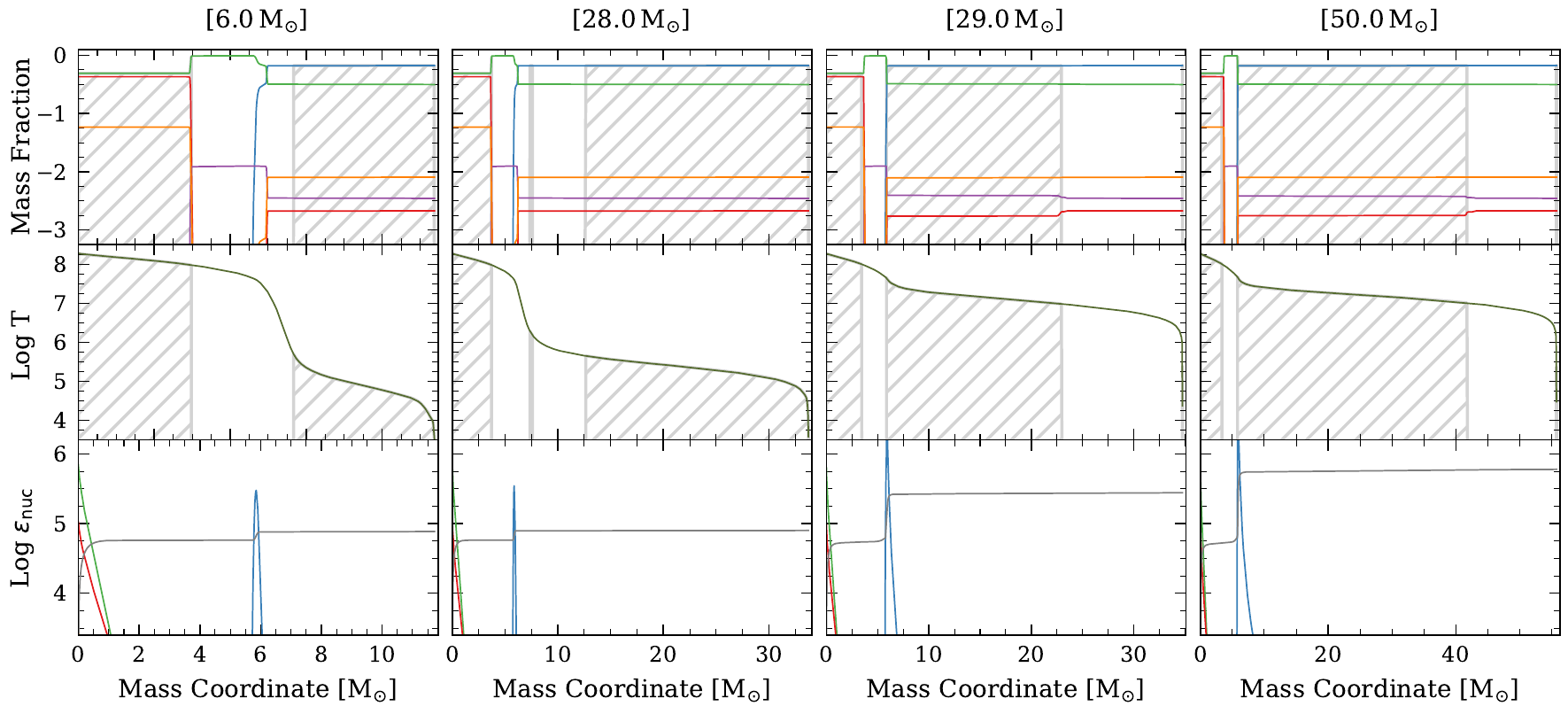}
  \caption{Internal structure profiles of core--He burning models with $\mcore = 5.7\msun$, $\yc = 0.50$, and a range of envelope masses (indicated at top of each panel) selected to represent the qualitative behaviour of the models. See caption of Fig. \ref{fig:interior_example_C4D1} for further details.}
 \label{fig:appendix:mass175}
 \end{figure*}

 \begin{figure*}
  \centering
  \includegraphics[width=\hsize]{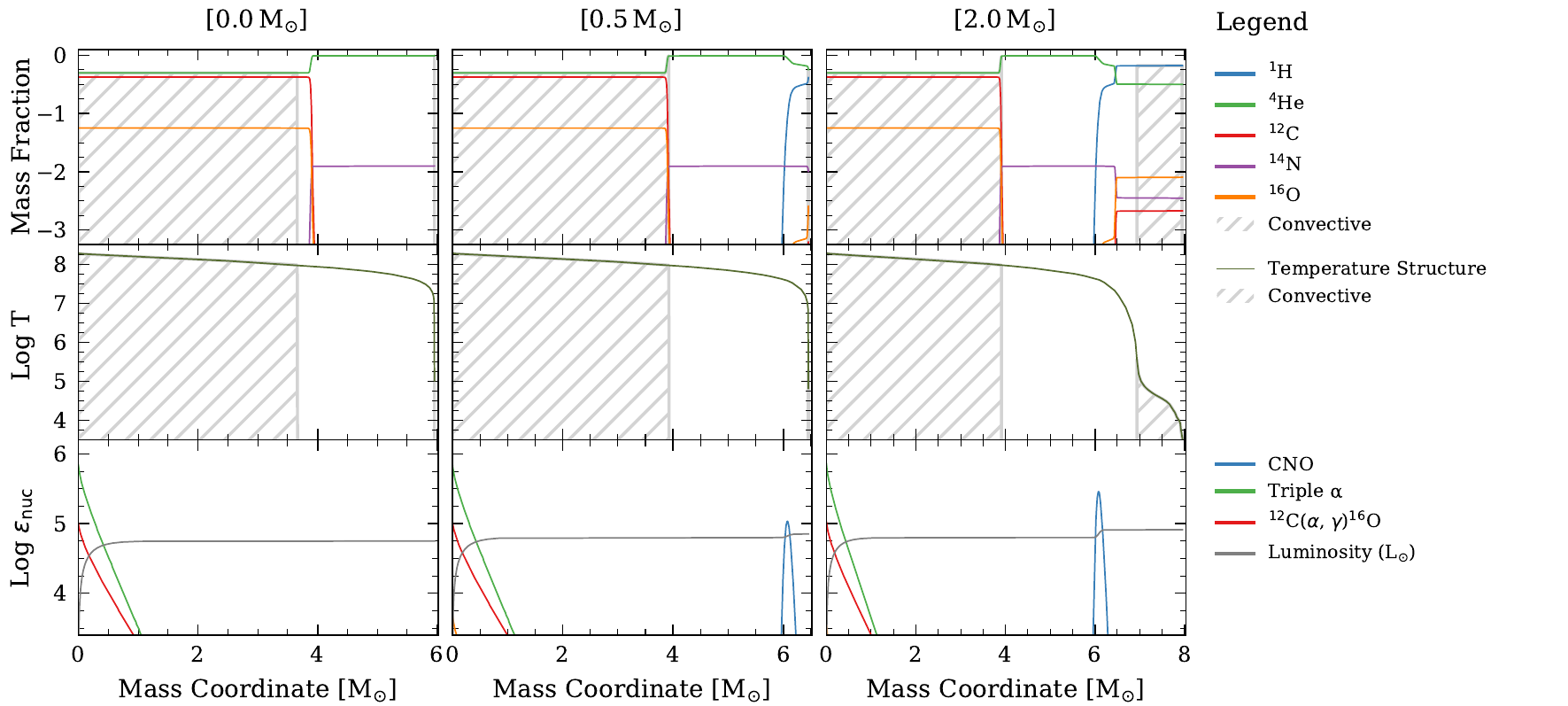}
  \includegraphics[width=\hsize]{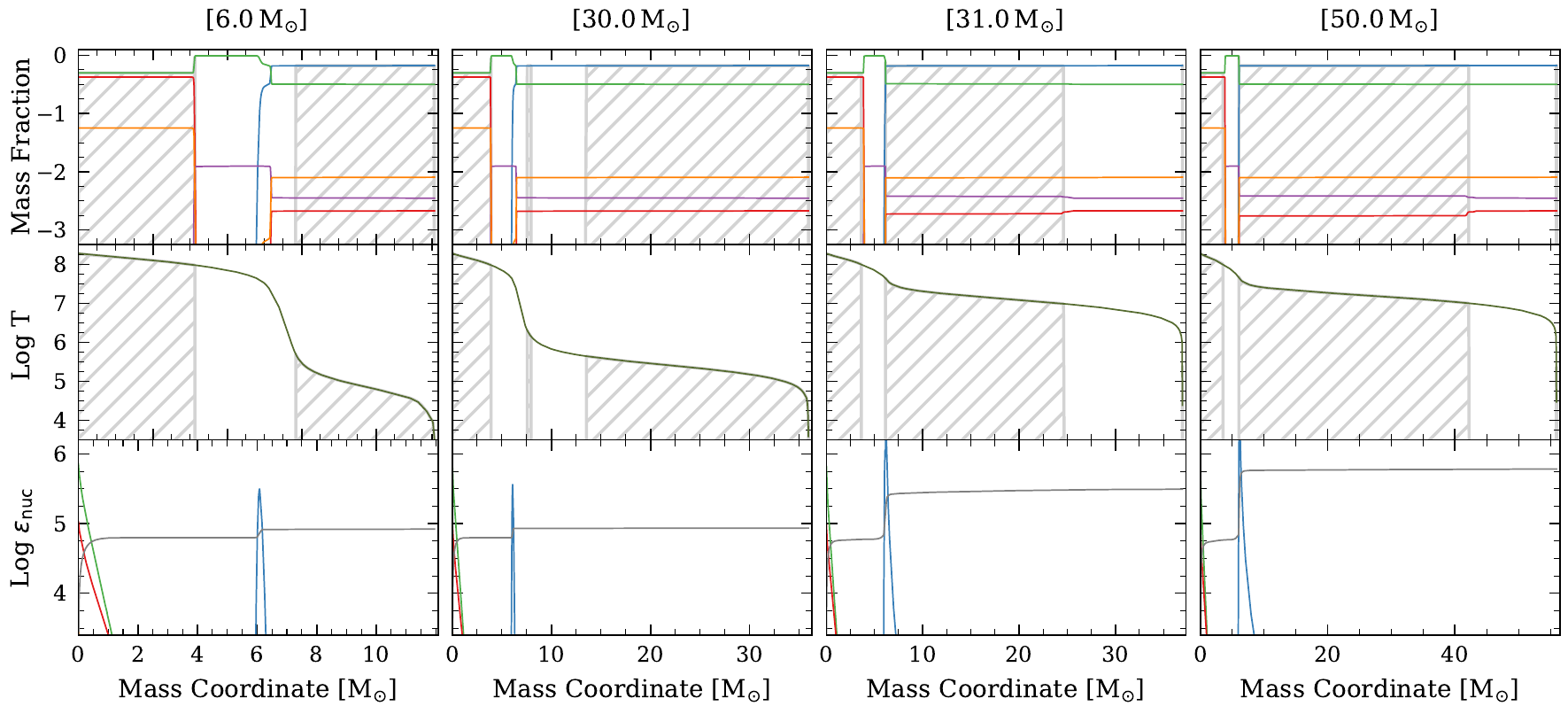}
  \caption{Internal structure profiles of core--He burning models with $\mcore = 6.0\msun$, $\yc = 0.50$, and a range of envelope masses (indicated at top of each panel) selected to represent the qualitative behaviour of the models. See caption of Fig. \ref{fig:interior_example_C4D1} for further details.}
 \label{fig:appendix:mass18}
 \end{figure*}

 \begin{figure*}
  \centering
  \includegraphics[width=\hsize]{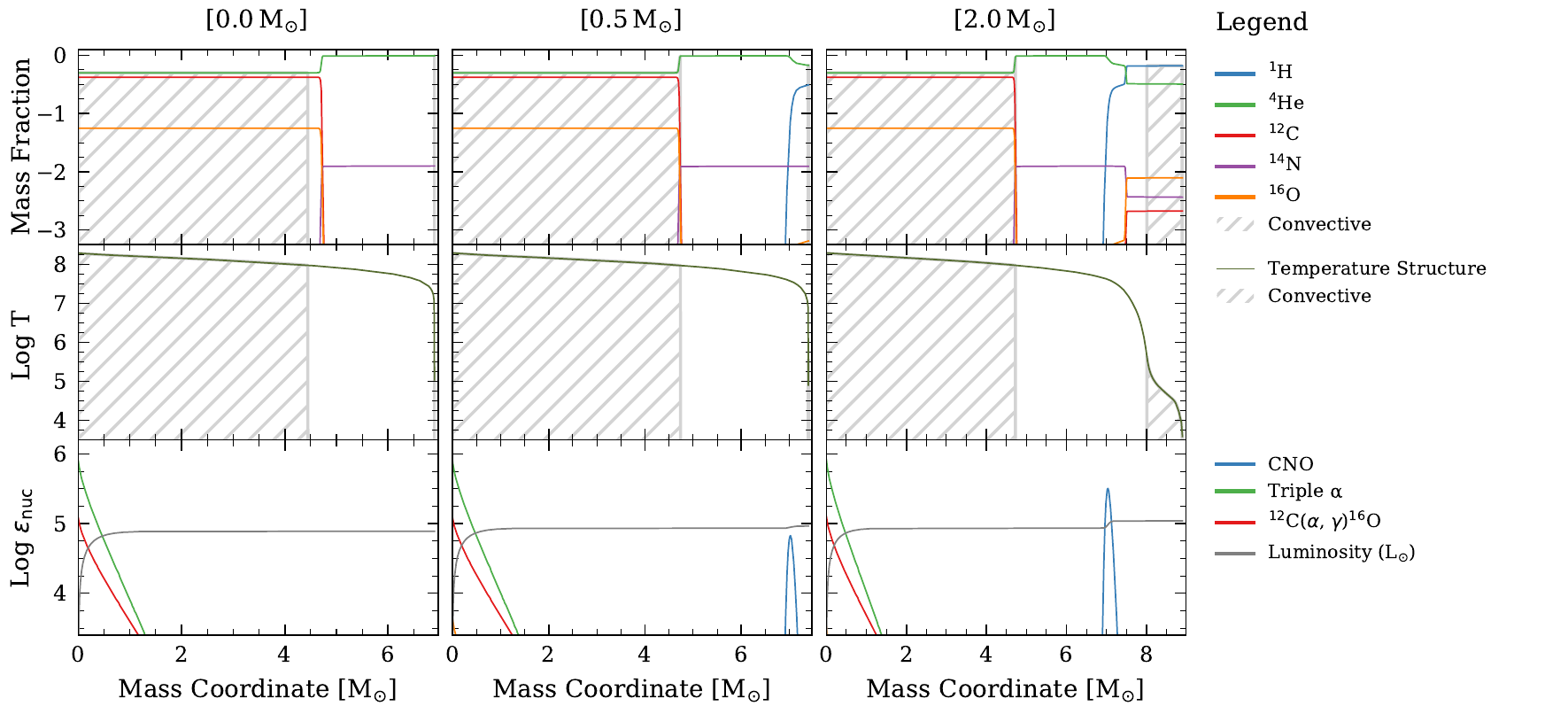}
  \includegraphics[width=\hsize]{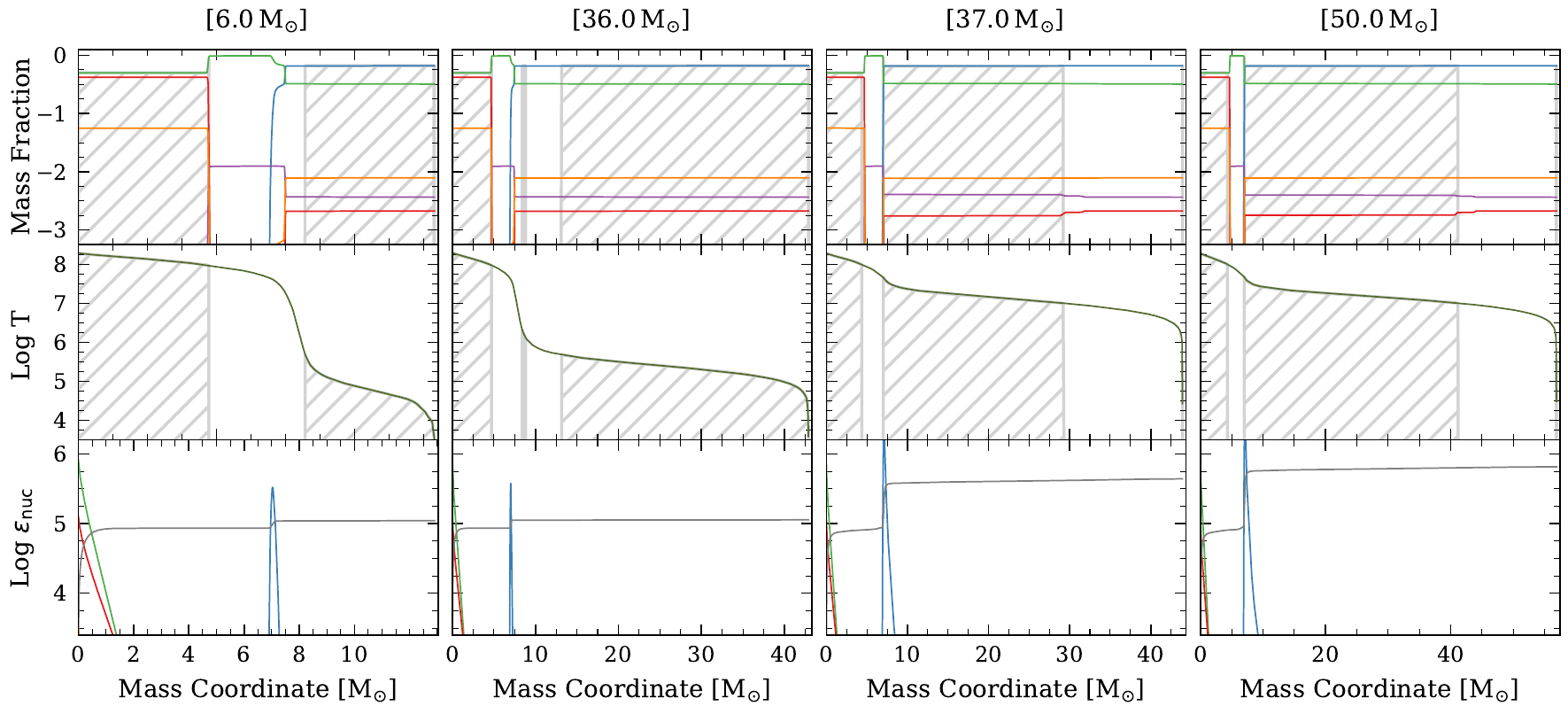}
  \caption{Internal structure profiles of core--He burning models with $\mcore = 6.9\msun$, $\yc = 0.50$, and a range of envelope masses (indicated at top of each panel) selected to represent the qualitative behaviour of the models. See caption of Fig. \ref{fig:interior_example_C4D1} for further details.}
 \label{fig:appendix:mass20}
 \end{figure*}

 \begin{figure*}
  \centering
  \includegraphics[width=\hsize]{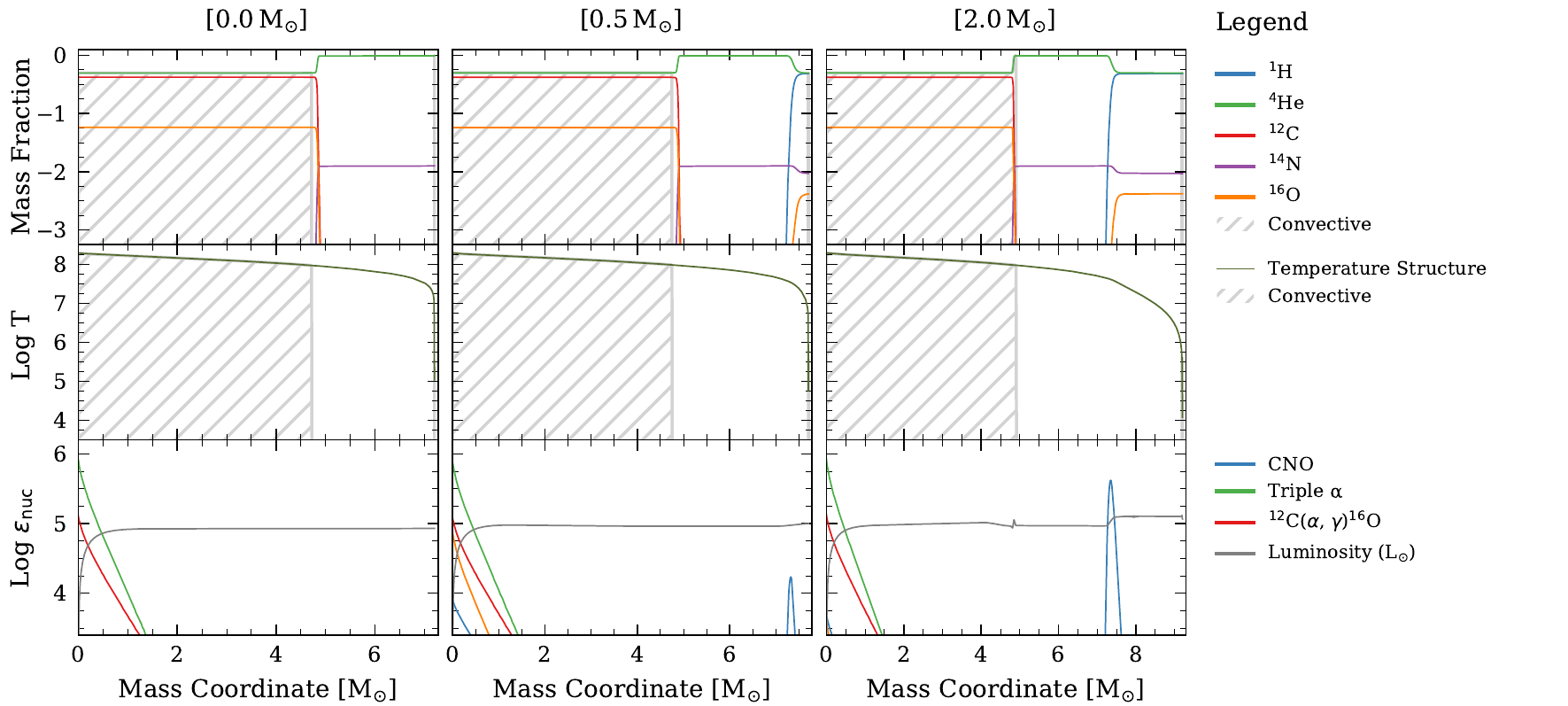}
  \includegraphics[width=\hsize]{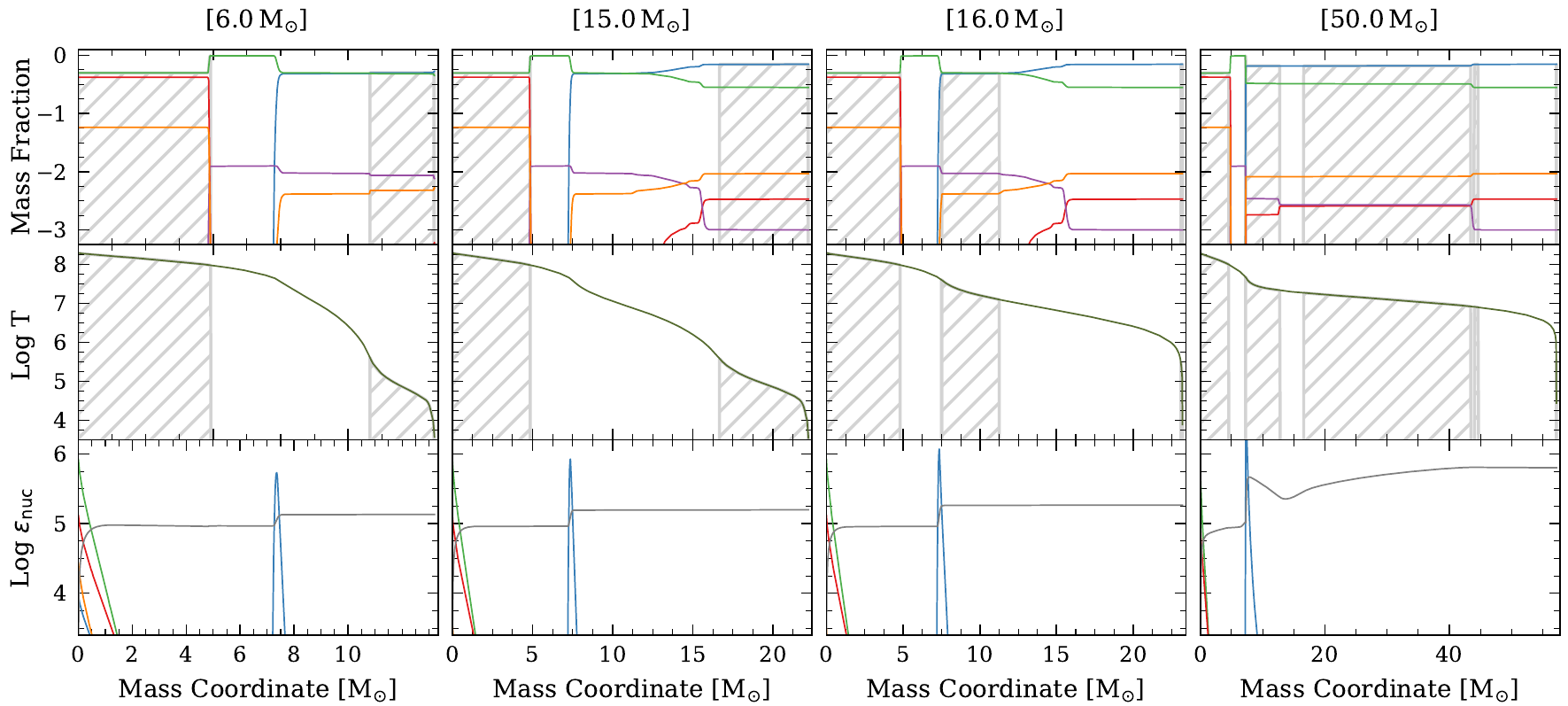}
  \caption{Internal structure profiles of core--He burning models with $\mcore = 7.2\msun$, $\yc = 0.50$, and a range of envelope masses (indicated at top of each panel) selected to represent the qualitative behaviour of the models. See caption of Fig. \ref{fig:interior_example_C4D1} for further details.}
 \label{fig:appendix:mass22}
 \end{figure*}

 \begin{figure*}
  \centering
  \includegraphics[width=\hsize]{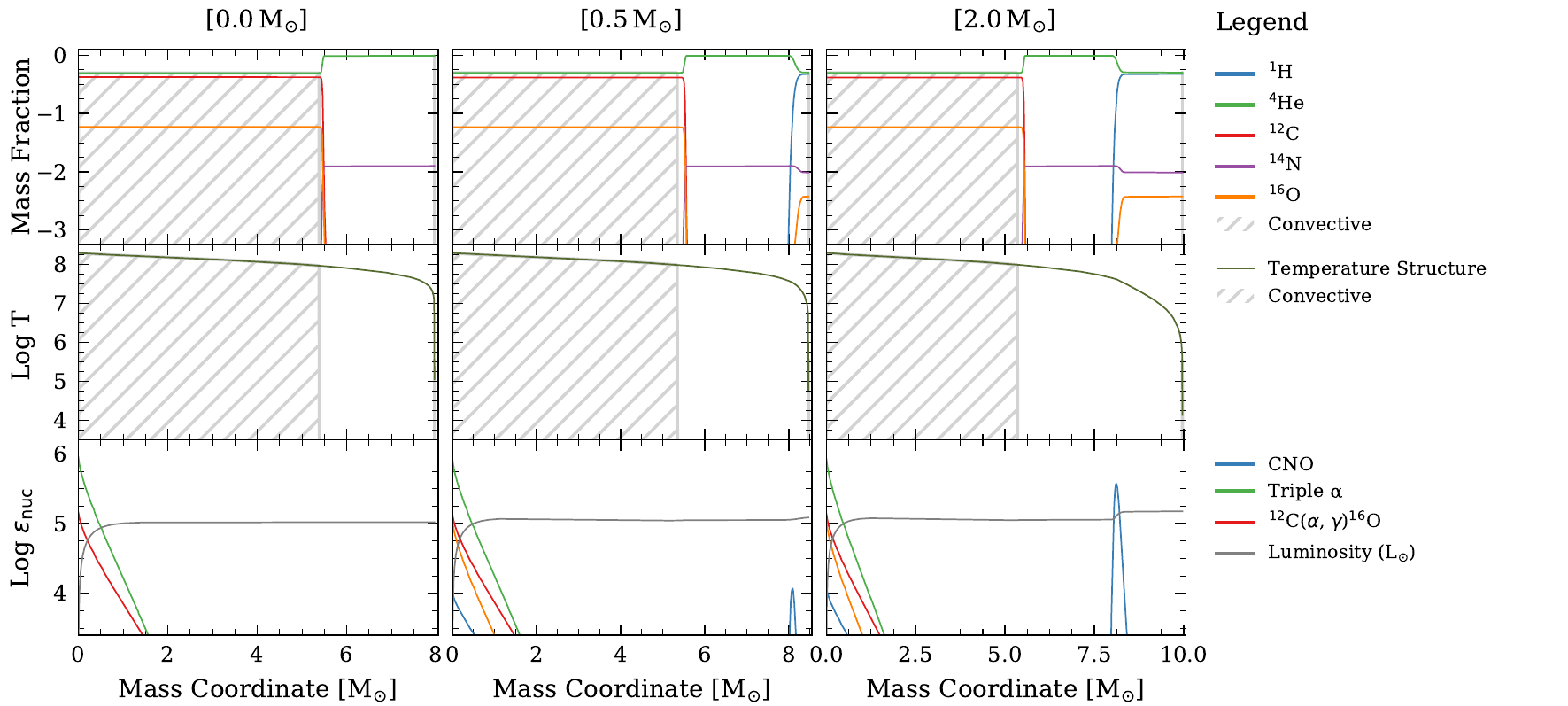}
  \includegraphics[width=\hsize]{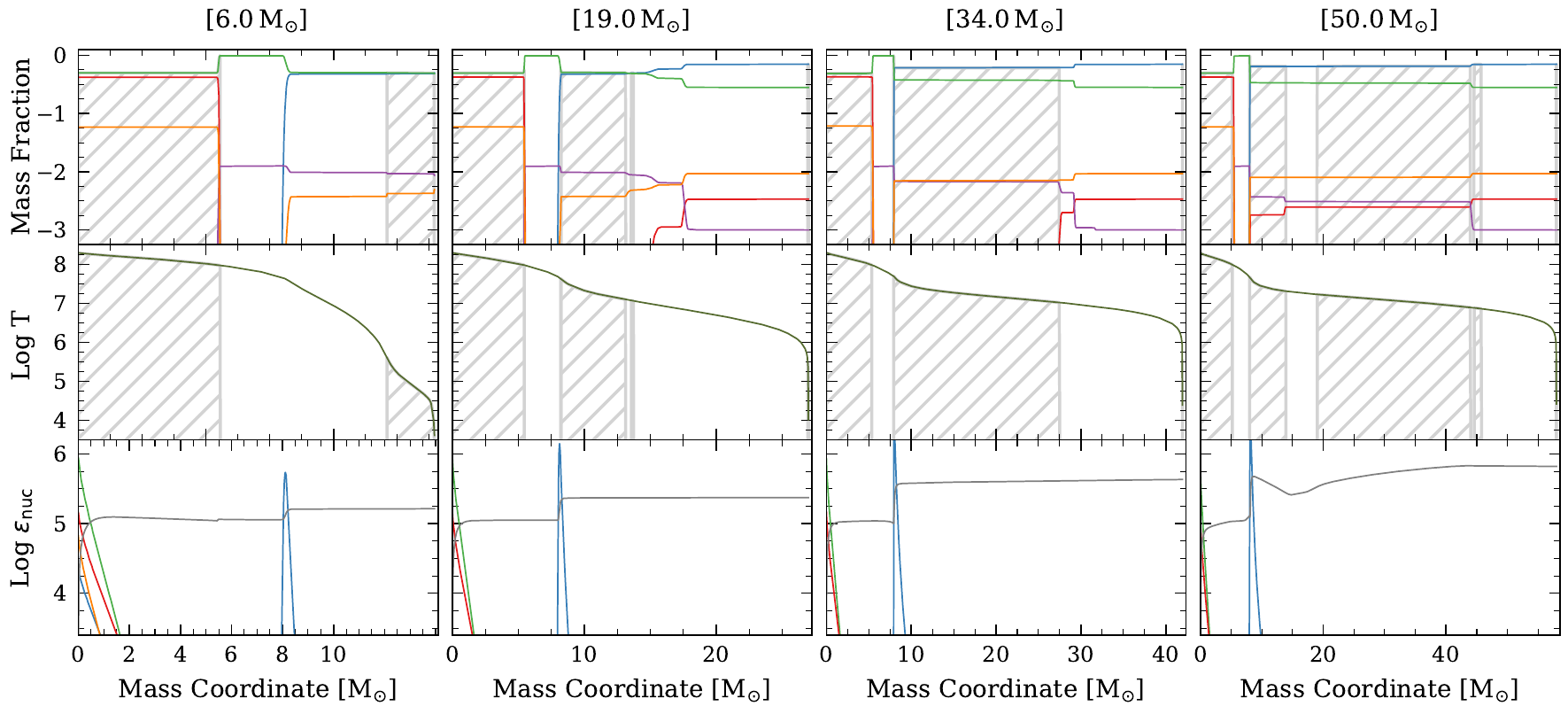}
  \caption{Internal structure profiles of core--He burning models with $\mcore = 8.0\msun$, $\yc = 0.50$, and a range of envelope masses (indicated at top of each panel) selected to represent the qualitative behaviour of the models. See caption of Fig. \ref{fig:interior_example_C4D1} for further details.}
 \label{fig:appendix:mass24}
 \end{figure*}

 \begin{figure*}
  \centering
  \includegraphics[width=\hsize]{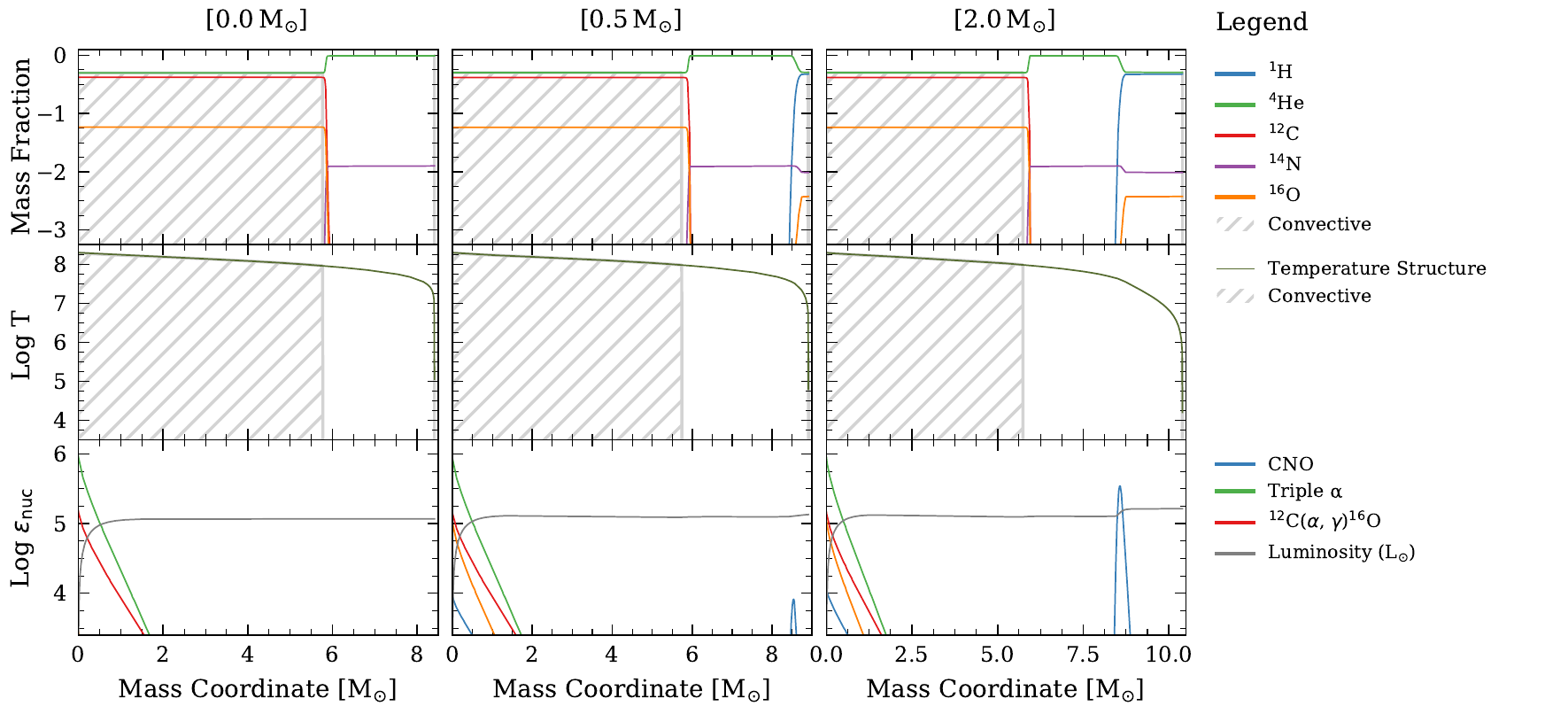}
  \includegraphics[width=\hsize]{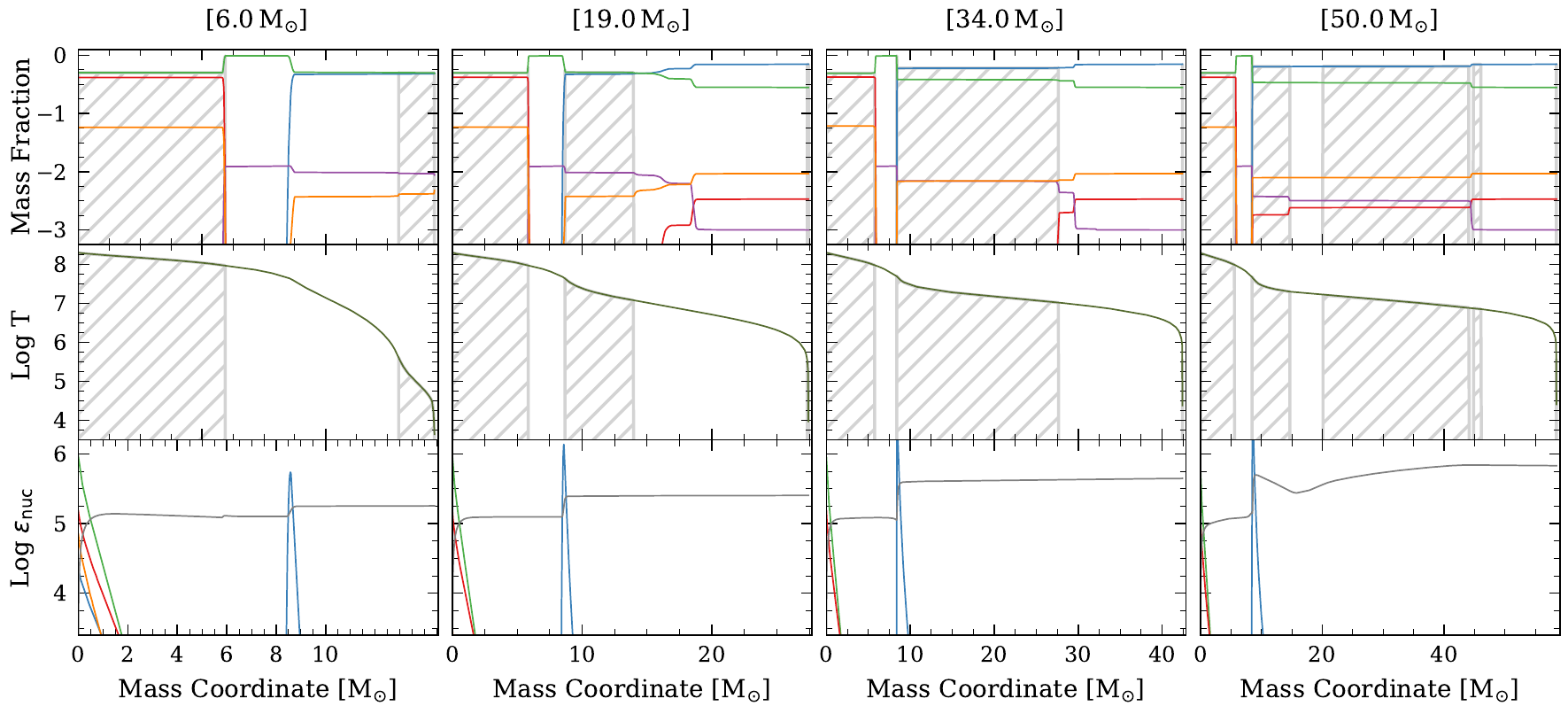}
  \caption{Internal structure profiles of core--He burning models with $\mcore = 8.4\msun$, $\yc = 0.50$, and a range of envelope masses (indicated at top of each panel) selected to represent the qualitative behaviour of the models. See caption of Fig. \ref{fig:interior_example_C4D1} for further details.}
 \label{fig:appendix:mass25}
 \end{figure*}

 \section{Kippenhahn-like figures}

\begin{figure}
  \includegraphics[width=\hsize]{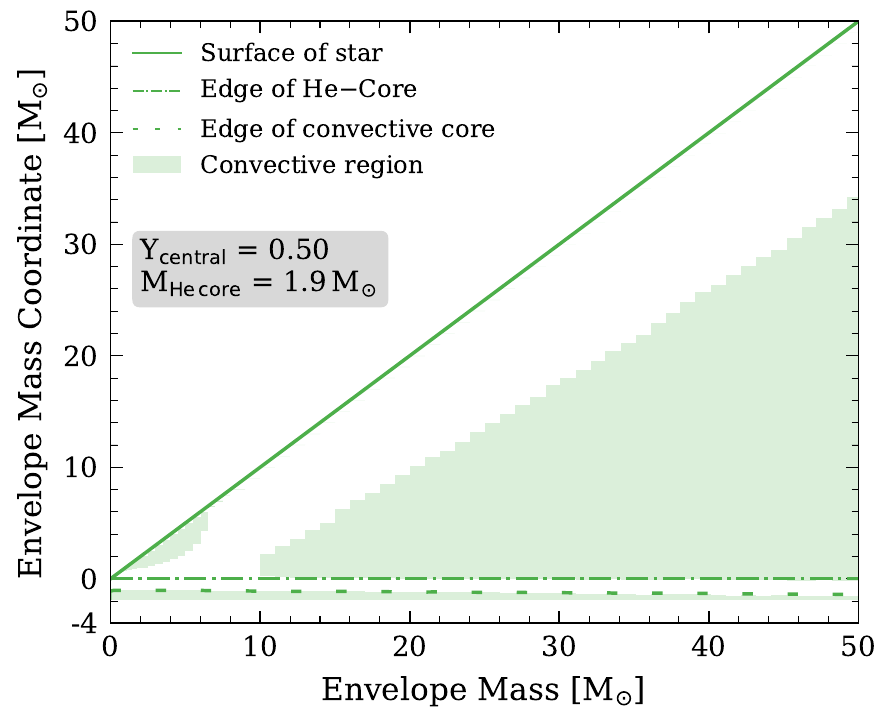}
  \caption{Kippenhahn-like diagram for core--He burning models with a constant $\mcore = 1.9\msun$ and varying envelope mass. See caption of Fig. \ref{fig:kippenhahn_example_C4D1} for further details.}
 \label{fig:appendix:kipp1.9_1}
 \end{figure}

 \begin{figure}
  \includegraphics[width=\hsize]{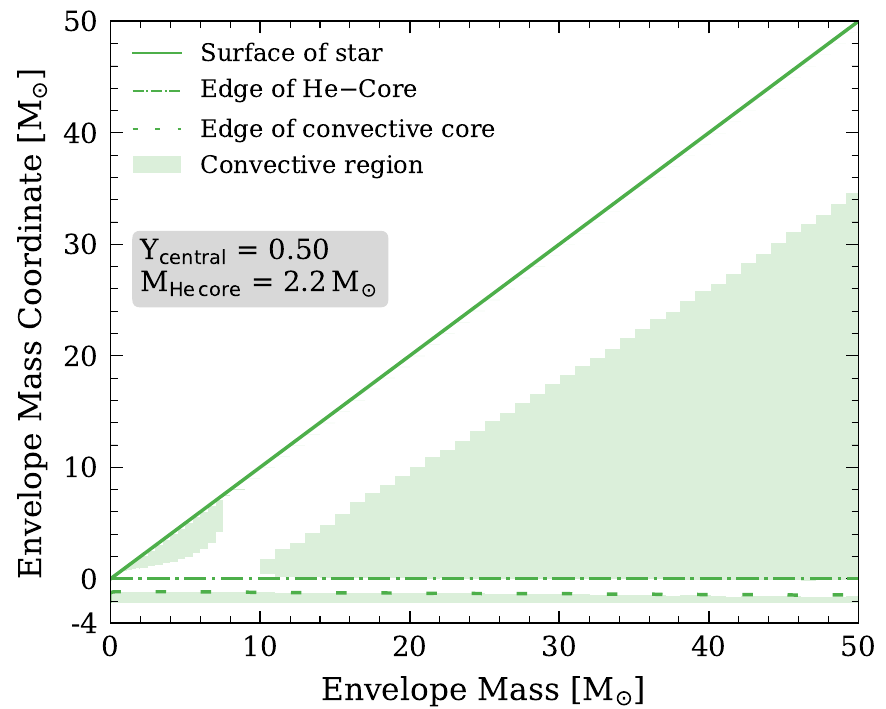}
  \caption{Kippenhahn-like diagram for core--He burning models with a constant $\mcore = 2.2\msun$ and varying envelope mass. See caption of Fig. \ref{fig:kippenhahn_example_C4D1} for further details.}
 \label{fig:appendix:kipp2.2_1}
 \end{figure}

 \begin{figure}
  \includegraphics[width=\hsize]{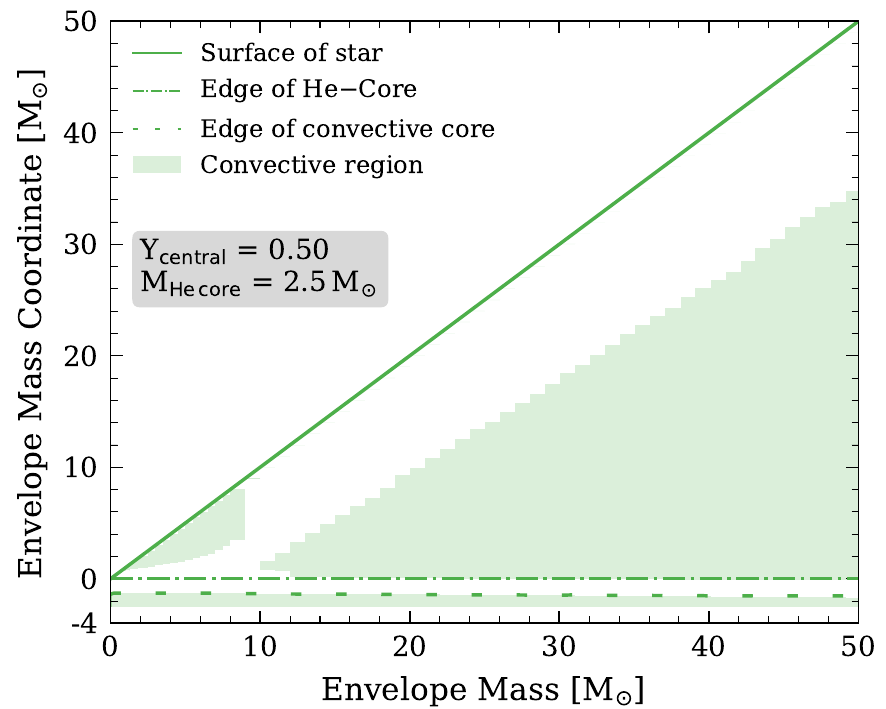}
  \caption{Kippenhahn-like diagram for core--He burning models with a constant $\mcore = 2.5\msun$ and varying envelope mass. See caption of Fig. \ref{fig:kippenhahn_example_C4D1} for further details.}
 \label{fig:appendix:kipp2.5_1}
 \end{figure}

 \begin{figure}
  \includegraphics[width=\hsize]{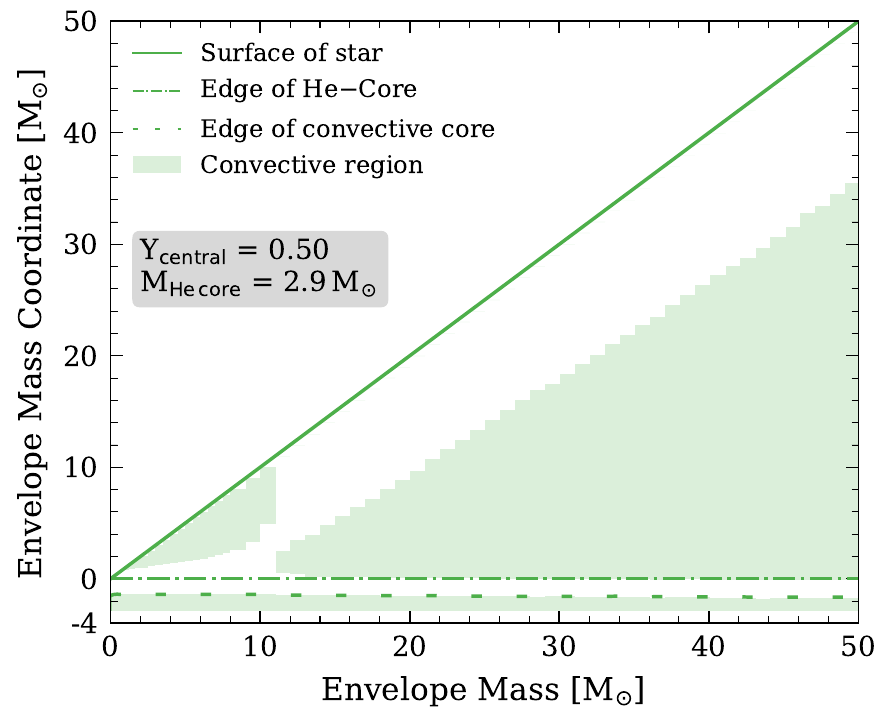}
  \caption{Kippenhahn-like diagram for core--He burning models with a constant $\mcore = 2.9\msun$ and varying envelope mass. See caption of Fig. \ref{fig:kippenhahn_example_C4D1} for further details.}
 \label{fig:appendix:kipp2.9_1}
 \end{figure}

 \begin{figure}
  \includegraphics[width=\hsize]{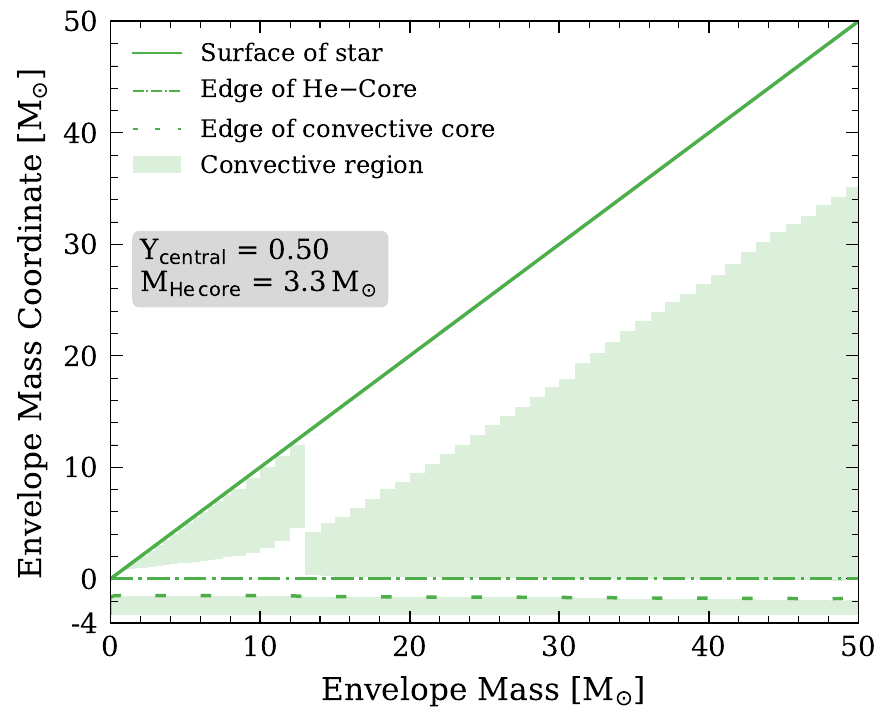}
  \caption{Kippenhahn-like diagram for core--He burning models with a constant $\mcore = 3.3\msun$ and varying envelope mass. See caption of Fig. \ref{fig:kippenhahn_example_C4D1} for further details.}
 \label{fig:appendix:kipp3.3_1}
 \end{figure}

 \begin{figure}
  \includegraphics[width=\hsize]{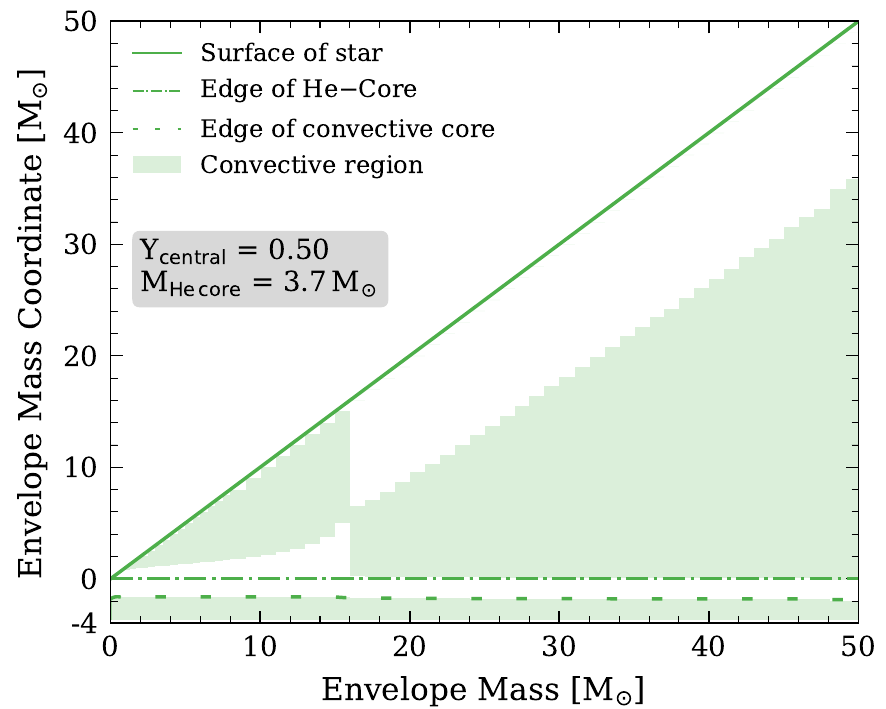}
  \caption{Kippenhahn-like diagram for core--He burning models with a constant $\mcore = 3.7\msun$ and varying envelope mass. See caption of Fig. \ref{fig:kippenhahn_example_C4D1} for further details.}
 \label{fig:appendix:kipp3.7_1}
 \end{figure}

 \begin{figure}
  \includegraphics[width=\hsize]{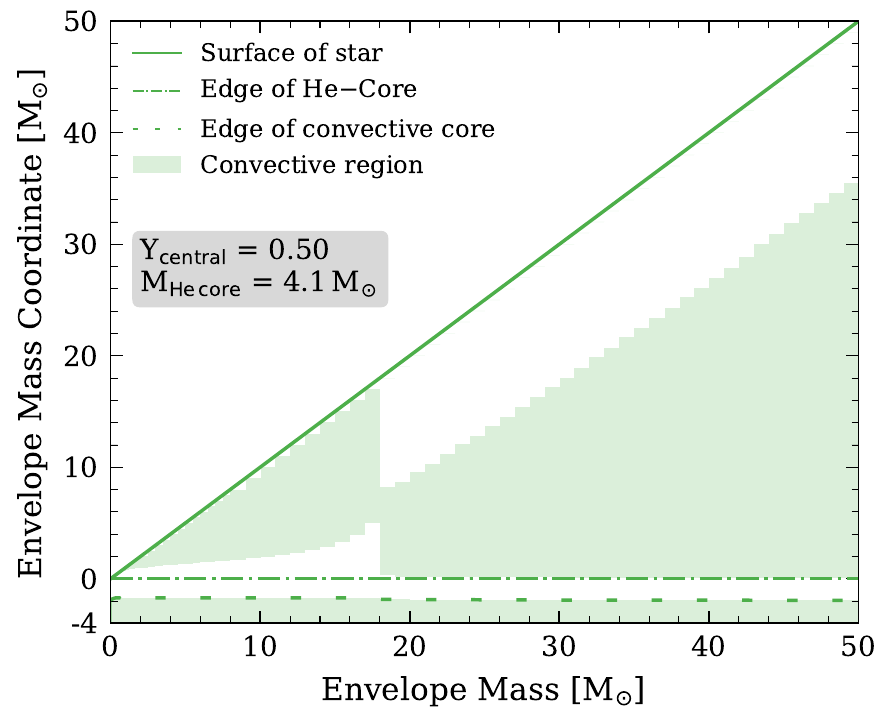}
  \caption{Kippenhahn-like diagram for core--He burning models with a constant $\mcore = 4.1\msun$ and varying envelope mass. See caption of Fig. \ref{fig:kippenhahn_example_C4D1} for further details.}
 \label{fig:appendix:kipp4.1_1}
 \end{figure}

 \begin{figure}
  \includegraphics[width=\hsize]{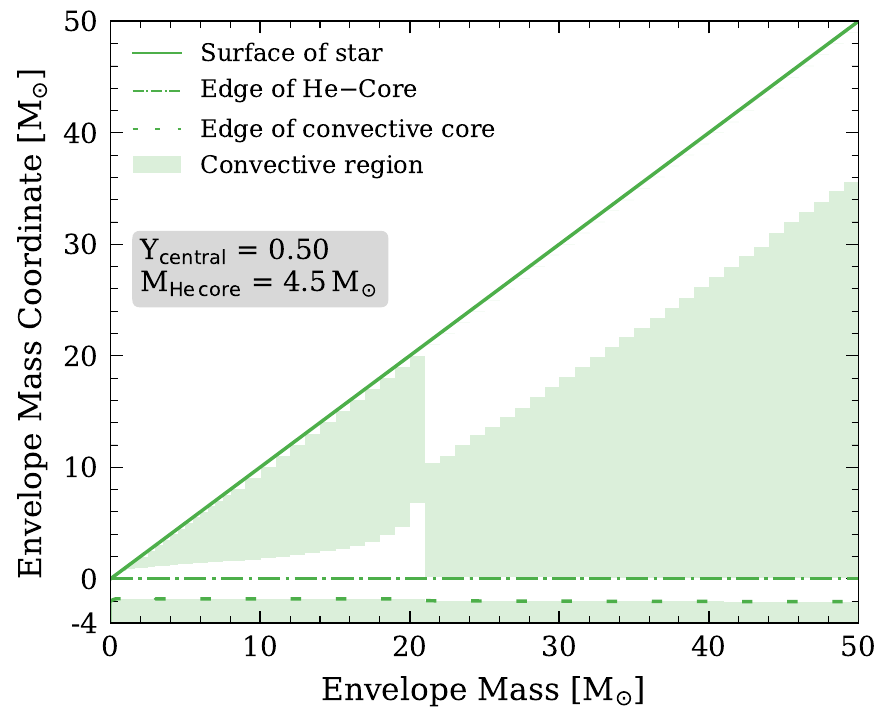}
  \caption{Kippenhahn-like diagram for core--He burning models with a constant $\mcore = 4.5\msun$ and varying envelope mass. See caption of Fig. \ref{fig:kippenhahn_example_C4D1} for further details.}
 \label{fig:appendix:kipp4.5_1}
 \end{figure}

 \begin{figure}
  \includegraphics[width=\hsize]{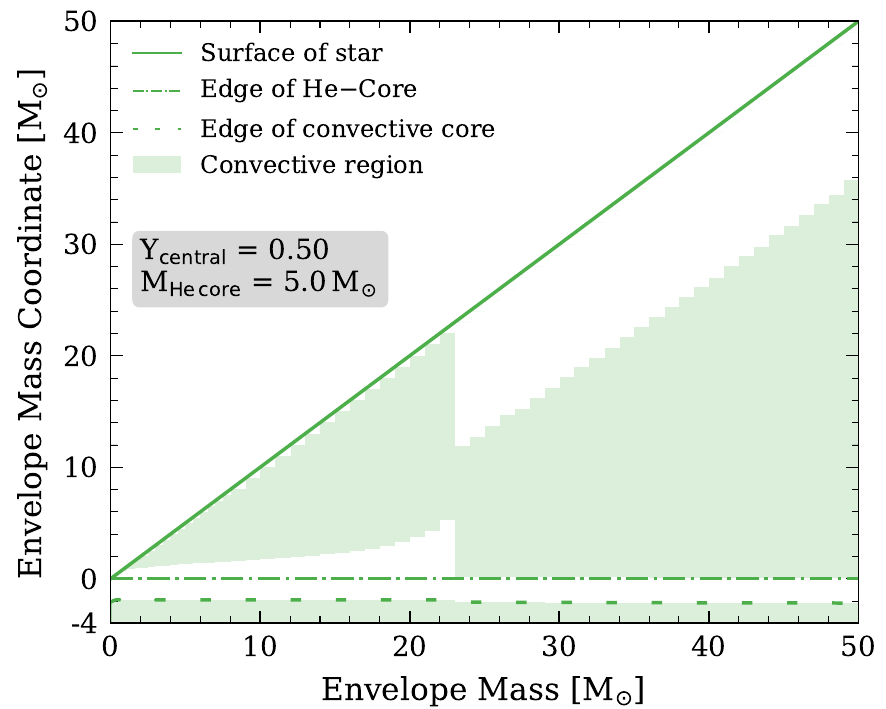}
  \caption{Kippenhahn-like diagram for core--He burning models with a constant $\mcore = 5.0\msun$ and varying envelope mass. See caption of Fig. \ref{fig:kippenhahn_example_C4D1} for further details.}
 \label{fig:appendix:kipp5.0_1}
 \end{figure}

 \begin{figure}
  \includegraphics[width=\hsize]{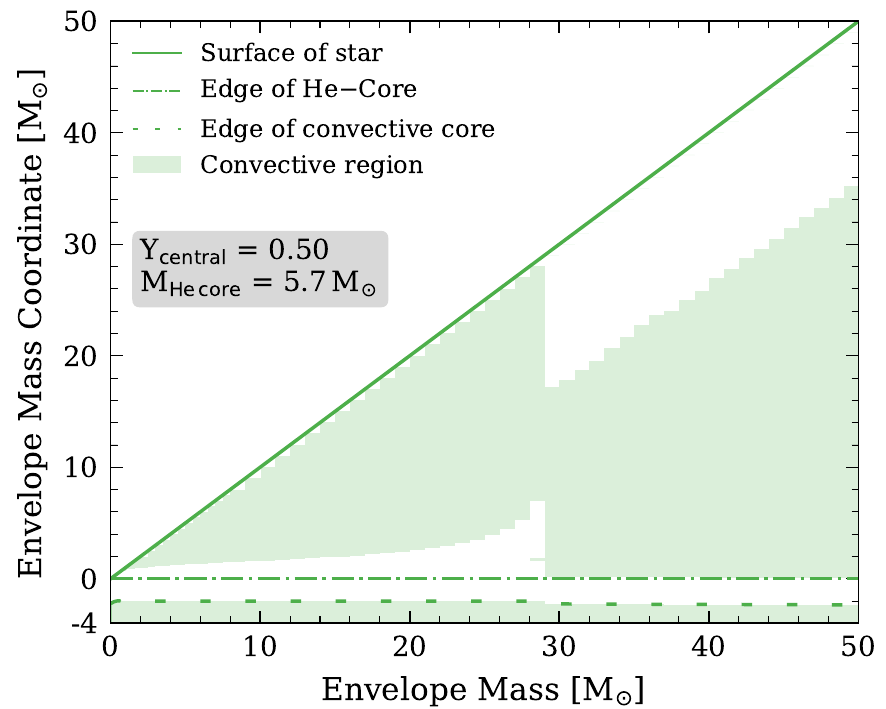}
  \caption{Kippenhahn-like diagram for core--He burning models with a constant $\mcore = 5.7\msun$ and varying envelope mass. See caption of Fig. \ref{fig:kippenhahn_example_C4D1} for further details.}
 \label{fig:appendix:kipp5.7_1}
 \end{figure}

 \begin{figure}
  \includegraphics[width=\hsize]{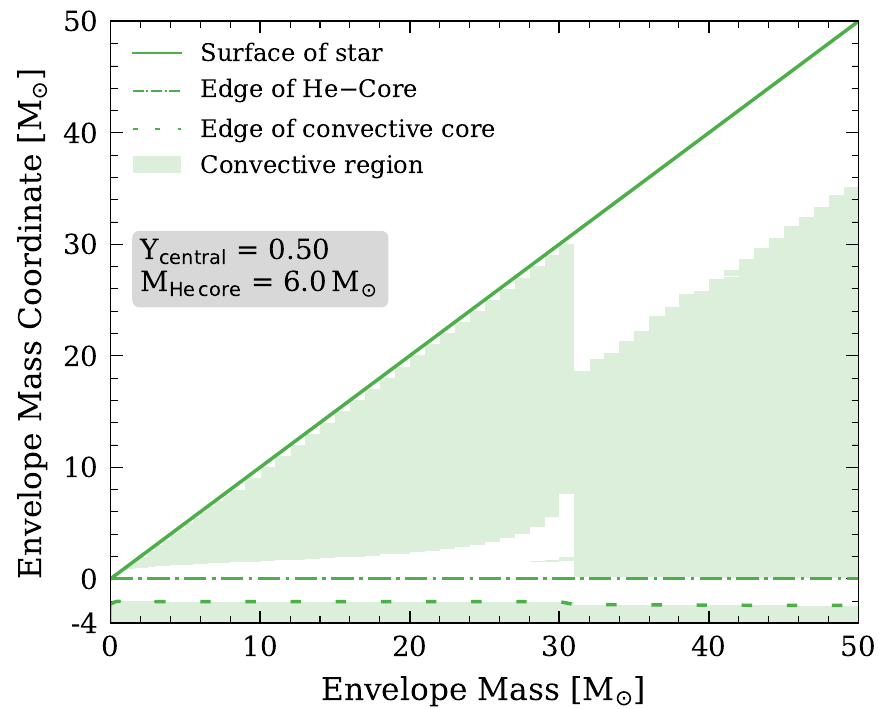}
  \caption{Kippenhahn-like diagram for core--He burning models with a constant $\mcore = 6.0\msun$ and varying envelope mass. See caption of Fig. \ref{fig:kippenhahn_example_C4D1} for further details.}
 \label{fig:appendix:kipp6.0_1}
 \end{figure}

 \begin{figure}
  \includegraphics[width=\hsize]{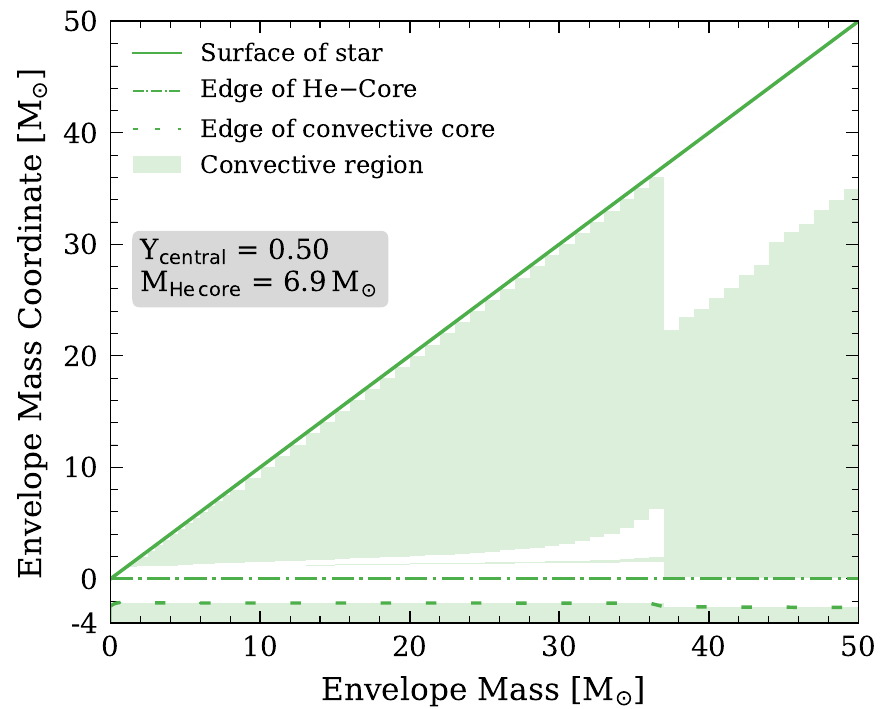}
  \caption{Kippenhahn-like diagram for core--He burning models with a constant $\mcore = 6.9\msun$ and varying envelope mass. See caption of Fig. \ref{fig:kippenhahn_example_C4D1} for further details.}
 \label{fig:appendix:kipp6.9_1}
 \end{figure}

 \begin{figure}
  \includegraphics[width=\hsize]{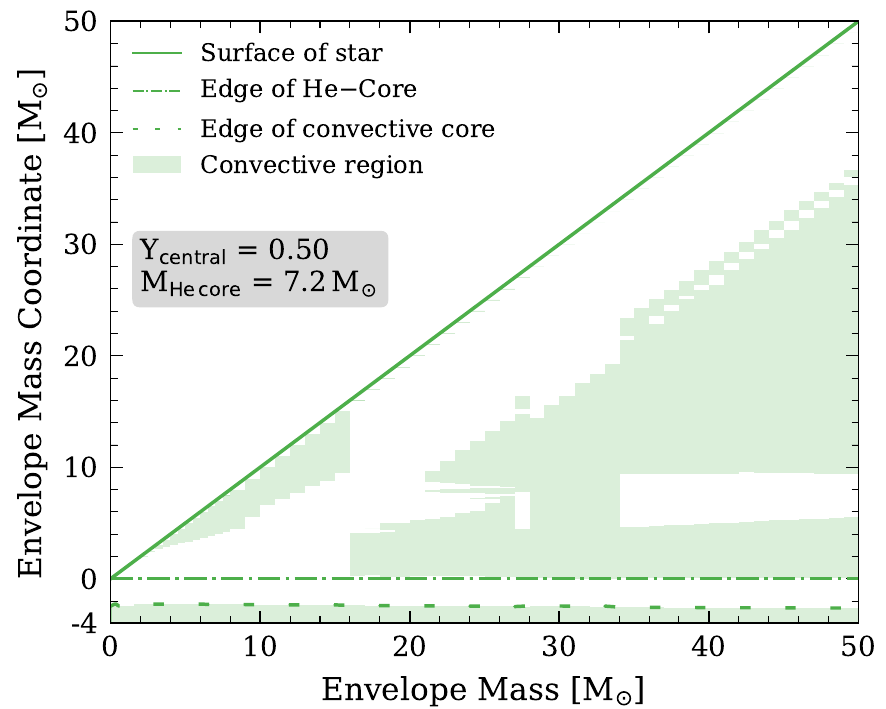}
  \caption{Kippenhahn-like diagram for core--He burning models with a constant $\mcore = 7.2\msun$ and varying envelope mass. See caption of Fig. \ref{fig:kippenhahn_example_C4D1} for further details.}
 \label{fig:appendix:kipp7.2_1}
 \end{figure}

 \begin{figure}
  \includegraphics[width=\hsize]{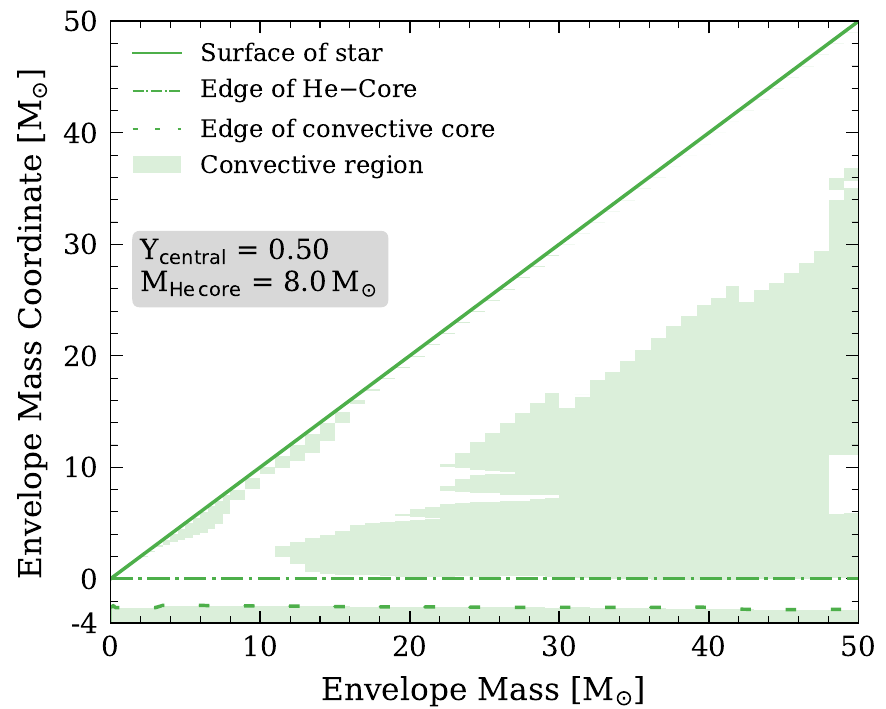}
  \caption{Kippenhahn-like diagram for core--He burning models with a constant $\mcore = 8.0\msun$ and varying envelope mass. See caption of Fig. \ref{fig:kippenhahn_example_C4D1} for further details.}
 \label{fig:appendix:kipp8.0_1}
 \end{figure}

 \begin{figure}
  \includegraphics[width=\hsize]{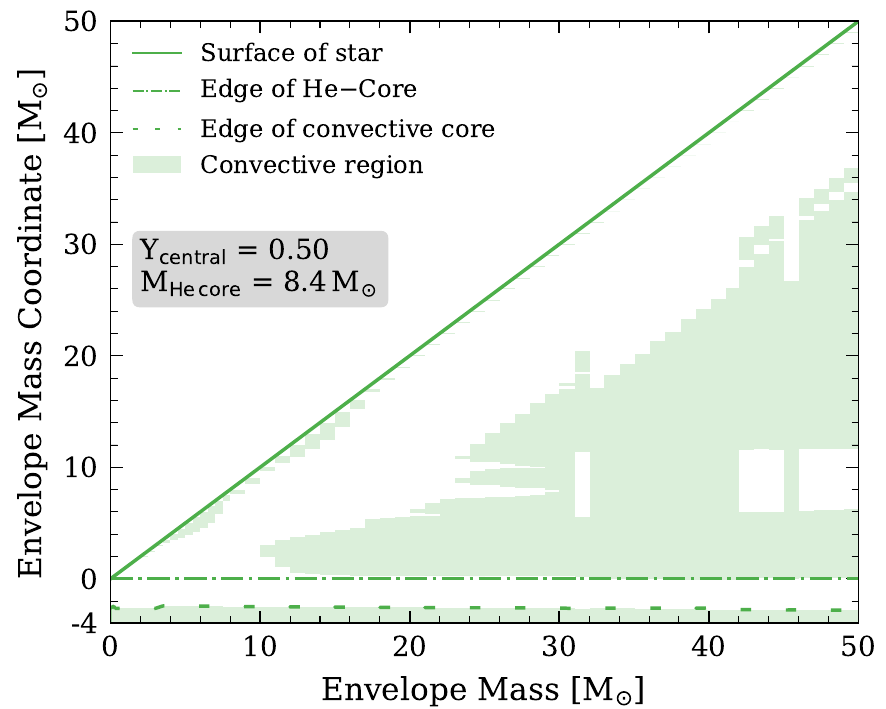}
  \caption{Kippenhahn-like diagram for core--He burning models with a constant $\mcore = 8.4\msun$ and varying envelope mass. See caption of Fig. \ref{fig:kippenhahn_example_C4D1} for further details.}
 \label{fig:appendix:kipp8.4_1}
 \end{figure}

 \begin{figure}
  \includegraphics[width=\hsize]{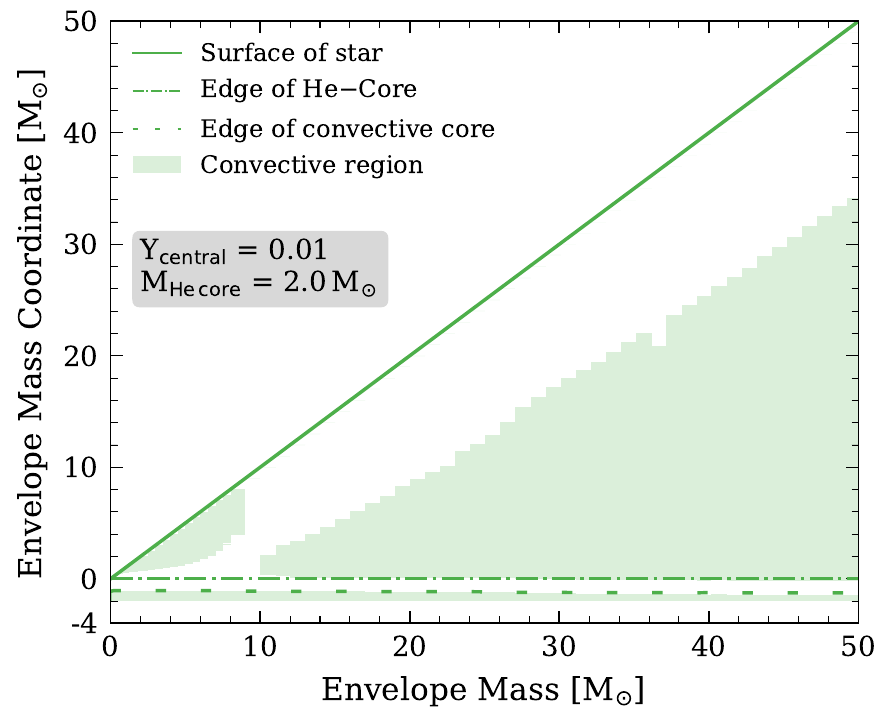}
  \caption{Kippenhahn-like diagram for core--He burning models with a constant $\mcore = 2.0\msun$ and varying envelope mass. See caption of Fig. \ref{fig:kippenhahn_example_C4D1} for further details.}
 \label{fig:appendix:kipp2.0_2}
 \end{figure}

 \begin{figure}
  \includegraphics[width=\hsize]{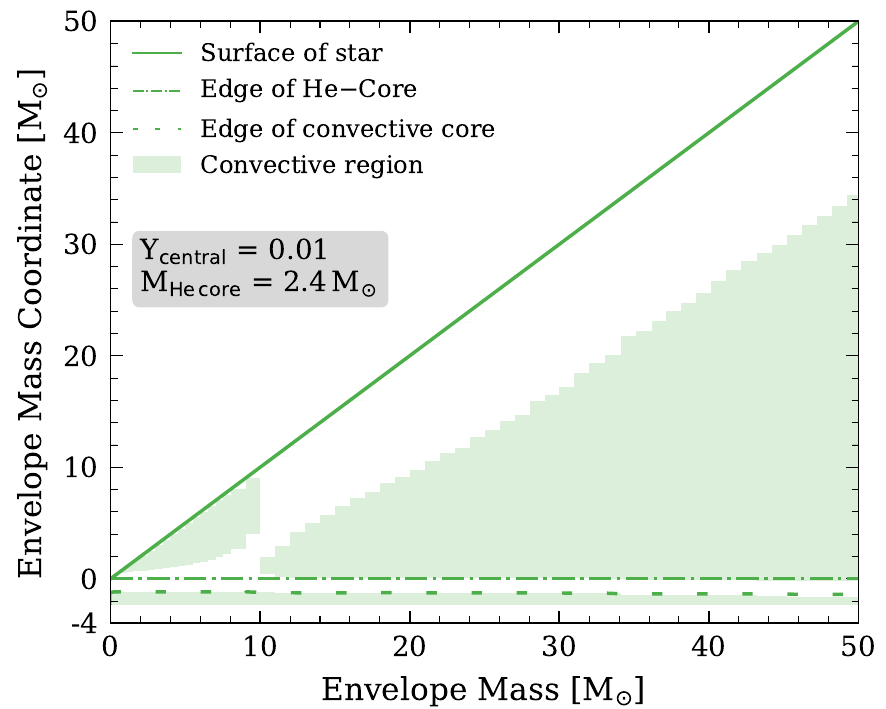}
  \caption{Kippenhahn-like diagram for core--He burning models with a constant $\mcore = 2.4\msun$ and varying envelope mass. See caption of Fig. \ref{fig:kippenhahn_example_C4D1} for further details.}
 \label{fig:appendix:kipp2.4_2}
 \end{figure}

 \begin{figure}
  \includegraphics[width=\hsize]{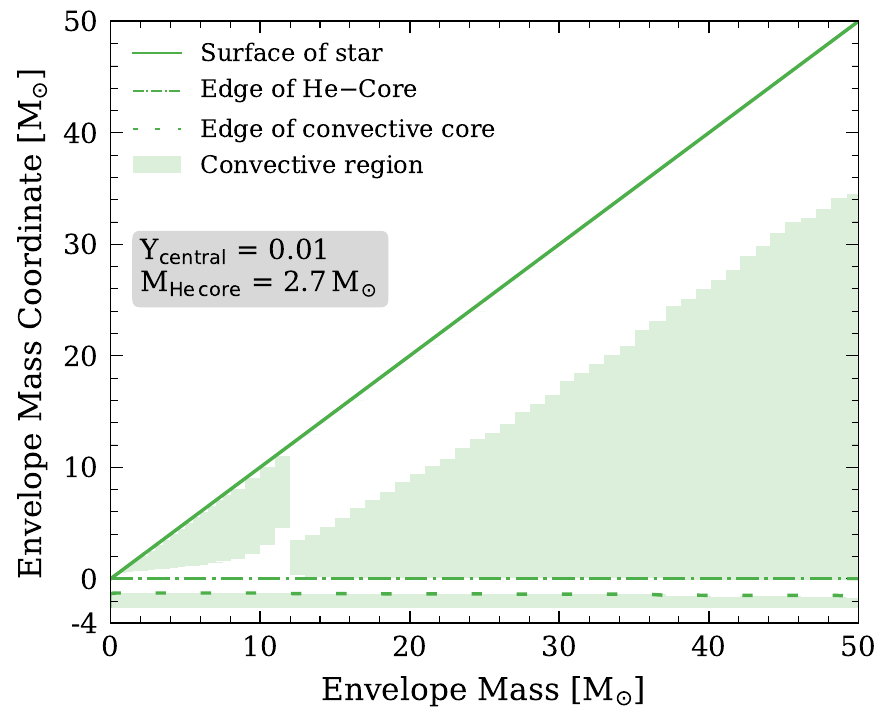}
  \caption{Kippenhahn-like diagram for core--He burning models with a constant $\mcore = 2.7\msun$ and varying envelope mass. See caption of Fig. \ref{fig:kippenhahn_example_C4D1} for further details.}
 \label{fig:appendix:kipp2.7_2}
 \end{figure}

 \begin{figure}
  \includegraphics[width=\hsize]{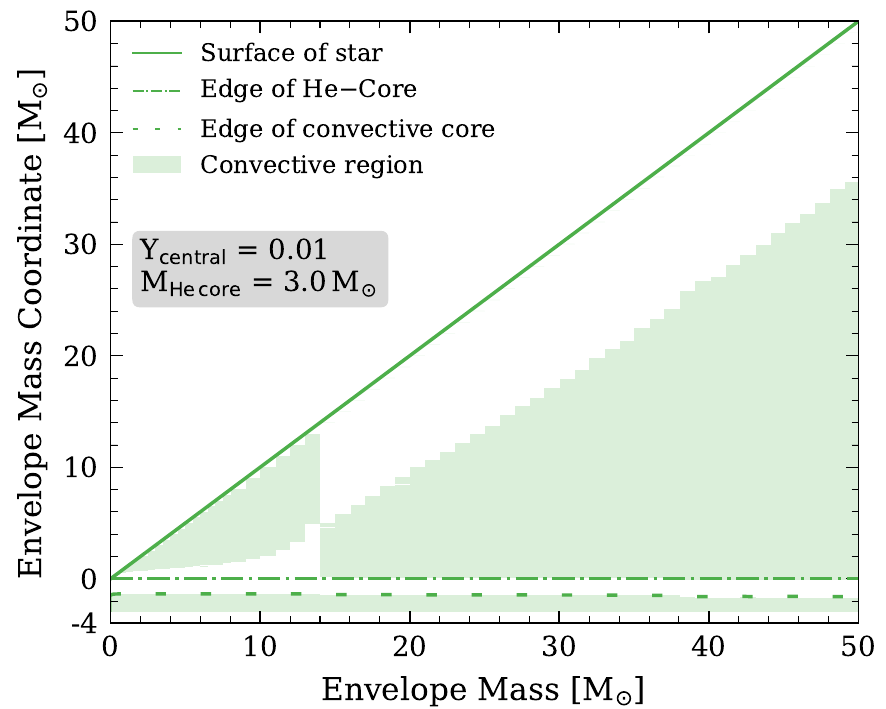}
  \caption{Kippenhahn-like diagram for core--He burning models with a constant $\mcore = 3.0\msun$ and varying envelope mass. See caption of Fig. \ref{fig:kippenhahn_example_C4D1} for further details.}
 \label{fig:appendix:kipp3.0_2}
 \end{figure}

 \begin{figure}
  \includegraphics[width=\hsize]{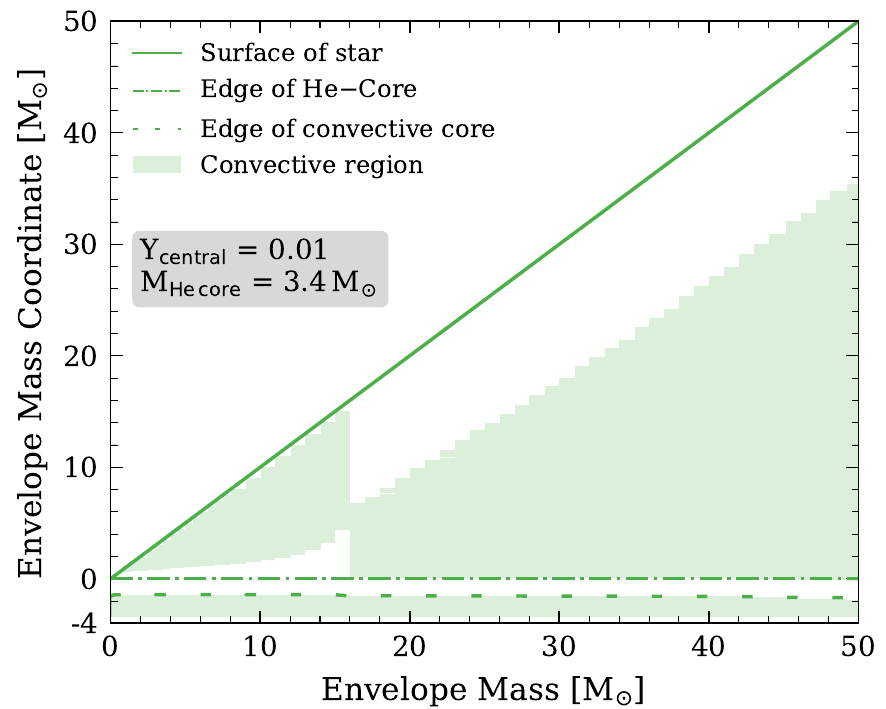}
  \caption{Kippenhahn-like diagram for core--He burning models with a constant $\mcore = 3.4\msun$ and varying envelope mass. See caption of Fig. \ref{fig:kippenhahn_example_C4D1} for further details.}
 \label{fig:appendix:kipp3.4_2}
 \end{figure}

 \begin{figure}
  \includegraphics[width=\hsize]{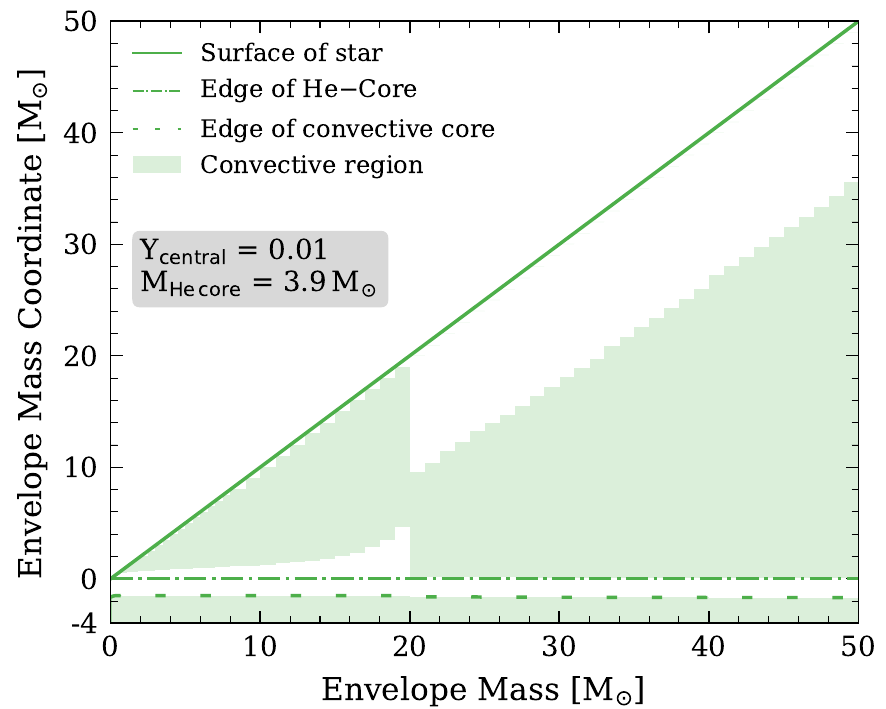}
  \caption{Kippenhahn-like diagram for core--He burning models with a constant $\mcore = 3.9\msun$ and varying envelope mass. See caption of Fig. \ref{fig:kippenhahn_example_C4D1} for further details.}
 \label{fig:appendix:kipp3.9_2}
 \end{figure}

 \begin{figure}
  \includegraphics[width=\hsize]{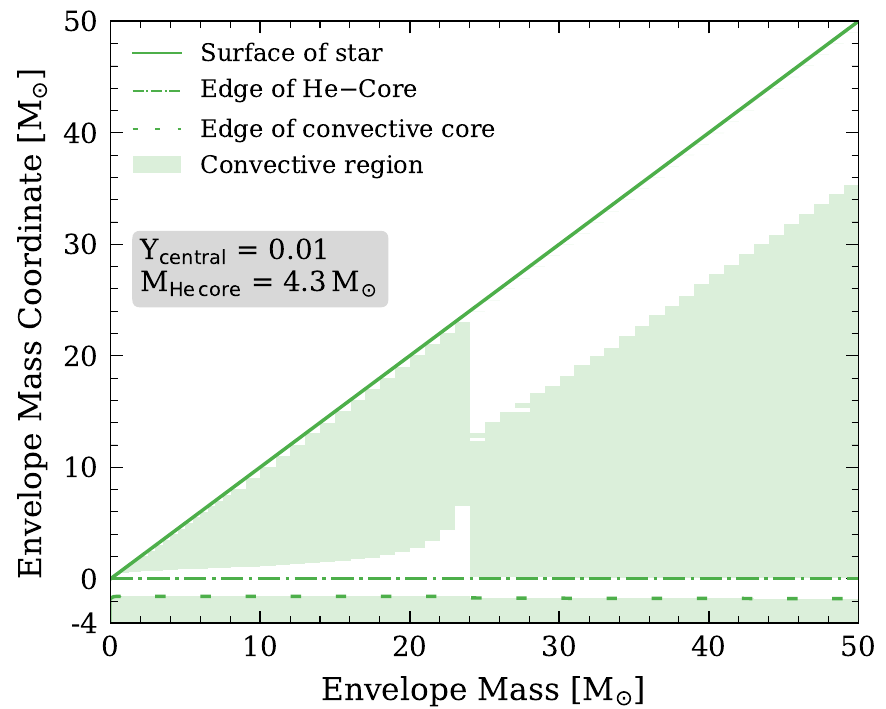}
  \caption{Kippenhahn-like diagram for core--He burning models with a constant $\mcore = 4.3\msun$ and varying envelope mass. See caption of Fig. \ref{fig:kippenhahn_example_C4D1} for further details.}
 \label{fig:appendix:kipp4.3_2}
 \end{figure}

 \begin{figure}
  \includegraphics[width=\hsize]{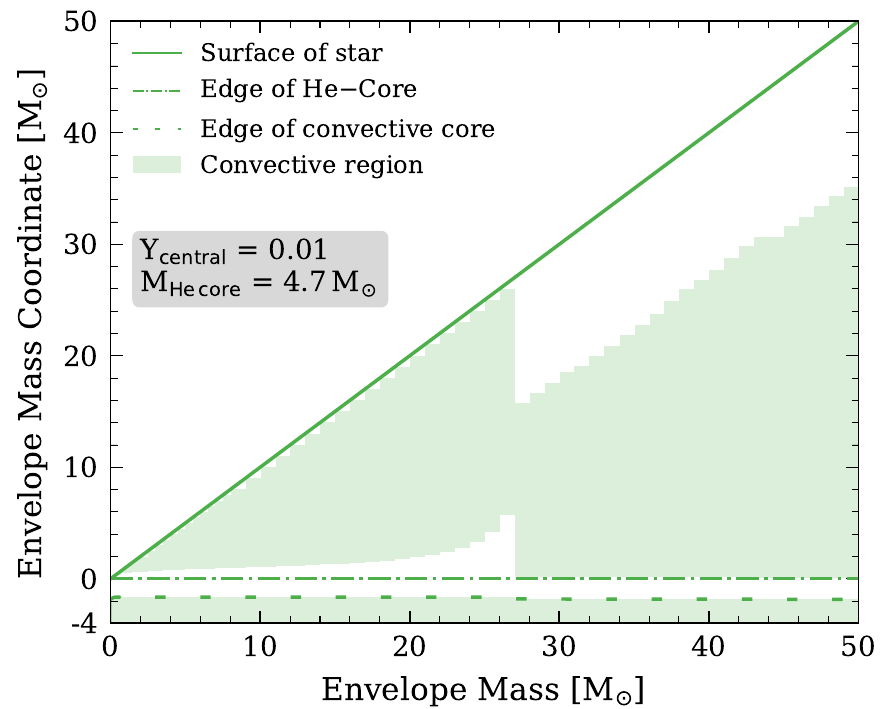}
  \caption{Kippenhahn-like diagram for core--He burning models with a constant $\mcore = 4.7\msun$ and varying envelope mass. See caption of Fig. \ref{fig:kippenhahn_example_C4D1} for further details.}
 \label{fig:appendix:kipp4.7_2}
 \end{figure}

 \begin{figure}
  \includegraphics[width=\hsize]{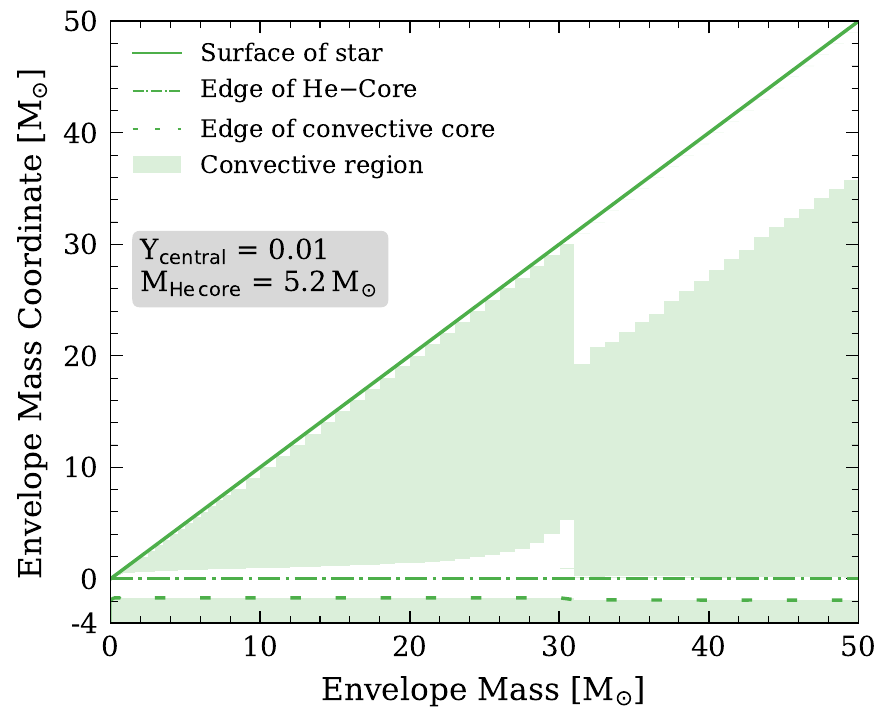}
  \caption{Kippenhahn-like diagram for core--He burning models with a constant $\mcore = 5.2\msun$ and varying envelope mass. See caption of Fig. \ref{fig:kippenhahn_example_C4D1} for further details.}
 \label{fig:appendix:kipp5.2_2}
 \end{figure}

 \begin{figure}
  \includegraphics[width=\hsize]{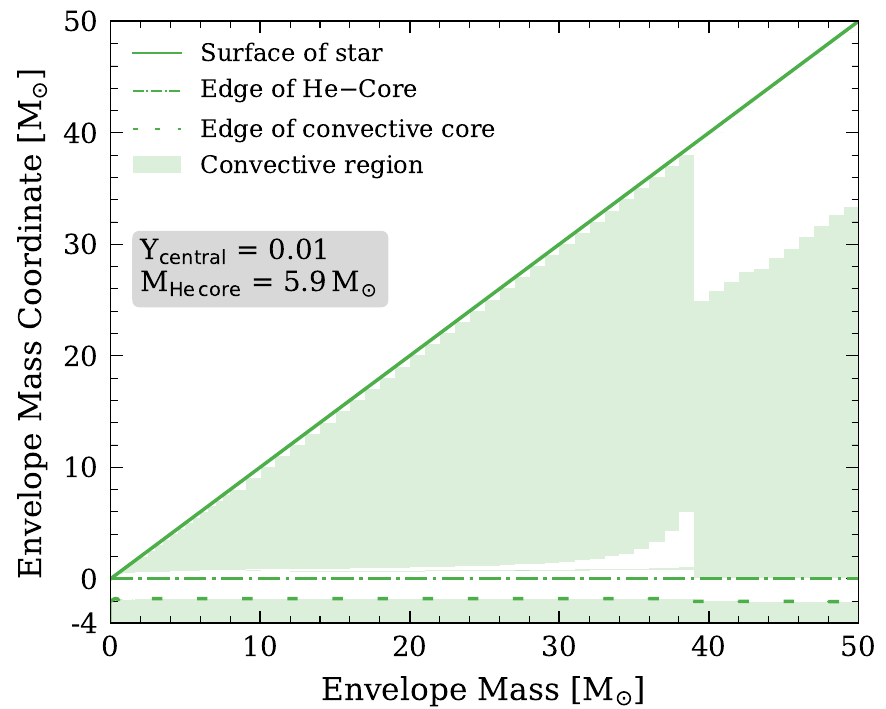}
  \caption{Kippenhahn-like diagram for core--He burning models with a constant $\mcore = 5.9\msun$ and varying envelope mass. See caption of Fig. \ref{fig:kippenhahn_example_C4D1} for further details.}
 \label{fig:appendix:kipp5.9_2}
 \end{figure}

 \begin{figure}
  \includegraphics[width=\hsize]{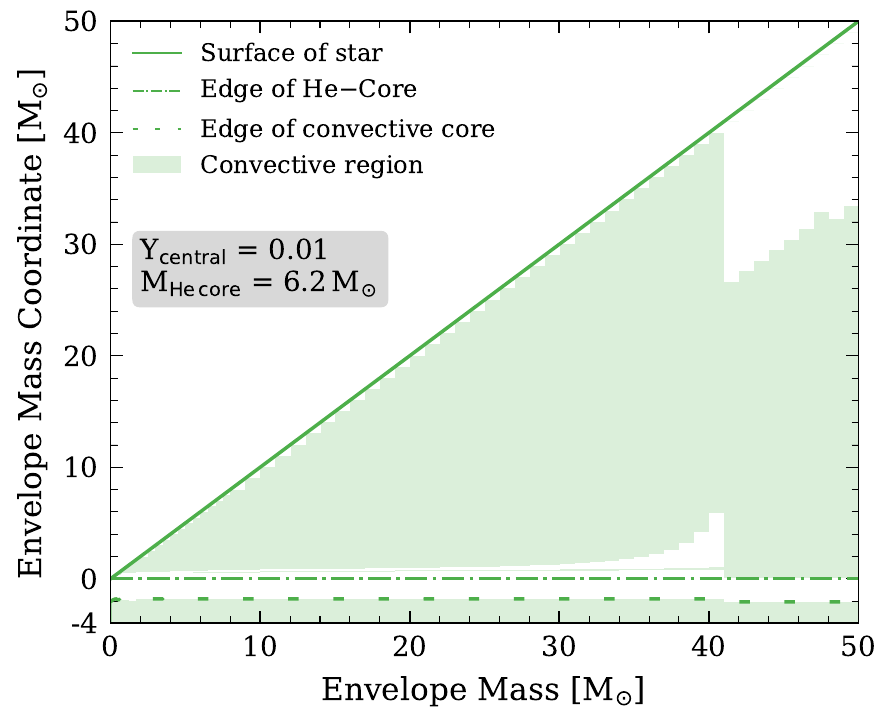}
  \caption{Kippenhahn-like diagram for core--He burning models with a constant $\mcore = 6.2\msun$ and varying envelope mass. See caption of Fig. \ref{fig:kippenhahn_example_C4D1} for further details.}
 \label{fig:appendix:kipp6.2_2}
 \end{figure}

 \begin{figure}
  \includegraphics[width=\hsize]{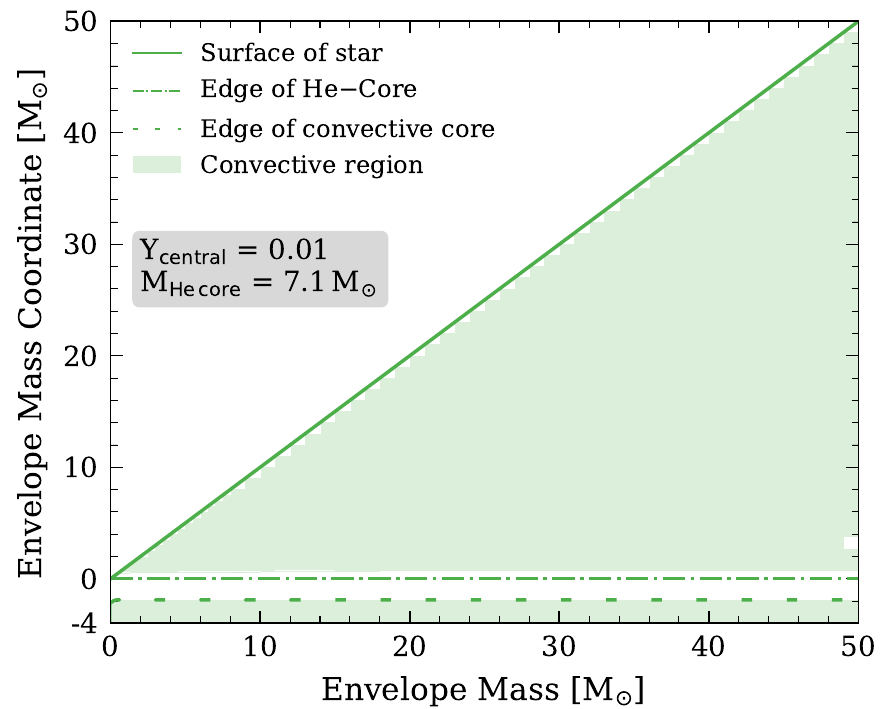}
  \caption{Kippenhahn-like diagram for core--He burning models with a constant $\mcore = 7.1\msun$ and varying envelope mass. See caption of Fig. \ref{fig:kippenhahn_example_C4D1} for further details.}
 \label{fig:appendix:kipp7.1_2}
 \end{figure}

 \begin{figure}
  \includegraphics[width=\hsize]{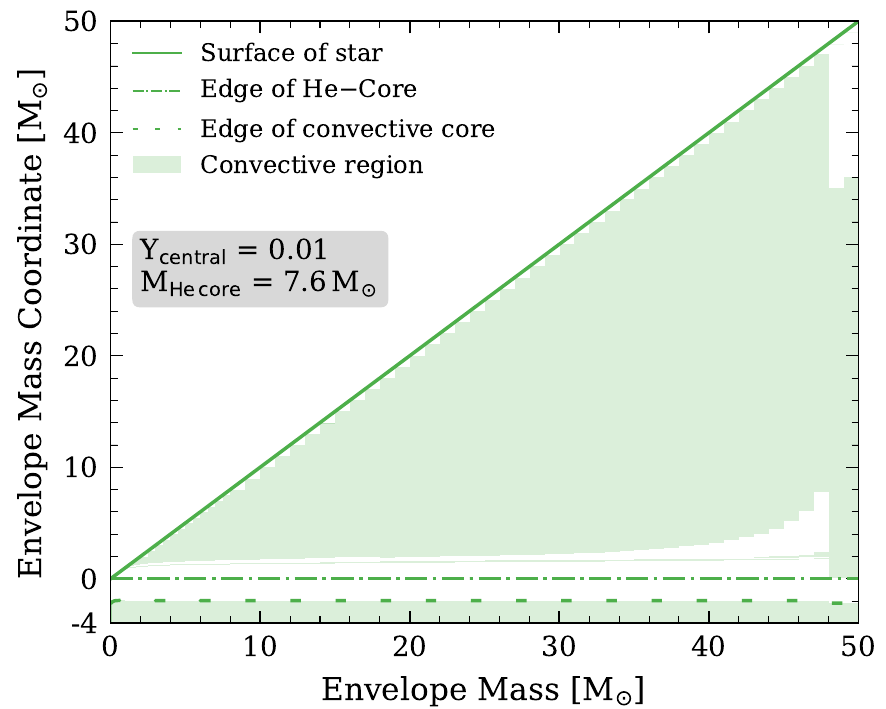}
  \caption{Kippenhahn-like diagram for core--He burning models with a constant $\mcore = 7.6\msun$ and varying envelope mass. See caption of Fig. \ref{fig:kippenhahn_example_C4D1} for further details.}
 \label{fig:appendix:kipp7.6_2}
 \end{figure}

 \begin{figure}
  \includegraphics[width=\hsize]{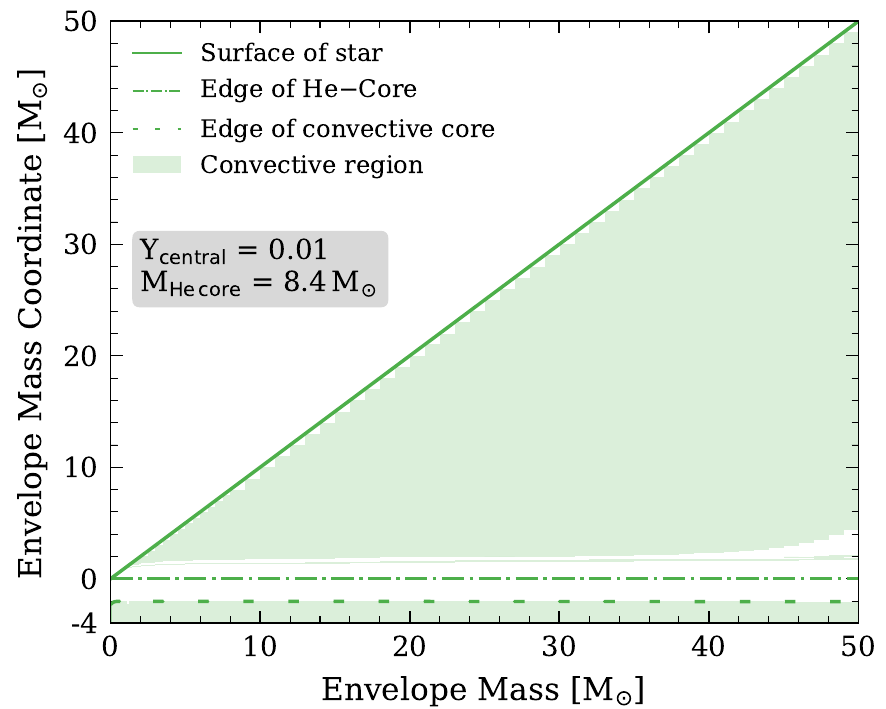}
  \caption{Kippenhahn-like diagram for core--He burning models with a constant $\mcore = 8.4\msun$ and varying envelope mass. See caption of Fig. \ref{fig:kippenhahn_example_C4D1} for further details.}
 \label{fig:appendix:kipp8.4_2}
 \end{figure}

 \begin{figure}
  \includegraphics[width=\hsize]{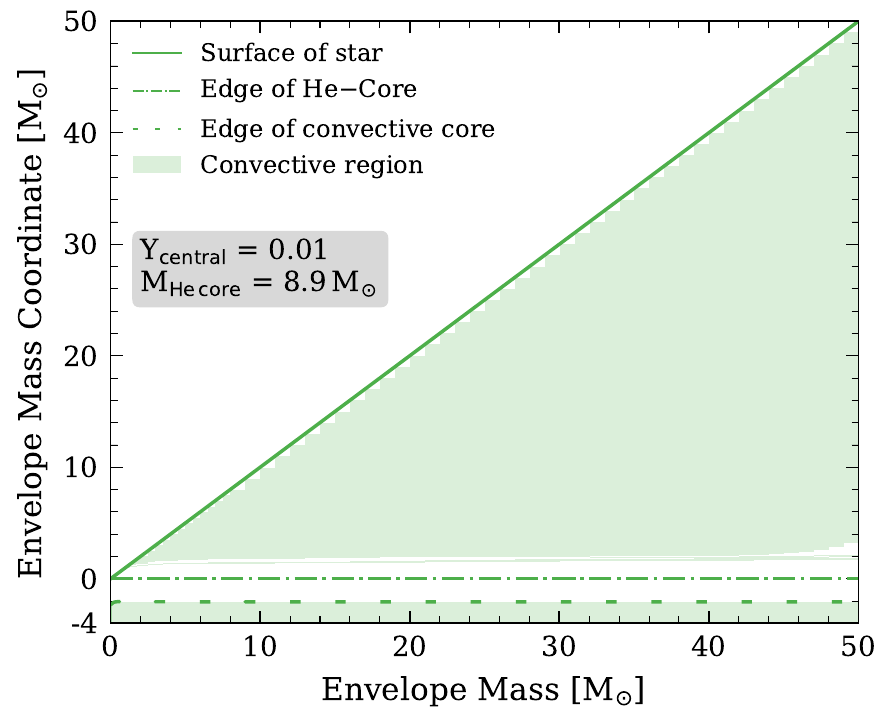}
  \caption{Kippenhahn-like diagram for core--He burning models with a constant $\mcore = 8.9\msun$ and varying envelope mass. See caption of Fig. \ref{fig:kippenhahn_example_C4D1} for further details.}
 \label{fig:appendix:kipp8.9_2}
 \end{figure}

\label{lastpage}
\end{document}